%% file: OSWS_resubmit.tex
\documentclass[twocolumn,aps,prx,superscriptaddress,reprint,longbibliography]{revtex4-1}
\usepackage[latin9]{inputenc}
\usepackage{amsmath,tikz,graphicx}
\usepackage{esint}

\usetikzlibrary{matrix}

\makeatletter
\usepackage{H1}

\newcommand{\bra}[1]{ \langle{{#1}}|}
\renewcommand{\ket}[1]{|{{#1}}\rangle}
\renewcommand{\textemdash}{\leavevmode\unskip\kern0.8pt\rule[0.19\baselineskip]{8pt}{0.4pt}\kern1pt\ignorespaces}

\makeatother

\begin{document}

\title{Diffusive hydrodynamics of out-of-time-ordered correlators with charge conservation}

\author{Tibor Rakovszky}

\affiliation{Department of Physics, T42, Technische Universit{ä}t M{ü}nchen, James-Franck-Straße 1, D-85748 Garching, Germany}

\author{Frank Pollmann}

\affiliation{Department of Physics, T42, Technische Universit{ä}t M{ü}nchen, James-Franck-Straße 1, D-85748 Garching, Germany}

\author{C.W.~von~Keyserlingk}

\affiliation{University of Birmingham, School of Physics \& Astronomy, B15 2TT,
UK}

\begin{abstract}

The scrambling of quantum information in closed many-body systems, as measured by out-of-time-ordered correlation functions (OTOCs), has lately received considerable attention. Recently, a hydrodynamical description of OTOCs has emerged from considering random local circuits. Numerical work suggests that aspects of this description are universal to ergodic many-body systems, even \textit{without} randomness; a conjectured explanation for this is that while the random circuits have noise built into them,  deterministic quantum systems, much like classically chaotic ones, ``generate their own noise'' and look effectively random on sufficient length and time scales. In this paper we extend this approach to systems with locally conserved quantities (e.g., energy). We do this by considering local random unitary circuits with a conserved U$(1)$ charge and argue, with numerical and analytical evidence, that the presence of a conservation law slows relaxation in both time ordered {\textit{and}} out-of-time-ordered correlation functions; both can have a diffusively relaxing component or ``hydrodynamic tail'' at late times. We verify the presence of such tails also in a deterministic, peridocially driven system. We show that for OTOCs, the combination of diffusive and ballistic components leads to a wave front with a specific asymmetric shape, decaying as a power law behind the front. These results also explain existing numerical investigations in non-noisy ergodic systems with energy conservation. Moreover, we consider OTOCs in Gibbs states, parametrized by a chemical potential $\mu$, and apply perturbative arguments to show that for $\mu\gg 1$ the ballistic front of information-spreading can only develop at times exponentially large in $\mu$ -- with the information traveling diffusively at earlier times. We also develop a new formalism for describing OTOCs and operator spreading, which allows us to interpret the saturation of OTOCs as a form of thermalization on the Hilbert space of operators.

\end{abstract}

\maketitle

\section{Introduction}\label{s:intro}
The question of how quantum information spreads in a closed quantum system as it approaches equilibrium via unitary time evolution has appeared in various guises in the literature of the past decade~\cite{Deutsch91,Srednicki94,ETHreviewRigol16}. While many studies focus on the buildup of entanglement between spatially separated regions~\cite{CalabreseCardy05,Alba2017,HyungwonHuse,Nahum16}, in recent years a great deal of attention has focused on different measures of the ``scrambling'' of quantum information, coming from the fields of high energy physics, condensed matter physics and quantum information theory~\cite{Hayden07,Sekino08,Brown12}. The problem of scrambling is related to the spreading of operators in the Heisenberg picture, and to the definition of ``many-body quantum chaos'' as put forward by Refs. \onlinecite{Roberts2015,Kitaev15}. These effects are captured by so-called out-of-time-ordered correlation functions, or OTOCs~\cite{Larkin69}, defined as
\begin{equation*}
\mathcal{C}^{WV}(t)=\frac{1}{2}\left\langle\left[\hat V(t),\hat W\right]^{\dagger}\left[\hat V(t),\hat W\right]\right\rangle,
\end{equation*}
where $\hat V$, $\hat W$ are two appropriately chosen operators and the expectation value is usually taken in some equilibrium state. The OTOC exhibits an initial exponential time dependence, in analogy with the exponential divergence of trajectories which defines classical chaos, in certain weakly coupled field theories~\cite{Stanford15,Stanford2016,Aleiner16,Dubail16} (and their extrapolations to the strongly correlated regime~\cite{Swingle17}) and in the Sachdev-Ye-Kitaev model~\cite{Sachdev93,Kitaev15,Polchinski16,Maldacena2016SYK}. There is, however, no clear indication of such an exponent appearing  in generic local lattice systems~\cite{Bohrdt16,Luitz17,Prosen17,Nahum17,RvK17,Cheryne}. In Ref. \onlinecite{Maldacena2016} it was shown that the growth rate of the OTOC has an upper bound which is linearly increasing with temperature and which is saturated by models that are dual to black holes. Moreover, in cases where the dynamics is local, it was found that OTOCs show a ballistic spreading with the so-called ``butterfly velocity''.

While there is a profusion of valuable numerical work on these questions, and various, often uncontrolled forays in quantum field theory, exact results are few and far between. Recent work by the authors and others~\cite{RvK17,Nahum16,Nahum17} set to examine these questions in the context of local random unitary circuits, where a number of exact results can be derived for the average behavior of OTOCs and other relevant quantities. Most prominently, the OTOCs in these circuits were found to obey a ``hydrodynamic'' equation of motion, given in terms of a biased diffusion equation. The main prediction of this formalism is that the OTOC has a light-cone structure in space time, where the light-cone itself broadens diffusively as a function of time. This prediction has been shown to hold more generally in systems without randomness, for example in deterministic ergodic Hamiltonians~\cite{Leviatan2017,Cheryne} and Floquet unitaries \cite{RvK17}.

It is striking that the OTOC behavior of random circuits agrees well with that in deterministic systems. One possible explanation follows. Consider first a classical problem, the Brownian motion of a tracer particle moving in a background of hard spheres. Even though the microscopic motion of all the particles is fully deterministic, such systems are well described by the Langevin equation, which ignores the complicated motion of the background particles, and simply replaces them with a noisy forcing term consistent with their temperature. This approach of approximating a many-body background by inserting noise ``by hand'' is widely used in studies of classical hydrodynamics~\cite{dorfman1999introduction}. Our motivation for considering random unitary circuits is similar. Consider a deterministic ergodic system, and some bipartition of the degrees of freedom. Our program is based on the conjecture that one of the components of the system can behave as noisy environment for the other, effectively inducing noisy dynamics. This conjecture is consistent with some intriguing recent studies on the coarse graining of Keldysh quantum field theories~\cite{Loganayagam17}, where it is shown that integrating out the high energy degrees of freedom in an interacting quantum field theory can induce Lindblad dynamics on the lower energy degrees of freedom (see also Ref. \onlinecite{Polonyi15}). In other words, the high energy degrees of freedom can act as a noisy bath for the lower energy degrees of freedom. Indeed, diffusion of conserved quantities, typically associated with noisy stochastic dynamics, appears in many deterministic/non-noisy interacting quantum systems and is conjectured to be generic~\cite{BLOEMBERGEN1949,DEGENNES1958,KADANOFF1963} at sufficiently high temperatures. Thus, we are motivated to consider quantum systems with noisy dynamics, like random circuits, in the hope that these tractable models describe the long wavelength, long time scale physics of deterministic ergodic systems. Understanding how and when precisely such a noisy effective description is applicable is an important direction for future research.

The goal of this work is to understand how the OTOC is affected by the presence of a conserved quantity (e.g., energy) in generic ergodic lattice systems. Our plan of attack is to assume that the above hypothesis concerning self-generated noise holds, and therefore we study this problem in the context of a one-dimensional local random unitary circuit with a conserved U(1) charge. In this setting we are able to obtain long-time numerical results on the behavior of OTOCs, as well as analytical arguments explaining their behavior. We then provide numerical evidence that the same features are also present in a system without any randomness. Our random circuit predictions also help in explaining existing numerical results in time-independent ergodic spin chains~\cite{MBLOTOC1}.

Our key results are as follows. \textbf{i)} For random circuits, we prove that on-site observables relax slowly (diffusively) on average if they overlap with the conserved charge, while they relax instantly otherwise. The result that the charge undergoes diffusion, derived for random circuits, is in agreement with the expectation that conserved densities generically diffuse in interacting ergodic systems, even those with deterministic/non-random dynamics~\cite{BLOEMBERGEN1949,DEGENNES1958,KADANOFF1963,Rosch13,Leviatan2017}. We then map the calculation of the OTOC to the evaluation of a classical partition function, which in turn allows us to simulate the system up to long times and establish numerically that \textbf{ii)} on top of the usual light cone structure understood in the case without symmetries, OTOCs also have diffusive relaxation when either of the operators involved have overlap (which we define precisely below) with the local charge density, otherwise the relaxation is exponentially fast as was the case in circuits without conserved quantities. We also provide both an analytical justification for this result and strong numerical evidence that it continues to hold in systems without noise. \textbf{iii)} Those OTOCs with diffusive relaxation have a particular algebraic space-time structure, $\sim 1/\sqrt{v_B t - x}$, well behind the front $|x|\leq v_B t$. \textbf{v)} Considering OTOCs in a Gibbs ensemble with respect to the conserved charge, parametrized by a chemical potential $\mu$, we show evidence that \textbf{i)-iii)} remain valid at small chemical potentials, while for large $\mu$ and short times $t < e^{2\mu}$ the OTOC can show a space-time structure which is diffusive, rather than ballistic.

Some comments are in order. We can explain \textbf{ii)} at a general level by rewriting the OTOC as an expectation value on a doubled version of the original Hilbert space, which we also call ``operator space''. This novel language, expressed in terms of superoperators, makes it explicit that there are two (rather than one) charge densities relevant for the dynamics of the OTOCs, which we denote as $\mathcal{L}_Q,\mathcal{R}_Q$. Whether or not the OTOC has slow diffusive relaxation can can be attributed to the diffusion of one or both of these new charge densities. As a useful aside, we show that the superoperator formalism also gives a direct interpretation of the saturation of OTOCs as a measure of thermalization on operator space. Moreover, we show that the long time saturation value of the OTOC is determined by a Gibbs like ensemble on operator space, involving $\mathcal{L}_Q$ and $\mathcal{R}_Q$. We briefly provide an alternative view of \textbf{ii)} by considering the problem of operator spreading in the presence of symmetries. Explaining \textbf{iii)} requires a direct calculation, approximating the Haar averaged OTOC. To do this, we study a circuit with gates acting on $2M$ rather than just $2$ sites. This $M$ can be used as a large parameter which allows us to better control an analytical calculation. To attack \textbf{iv)} we apply our partition function method to the case of nonzero chemical potential $\mu$, focussing on the $\mu \gg 1$ limit. The most striking feature of this is a lack of a ballistically travelling front up to times $t\sim e^{2\mu}$, which we understand by developing a perturbative expansion for the OTOCs around the $\mu=\infty$ (``zero temperature'') limit. We show that in this limit certain OTOCs can exhibit a double plateau structure, saturating to a prethermal value on a $\mathcal{O}(1)$ time scale and only reaching their expected long time values at a time scale that diverges in the $\mu\to\infty$ limit.

The remainder of the paper is organized as follows. In \secref{Sec:Random-Local-Unitary} we introduce charge-conserving random circuits and then prove charge diffusion in ~\secref{Sec:Diffusion}. In Sec.~\ref{Sec:OTOC} we turn to the discussion of out-of-time-ordered correlators. We begin by showing how to map the computation of the Haar averaged OTOC onto the evaluation of a classical partition function which can more readily be computed numerically. We summarize these numerical results for the case of $\mu=0$ in \secref{Sec:otoc_partfunc_results}, supporting assertions \textbf{ii)} and \textbf{iii)} above. We complement our discussion of the random circuit model in~\secref{Sec:Floquet} with numerical data on a non-random spin-chain, showing the same long-time tails in the relaxation of OTOCs. \secref{Sec:mu0} provides a theoretical description of these hydrodynamical tails, first in the language of operator spreading in~\secref{ss:supoptails}, and then in a superoperator formalism which we develop in~\secref{Sec:supop_language}. Then in \secref{Sec:coarsegrainedanalyis} we describe analytically the detailed $\sim 1/\sqrt{v_B t - x}$ space-time structure of certain OTOCs at late times using a ``coarse-grained'' unitary circuit. Last in \secref{Sec:FiniteMu}, we consider the behavior of OTOCs in the nonzero $\mu$ case, starting with an analytical calculation of their long time saturation values \secref{Sec:Saturation}, which we verify in~\secref{Sec:musmall}, and then by expanding around the $\mu=\infty$ limit~\secref{Sec:mubig}. We conclude in~\secref{Sec:Conclusion} with a summary and discussion.

\paragraph*{Content of appendices:} In \appref{App:Identities} we derive formulae for the average effect of a single, charge-conserving random unitary operator, which are used for the derivation of the classical partition function in~\secref{Sec:partfunc}. We use these formulae also in \appref{App:SupOpDeriv} to derive the result~\eqref{eq:Zotocapprox} for the shape of the OTOC wave front in the coarse-grained version of the circuit (introduced in~\secref{Sec:coarsegrainedanalyis}). \appref{App:Opequilibration} explains how the long-time limit of OTOCs can be interpreted as a form of thermalization on the space of operators. In~\appref{App:2pwalk} and~\appref{App:Qp_lowT} we present additional details of the calculation of OTOCs in the low filling (large chemical potential) limit.

\section{Random local unitary dynamics with a conserved charge}\label{Sec:Random-Local-Unitary}

\begin{figure}
\includegraphics[width=0.65\columnwidth]{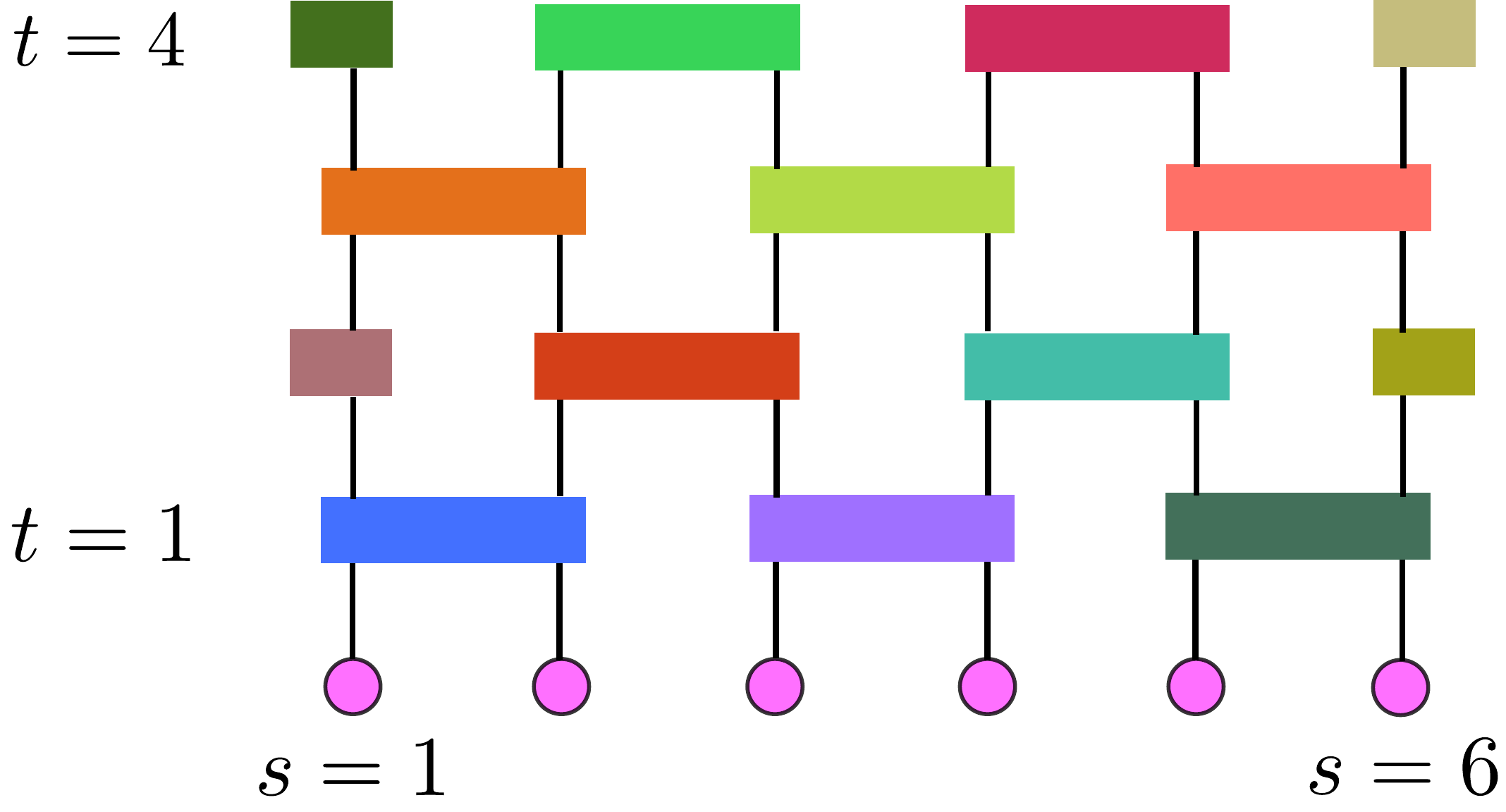}

\caption{Structure of the local unitary circuits. The
on-site Hilbert space dimension is $q$. Each two-site gate is an independently chosen $q^{2}\times q^{2}$ unitary matrix commuting with the U(1) charge $\hat Q$, defined in Eq.~\eqref{eq:charge_def}.\label{fig:circuit}}
\end{figure}

For most of the paper we focus on a random circuit with the geometry illustrated in Fig.~\ref{fig:circuit}, wherein two-site gates act in turns on the even and odd bonds of a one-dimensional spin chain. (Later in \secref{Sec:coarsegrainedanalyis} and its associated appendices we consider a slightly modified geometry, consisting of longer range unitary gates as shown by~\figref{fig:coarsegrained_circuit}). Each gate is independently chosen from an ensemble of random unitary operators, which are block-diagonal with respect to the total charge on the two sites, but Haar random within each block.  

Consider a spin system with $L$ sites and a $q$-dimensional on-site Hilbert space
$\mathcal{H}^\text{on-site}=\mathbb{C}^{q}$. We will think of these $q$ different states as corresponding to $q$ possible values of some charge, measured by the operator $\hat Q^\text{on-site} = \text{diag}\left(0,1,\ldots,q-1\right)$. We then define a global conserved charge $\hat Q$ as the sum over sites $r$ of the local charge density $\hat Q_r$, given by
\begin{align}\label{eq:charge_def}
\hat{Q} \equiv \sum_{r=1}^{L}\hat{Q}_{r}; & & \hat{Q}_{r} \equiv \bigotimes_{s \neq r} 1\!\!1^\text{on-site}_{s} \otimes \hat{Q}^\text{on-site}_r,
\end{align}
where $1\!\!1^\text{on-site}_{s}$ is a local identity operator acting on site $s$, and $\hat{Q}^\text{on-site}_r$ is the on-site charge operator on site $r$.

The random circuit model is defined as follows. Consider a
discrete time evolution, consisting of layers of two-site unitary
gates acting on pairs of neighboring sites in the chain. Odd numbered layers act
on all the odd bonds of the chain while even numbered layers act on
even bonds. Each two-site gate is chosen independently from the Haar
distribution over $q^{2}\times q^{2}$ unitary matrices which commute
with $\hat{Q}$. In practice, this means that the two site unitary
$U_{r,r+1}$, acting on sites $r,r+1$, is block diagonal with respect
to $\hat{Q}_{r}+\hat{Q}_{r+1}$, and each of the blocks is Haar random.
With the definition of $\hat{Q}$ given above, the block structure of
such a two-site unitary is $U_{r,r+1}=\bigoplus_{Q=0}^{2(q-1)}U_{Q}$
where $U_{Q}$ is a Haar random unitary acting on $\mathcal{H}_Q$, the $d_{Q}\equiv \text{dim}(\mathcal{H}_Q) = q-\left|Q+1-q\right|$ dimensional space of states on sites $r$, $r+1$ that have total charge $Q$. For example for $q=2$ it has the form
\begin{equation*}
U_{r,r+1}=
\begin{tikzpicture}[baseline=(current bounding box.center)]%
\matrix[matrix of math nodes,
inner sep=0,
nodes={draw,outer sep=0,inner sep=2pt},
every left delimiter/.style={xshift=1ex},
every right delimiter/.style={xshift=-1ex},
left delimiter={(},right delimiter={)},
column sep=-\pgflinewidth,row sep=-\pgflinewidth] (r) {
	|[inner sep=1.5mm]|Q=0&&\\
	&|[inner sep=3mm]|Q=1&\\
	&&|[inner sep=1.5mm]|Q=2\\};
\end{tikzpicture},
\end{equation*}
where the first and last blocks are $1\times 1$ and the second block is a $2\times 2$ Haar-random untary.

The time evolution after an even number of $2t$ layers is given by
\begin{align}\label{eq:fullcircuit}
U(t) & =\prod_{\tau=1}^{2t}\prod_{x=1}^{L/2}U_{2x-1+n_{\tau},2x+n_{\tau}}(\tau),
\end{align}
where $n_{\tau}=\frac{1+(-1)^{\tau}}{2}$ and each of the unitaries $U_{r,r+1}(\tau)$, labeled by the pair of sites they act on as well as the layer/time label $\tau$, is an independent random matrix chosen from the charge conserving (i.e., block diagonal) random ensemble defined above. The product
$\prod_{\tau=1}^{2t}$ is defined to be time ordered. The geometry of such a circuit is graphically illustrated in Fig.~\ref{fig:circuit}. We denote averages over the different circuit realizations by $\overline{(\ldots)}$.

\section{Charge diffusion and time ordered correlators}\label{Sec:Diffusion}

Before attacking the problem of OTOCs, we begin our study of the charge-conserving random circuit by showing rigorously that the density of conserved charge diffuses in this system when averaged over many realizations of the circuit. This is analogous to the diffusive behavior observed~\cite{Rosch13,Huse06,Bohrdt16,Leviatan2017} in generic interacting many-body systems with a few (in our case a single) global conserved quantity in the regime of incoherent transport. We show the diffusive spreading by directly considering the time evolution of the local charge operator $\hat Q_r$ in the Heisenberg picture, and discuss how it appears in time-ordered correlation functions. 

To understand the dynamics of $\hat Q_r$ let us first understand how a single two-site gate, acting on sites $r$ and $r+1$, evolves a generic operator $\hat O$ acting on the same sites. After applying a single two-site random charge-conserving gate on these two sites the operator becomes, on average (see also~\appref{App:Identities})
\begin{align}\label{eq:onestep_opavg}
\overline{\hat O(\Delta\tau)} & =
 \overline{\sum_{Q,Q'}\hat P_{Q}^{\phantom{\dagger}}U_{Q}^{\phantom{\dagger}}\hat P_{Q}^{\phantom{\dagger}}\hat{O} \hat P_{Q'}^{\phantom{\dagger}}U_{Q'}^{\dagger}\hat P_{Q'}^{\phantom{\dagger}}} \nonumber\\
 & =\sum_{Q}\frac{1}{d_{Q}}\hat P_{Q}\text{tr}\left(\hat O\hat P_{Q}\right),
\end{align}
where $\hat P_{Q}$ projects onto the sector of the two site Hilbert space with $\hat{Q}=Q$ and we used the fact that $U$ decomposes into blocks $U_Q$, each of which is Haar random. We use $\Delta \tau$ as shorthand for time evolution with a single layer of the random circuit. The diffusion of charge density follows from this algebraic result, but a more elementary argument goes as follows.  Note that $\overline{\hat{Q}(\Delta \tau)}=\hat{Q}$ because the Haar ensemble commutes with the total charge on two sites. On the other hand, the ensemble of two-sites gates is invariant under multiplication by the operator swapping sites $r,r+1$, so $\overline{\hat{Q}}_{r}(\Delta \tau)=\overline{\hat{Q}}_{r+1} (\Delta \tau)$.  This allows us to write
\begin{equation}\label{eq:randomwalk}
\overline{\hat{Q}_{r}\left(\Delta \tau\right)} = \frac{1}{2}\left(\hat{Q}_{r}+\hat{Q}_{r+1}\right).
\end{equation}

Let us iterate the above formula for a series of two-site gates arranged in the regular gate geometry shown in Fig.~\ref{fig:circuit}. The local charge operator performs a random walk, such that at each application of a two-site gate it ends up on either of the two sites with equal probabilities. It is readily verified that after an even number
$2t$ layers of the circuit it becomes
\begin{equation}\label{eq:diffusiononlattice}
\overline{\hat{Q}_{r}(t)}=\frac{1}{2^{2t}}\sum_{k=0}^{2t-1}{2t-1 \choose k}\left(\hat{Q}_{2j-2t+2k}+\hat{Q}_{2j+1-2t+2k}\right),
\end{equation}
where $j=\left\lfloor \frac{r+1}{2}\right\rfloor $. At large times,
the right hand side behaves like an unbiased diffusion kernel. Note that summing the equation over all $r$ gives $\hat{Q}(t)=\hat{Q}(0)$, which is the global conservation law. 

An approximate continuum formulation of the above discrete operator
equation is 
\begin{equation}\label{eq:diffusion}
\partial_{t}\overline{\hat{Q}}(x,t)=D\partial_{x}^{2}\overline{\hat{Q}}(x,t),
\end{equation}
where $D$ is a constant independent of $q$~\cite{DiffusionNote}. Hence, on average, the
local charge density obeys diffusive dynamics. In this sense our random circuit model can be thought of as a toy model for a many-body system in the regime of incoherent, diffusive transport. Such behavior is expected also in clean systems at times longer than the coherence time of charged quasi-particle excitations~\cite{BLOEMBERGEN1949,DEGENNES1958,KADANOFF1963} (which can be very short, for example at high temperatures~\cite{Bohrdt16}), or in systems that do not possess well defined quasi-particles at all~\cite{Hartnoll2016}. Note that in our case the diffusion of charge appears directly at the level of operators, without having to refer to any particular state, indicating incoherent charge transport over all time scales. This is consistent with the behavior of the single-particle Green's function, $\langle \hat \sigma^{-}_0(t) \hat \sigma^{+}_r\rangle$, where $\hat \sigma_r^+$ is the operator creating a single charge on site $r$. Applying formula \eqnref{eq:onestep_opavg} shows that this Green's function vanishes on average after only a single time step, independently of the state chosen, which is another way of saying that there is no coherent charge transport.

We have shown that the local charge density relaxes diffusively. As a result, at the longest times ($t>L^2/D$) the charge density becomes uniform in the system. Off-diagonal operators, on the other hand, equilibrate immediately to zero on average. Both of these statements imply that for the purposes of calculating on-site expectation values, the system thermalizes to a Gibbs ensemble of the form 
\begin{equation}\label{eq:Gibbs_def}
\hat\rho_{\mu}= e^{-\mu \hat Q} / \text{tr}\left(e^{-\mu \hat Q}\right),
\end{equation} 
where $\mu$ is determined by the charge density of the initial state. It is similarly possible to argue that more complicated many-body operators eventually equilibrate to a value deterined by same ensemble on average (we leave the proof of this to future work). Using this ensemble we can also make contact with more conventional definitions of the diffusion constant~\cite{Bohrdt16}, given by the autocorrelation function $\left\langle \hat{Q}_{r}(t)\hat{Q}_{r}\left(0\right)\right\rangle_\mu - \langle \hat{Q}_{r}\rangle^2_\mu$ in the above Gibbs state. This correlator captures the relaxation of charge to the equilibrium value. Applying the solution Eq.~\eqref{eq:diffusiononlattice} we find that it behaves at long times as
\begin{equation*}
\overline{\braket{\hat{Q}_{r}(t)\hat{Q}_{r}(0)}_\mu - \braket{\hat{Q}_{r}}^2_\mu} \approx \frac{1}{\sqrt{\pi t}}\frac{1}{\left(2\cosh\frac{\mu}{2}\right)^{2}}=\frac{1}{\sqrt{\pi D(\mu)t}}.
\end{equation*}
The last equation defines an effective diffusion constant $D(\mu)$ which singles out an effective time scale for charge relaxation, $t_D \propto 1/D(\mu)$ with $D(\mu) = 4 \cosh^4{\mu/2}$.

\section{Out-of-time-ordered correlators}\label{Sec:OTOC}

We now turn to the description of out-of-time-ordered correlators (OTOCs) in the charge conserving random circuit. Such quantities are a measure of the spreading of quantum information in many-body systems~\cite{Larkin69,Shenker13,Kitaev15,Stanford15,Stanford2016,Swingle17,Aleiner16,Dubail16,Bohrdt16, Luitz17}.
For translation invariant systems they have been studied in
weakly coupled~\cite{WeaklyCoupledNote} local quantum field theories ~\cite{Stanford2016,Swingle17,Patel17,
Aleiner16,Kivelson17}, in models for black hole scrambling~\cite{Kitaev15,Hosur2016,Maldacena2016} and more recently in local random circuits~\cite{Nahum17,RvK17}. In all these studies it was found that the OTOC exhibits ballistic behavior with a linearly moving front, behind which it saturates to an $\mathcal{O}(1)$ value, even in cases where conventional (i.e., time-ordered) correlators behave diffusively. In this regard the OTOC is more similar to measures of quantum information, such as entanglement~\cite{HyungwonHuse}, rather than to usual correlation functions. 

In lattice systems the OTOC can be understood as a measure of `operator spreading', i.e., how simple product-operators become superpositions of many such operators under time evolution (also resulting in the growth of operator entanglement~\cite{Prosen07,DavidNote}). In Refs.  \onlinecite{Nahum17,RvK17} it was shown that the behavior of the OTOC in spin chains can be understood in terms of a hydrodynamic description, taking the form of a biased diffusion equation in 1D, which gives rise to the aforementioned ballistic front, albeit with a front that itself broadens in time diffusively. This description was shown to hold exactly on average for random circuits without symmetries and it was conjectured to remain valid as an effective description in other chaotic systems at sufficiently large time and length scales, evidence of which has been observed numerically~\cite{RvK17,Leviatan2017}.

One question of great interest is how the behavior of OTOCs changes with temperature. On general grounds it is expected that at higher temperatures many-body systems behave more chaotically as there is effectively more states to scramble over. In Ref.  \onlinecite{Maldacena2016} a temperature dependent upper bound was derived for the growth rate of OTOCs which is known to be saturated in certain holographic models. In more generic systems, however, not much is known about the detailed dependence of out-of-time-ordered correlators on temperature~\cite{Bohrdt16}.

While temperature is not well-defined for the random circuits we study, due to lack of energy conservation, it is plausible that the chemical potential $\mu$ can play a similar role, setting the equilibrium entropy density of the system and thus effectively limiting the size of the Hilbert space available for the dynamics. For example the $\mu\to\infty$ projects it down to a single stationary state, analogous to $T\to 0$ in conventional systems. On the other hand, $\mu\to 0$ is equivalent to the $T\to \infty$ infinite temperature limit. We therefore define the out-of-time-ordered correlator between operators $\hat V$ and $\hat W$ as
\begin{equation}\label{eq:otoc_def}
\mathcal{C}^{WV}_\mu(t)=\frac{1}{2}\text{tr}\left(\hat\rho_\mu\left[\hat V(t),\hat W\right]^{\dagger}\left[\hat V(t),\hat W\right]\right),
\end{equation}
where $\hat\rho_{\mu}=e^{-\mu \hat Q}/\text{tr}\left(e^{-\mu \hat Q}\right)$ is the Gibbs state defined in Eq.~\eqref{eq:Gibbs_def}. By expanding the commutators we get
\begin{align*}\label{eq:otoc_expand}
\mathcal{C}^{WV}_\mu(t) = & \frac{\langle \hat W^\dagger \hat V^\dagger(t)\hat V(t) \hat W \rangle_\mu + \langle \hat V^\dagger(t) \hat W^\dagger \hat W \hat V(t) \rangle_\mu}{2} - \nonumber \nonumber \\
- & \text{Re}\langle \hat V^\dagger(t) \hat W^\dagger \hat V(t) \hat W \rangle_\mu.
\end{align*}
We will refer to the last term as the out-of-time-ordered part of the OTOC and to the first two terms as its time-ordered part. The interesting physics of the OTOC are captured by the out-of-time-ordered part~\cite{TimeOrderedNote}, which we denote
\begin{equation}\label{eq:def_oto_part}
\mathcal{F}^{VW}_\mu(t) \equiv \text{Re}\langle \hat V^\dagger(t) \hat W^\dagger \hat V(t) \hat W \rangle_\mu.
\end{equation}
In the following we will mostly focus on this quantity (a notable exception in Sec.~\ref{Sec:Saturation} where we discuss the long-time limit of the full OTOC, which is mostly dominated by its time-ordered part). Unless stated otherwise, we assume that $\hat V$ and $\hat W$ both have trace zero.

It will be convenient to consider operators $\hat V, \hat W$ with particular charges $\lambda_V, \lambda_W$ under the adjoint action, i.e., $[\hat Q,\hat V]=\lambda_V \hat V$. For example, in the $q=2$ case which we focus on, the one-site operators
$\hat \sigma^{+},\hat \sigma^{-},\hat Z$ have charges $+1,-1,0$, respectively (in the following, $\hat Z_r$ denotes the Pauli $z$ operator on site $r$, while the operators $\hat \sigma_r^\pm$ increase/decrease the local charge by one). As we show below, the behavior of the OTOC can depend strongly on the charges $\lambda_V$ and $\lambda_W$. It is particularly interesting to consider operators with charge $\lambda_V = 0$, which can have a non-vanishing overlap with the conserved quantity, $\text{tr}(\hat Q \hat V) \neq 0$. As we argue below, for such operators the diffusion of charge implies a) slow relaxation of the OTOCs and b) non-trivial long-time saturation values at finite $\mu$ (see Sections~\ref{Sec:otoc_partfunc_results},~\ref{Sec:mu0} and~\ref{Sec:Saturation} in particular)~\cite{NonLocalOpNote}.

We can reduce the number of distinct OTOCs we need to consider by noting that there are certain relations between them. For example note that
\begin{equation*}
\mathcal{F}^{V^\dagger W^\dagger}_\mu(t) = e^{-\mu(\lambda_V + \lambda_W)} \mathcal{F}^{WV}_\mu(t)
\end{equation*}
holds on general grounds, decreasing the number of independent OTOCs. Moreover, in the $q=2$ case we discuss below, we can also make use of the relation
\begin{equation}\label{eq:relatepm}
\mathcal{F}^{\sigma^+_0 \sigma^-_r}_\mu (t) = e^{-\mu} \mathcal{F}^{\sigma^+_0 \sigma^+_r}_\mu (t).
\end{equation}
Therefore we will focus solely on the OTOCs between operators $\hat Z \hat Z$, $\hat Z \hat \sigma^+$ and $\hat \sigma^+ \hat \sigma^+$. Note that we can relate $\hat Z_r$ to the local charge density $\hat Q_r$ as 
\begin{equation}\label{eq:QtoZ}
\hat Z_r (t) = \hat{1\!\!1}_r - 2 \hat Q_r(t),
\end{equation}
which means that any correlator of the form~\eqref{eq:def_oto_part} has the same behavior if we replace all occurances of $\hat Z_r$ with $\hat Q_r$, up to some unimportant contributions that are either time-independent or decay diffusively, as in Eq.~\eqref{eq:diffusiononlattice}.

In the remainder of this section we first focus on OTOCs at zero chemical potential. We begin by showing that computing the average value of the OTOC in our random circuit problem is equivalent to evaluating a classical partition function. This allows us to compute the OTOC to significantly longer times than those available to direct numerical calculations. We find that at $\mu = 0$ all OTOCs spread in a ballistic wavefront, wherein the width of the front broadens in time, similarly to the case of random circuits without symmetries. The main new feature is that OTOCs involing the conserved operator $\hat Z$ exhibit a slow decay behind their wave front, which we confirm also for a non-random spin chain. We explain this behavior below, in~\secref{Sec:mu0}, by building on the results of~\secref{Sec:Diffusion} and detailing the different ways in which the diffusion of charge effects the dynamics of OTOCs. We give further support to our numerical results in~\secref{Sec:coarsegrainedanalyis} by considering a modified version of the random circuit where we are able to derive analytical predictions for the dynamics of different OTOCs. We return to the question of their behavior at finite chemical potential in~\secref{Sec:FiniteMu}.

\subsection{Mapping to a classical partition function}\label{Sec:partfunc}

We now outline how to compute OTOCs in the charge-conserving random circuit problem. The properties of the Haar distribution allow us to evaluate the average effect of a single 2-site gate on the OTOC exactly. As we show below, applying this averaging procedure to all the gates in the circuit transforms the problem of computing the average OTOC to the evaluation of a particular classical partition function, similar to what has been achieved in random circuits without symmetries in Refs. \onlinecite{Nahum17,RvK17}. While the classical model we obtain has much more structure than the non-symmetric case, and does not allow for an exact closed form solution, it serves as the basis of both numerical calculations and analytical approximations which we present throughout the rest of the paper.

\begin{figure}
	\includegraphics[width=0.35\columnwidth]{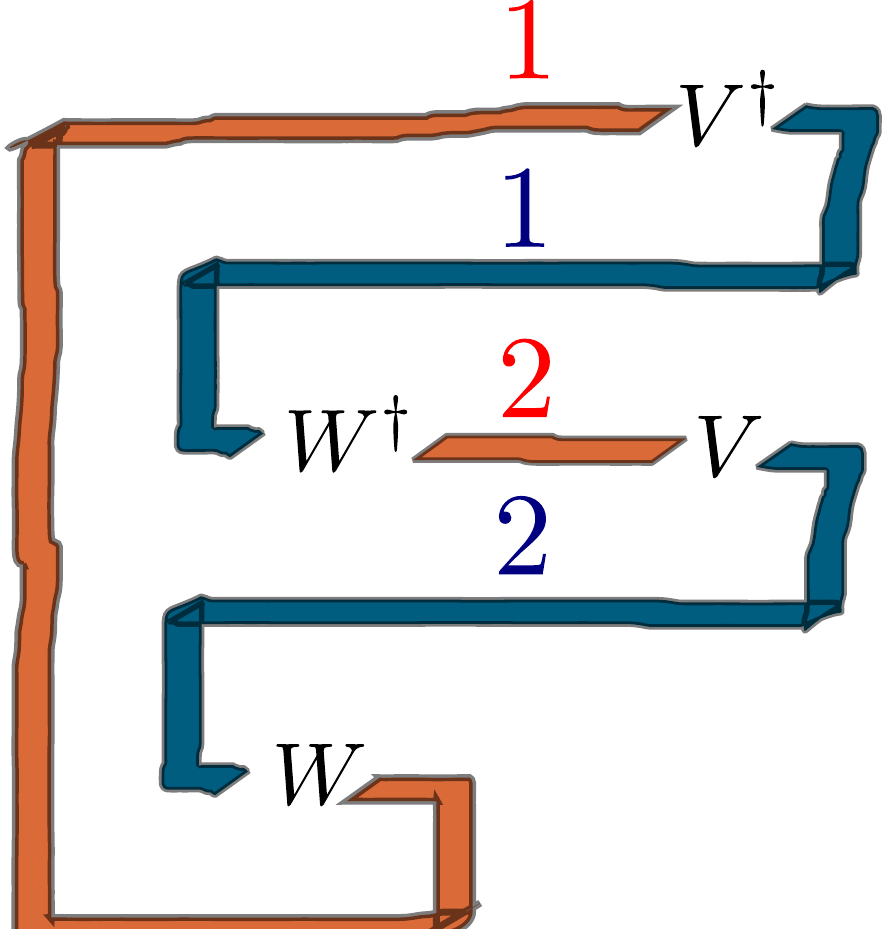}
	\caption{Representation of the OTOC $\braket{\hat V^\dagger(t)\hat W^\dagger \hat V(t) \hat W}$, as a `path integral' involving four layers. Each layer corresponds to one of the unitary time evolution operators (blue: $U$; red: $U^\dagger$) appearing in the correlator. These unitaries are given by a realization of the random circuit, and averaging over them gives rise to interactions between different layers.}
	\label{fig:keldysh}
\end{figure}

We begin by observing that one can write the $\mu=0$ OTOC as (the generalization to finite $\mu$ is straightforward, as we describe later in ~\secref{Sec:FiniteMu})
\begin{multline}\label{eq:otoc_as_superop}
\mathcal{F}^{VW}_{\mu=0}(t) \propto U^*_{\beta\alpha} V^*_{\gamma\beta} U_{\gamma\delta} W^*_{\mu\delta} U^*_{\nu\mu} V_{\nu\lambda} U_{\lambda\eta} W_{\eta\alpha} \\
= V^*_{\gamma\beta} V_{\nu\lambda} (U^* \otimes U \otimes U^* \otimes U)_{(\beta\gamma\nu\lambda)(\alpha\delta\mu\eta)}  W^*_{\mu\delta} W_{\eta\alpha},
\end{multline}
where $U$ is defined in \eqnref{eq:fullcircuit}. Therefore the central quantity one needs to compute in order to obtain the average OTOC is $\overline{U^* \otimes U \otimes U^* \otimes U}$. This is an operator acting on four copies of the original Hilbert space. We can think of this construction as a generalization of the Keldysh contour~\cite{Aleiner16}, involving four `layers', as illustrated in Fig.~\ref{fig:keldysh}. Each of the four operators appearing in the definition of the OTOC connects two of these layers. 

Since every gate in the random circuit is independently chosen, we can evaluate the Haar-average for all of the gates individually. For each gate, the Haar averaging results in a four-leg tensor with two incoming and two outgoing legs, one for both of the sites the gate acts on. One leg of this averaged tensor corresponds to four copies of the original Hilbert space. For our $q=2$ case this would mean in principle 16 states per site of the form $\ket{\alpha\beta\gamma\delta}$ for $\alpha,\beta,\gamma,\delta=0,1$. As we argue below, only $6$ of these $16$ appear in the averaged circuit and therefore the average OTOC can be calculated in terms of an effective description involving $6$ states per site. 

In particular, as we show in~\appref{App:Identities}, the average for \emph{a single two-site gate} takes the form
\begin{multline}\label{eq:onegate_avg_4layers}
\overline{U^* \otimes U \otimes U^* \otimes U} = \sum_{s=\pm} \sum_{Q_1 \neq Q_2} \frac{1}{d_{Q_1}d_{Q_2}} \ket{\mathcal{I}^s_{Q_1Q_2}}\bra{\mathcal{I}^s_{Q_1Q_2}} \\
+ \sum_{s=\pm} \sum_Q \frac{1}{d_Q^2 -1} \left[ \ket{\mathcal{I}^s_{QQ}}\bra{\mathcal{I}^s_{QQ}} - \frac{1}{d_Q} \ket{\mathcal{I}^s_{QQ}}\bra{\mathcal{I}^{-s}_{QQ}} \right],
\end{multline}
where $\ket{\mathcal{I}^\pm_{Q_1Q_2}}$ are states from four copies of the two-site Hilbert space, defined as 
\begin{align}\label{eq:def_gate_projectors}
\ket{\mathcal{I}^+_{Q_1Q_2}} \equiv  \sum_{\substack{\alpha \in \mathcal{H}_{Q_1} \\ \beta \in \mathcal{H}_{Q_2}}}\ket{\alpha\alpha\beta\beta}; & & \ket{\mathcal{I}^-_{Q_1Q_2}} \equiv \sum_{\substack{\alpha \in \mathcal{H}_{Q_1} \\ \beta \in \mathcal{H}_{Q_2}}} \ket{\alpha\beta\beta\alpha}.
\end{align}
$\mathcal{H}_{Q}$ here is the sector of the two-site Hilbert space with total charge $Q$. 

\begin{figure}
	\centering
	\includegraphics[width=0.385\columnwidth]{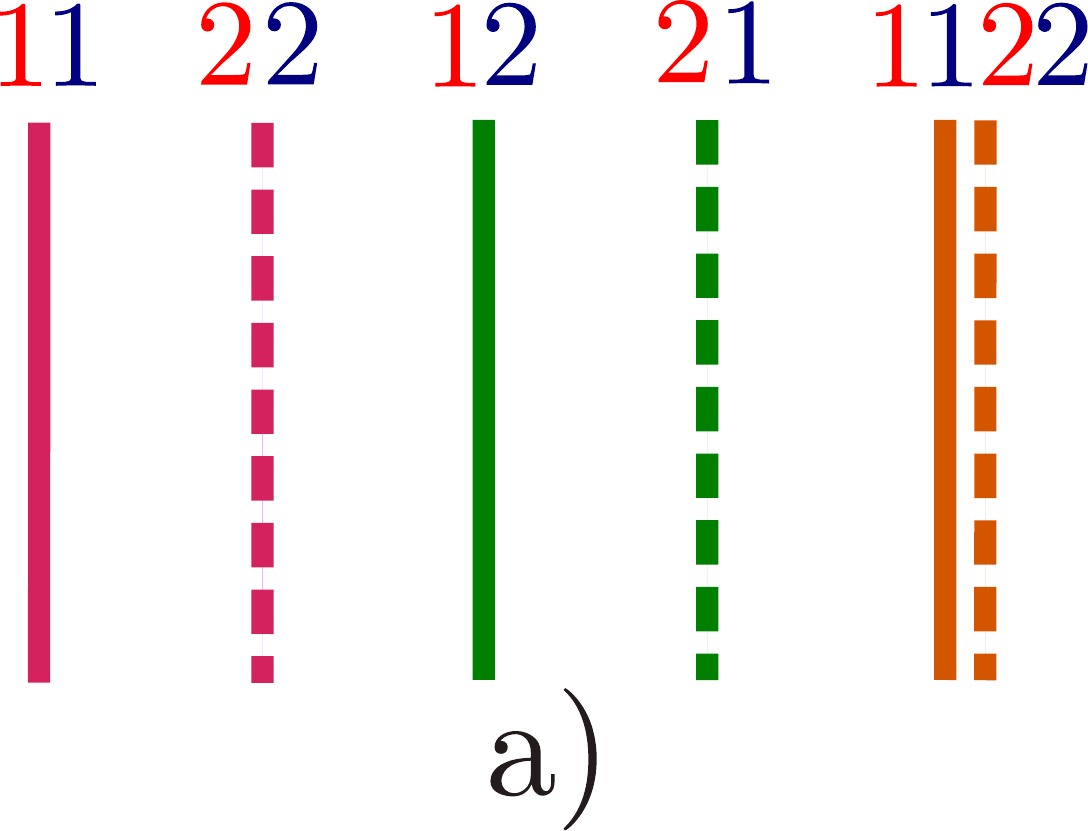}
	\hspace{0.4cm}
	\includegraphics[width=0.545\columnwidth]{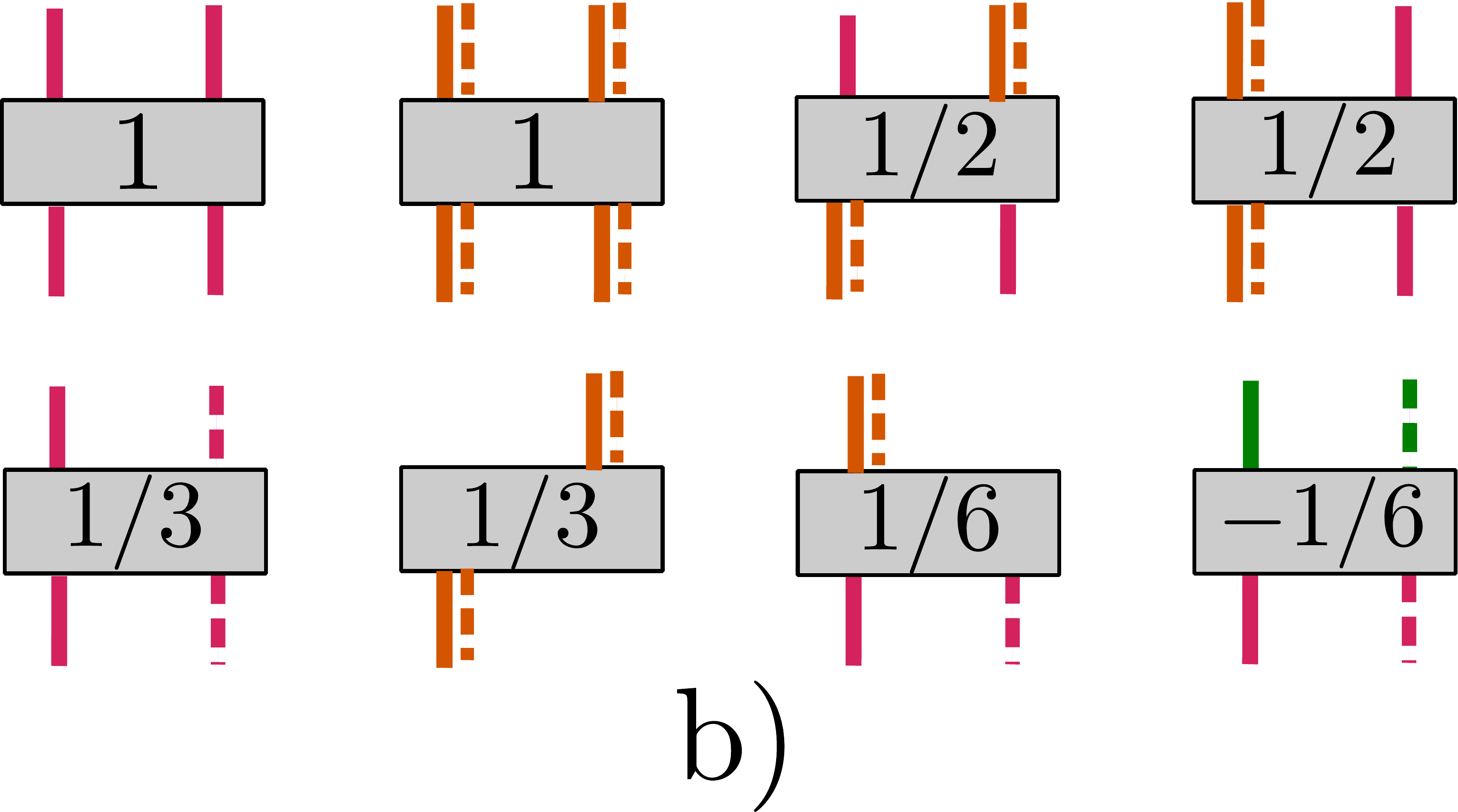}
	\caption{Interpretation of Eq.~\eqref{eq:onegate_avg_4layers} in terms of local states. a) Notation of the five different `particle types' that can occur: the first four correspond to exactly two of the four layers (shown in Fig.~\ref{fig:keldysh}) being occupied by a charge, while the last one is a bound state, either formed by the first two or the second two particle types. The empty state is not denoted. b) Some possible one- and two-particle processes generated by averaging over a single two-site gate.}
	\label{fig:linetypes}
\end{figure}

The states $\ket{\mathcal{I}^\pm_{Q_1Q_2}}$ cannot be written as products of states on (four copies of) the individual sites~\cite{FactorizeNote}. Nevertheless, as we detail in~\appref{App:Identities}, they can all be written in terms of the following six states, living on four copies of a \emph{single site:} $\ket{0000}$, $\ket{1100}$, $\ket{0011}$, $\ket{1001}$, $\ket{0110}$, and $\ket{1111}$ (e.g., $\ket{\mathcal{I}^-_{12}} = \ket{1111}_1\ket{0110}_2 + \ket{0110}_1\ket{1111}_2$). The first of these is an `empty' state, wherein all four layers are unoccupied at a given site. We will refer to the states with exactly two layers occupied as having a single `particle' on a given site, which can belong to four different species, as illustrated in Fig.~\ref{fig:linetypes} a). The last state then can be thought of as a site being occupied by two particles. In terms of these six states Eq.~\eqref{eq:onegate_avg_4layers} defines a four-leg tensor that maps each of the $36$ possible states on the two sites to a linear combination of the same $36$ states with some particular (real, but possibly negative) coefficients. Some of these possible processes are shown in Fig.~\ref{fig:linetypes} b), while the other non-zero coefficients can be obtained by swapping the two sites (either on the bottom or the top of the gate) or permuting the different particle types.

To compute the full time evolution we need to contract the four-leg tensors, defined above, according to the geometry of the circuit seen in Fig.~\ref{fig:circuit}. Thus every layer of the random circuit acts as a transfer matrix, evolving a configuration of particles (in the sense defined above) to a linear combination of different configurations. Finally, in the first and last layers, we need to contract the remaining legs with those of the operators $\hat V$ and $\hat W$ appearing in the OTOC formula~\eqref{eq:otoc_as_superop}. We can write the result as the matrix element
\begin{equation}\label{eq:otoc_as_partfunc}
\overline{\mathcal{F}^{VW}_{\mu=0}}(t) \propto \braket{\mathcal{P}_V | \overline{U^* \otimes U \otimes U^* \otimes U} |\mathcal{D}_W},
\end{equation}
where $\ket{\mathcal{P}_V}$ and $\ket{\mathcal{D}_W}$ are states in the four-copy Hilbert space, defined by $\braket{\alpha\beta\gamma\delta|\mathcal{P}_V} = V_{\beta\alpha} V^*_{\gamma\delta}$ and $\braket{\alpha\beta\gamma\delta|\mathcal{D}_W} = W_{\delta\alpha} W^*_{\gamma\beta}$ (we give an interpretation of these quantities in terms of superoperators in~\secref{Sec:supop_language}). In this formula $U$ is the full unitary circuit of \eqnref{eq:fullcircuit}. Eq.~\eqref{eq:otoc_as_partfunc}, together with Eq.~\eqref{eq:onegate_avg_4layers}, can be interpreted as a classical partition function on a two-dimensional lattice, where every site has six possible states. The gates in one layer of the circuit form a transfer matrix of this classical spin problem while the operators $\hat V$ and $\hat W$ in the definition of the OTOC appear through the boundary conditions, at times $0$ and $t$, respectively. While we are not able to evaluate this partition function exactly (unlike the case with no symmetries), it allows for efficient numerical computations, much beyond the time scales attainable otherwise, as well as for some analytical approximations, which we detail below.

\subsection{Hydrodynamic tails and the shape of the OTOC wavefront in the random circuit model}\label{Sec:otoc_partfunc_results}

Using the formalism developed in the previous section for computing circuit-averaged OTOCs as classical partition functions, we are able to investigate their dynamics at time scales much larger than what is obtainable using real time evolution. We evaluate this partition function numerically, representing it as a two-dimensional tensor network. We present the results below and find evidence of both the ballistically propagating, diffusively broadening wave front, found previously for random circuits without symmetries, as well as the late-time power law tails with characteristic space-time dependence $\propto (v_\text{B}t - r)^{-1/2}$ mentioned in the introduction. We present our analytical understanding of these result in \secref{Sec:mu0}.

Armed with the mapping the classical partition function described in~\secref{Sec:partfunc}, we evaluate the OTOC up to time $t\approx 40$ (80 layers of the random circuit). We do this by representing the partition function as a two-dimensional tensor network, built out of the four-legged tensors, defined in Eq.~\eqref{eq:onegate_avg_4layers}, that arise when averaging a single gate. We contract these together by representing the boundary condition $\ket{\mathcal{D}_W}$ at $t=0$ as a matrix product state, which we then propagate forward layer-by-layer, using a method analogous to the well-known time-evolving block decimation (TEBD) algorithm.~\cite{VidalTEBD,MurgReview} The OTOC is then computed by taking the overlap of this MPS with another one that represents the boundary condition $\bra{\mathcal{P}_V}$ at time $t$. 

\begin{figure}[t]
	\includegraphics[width=0.49\columnwidth]{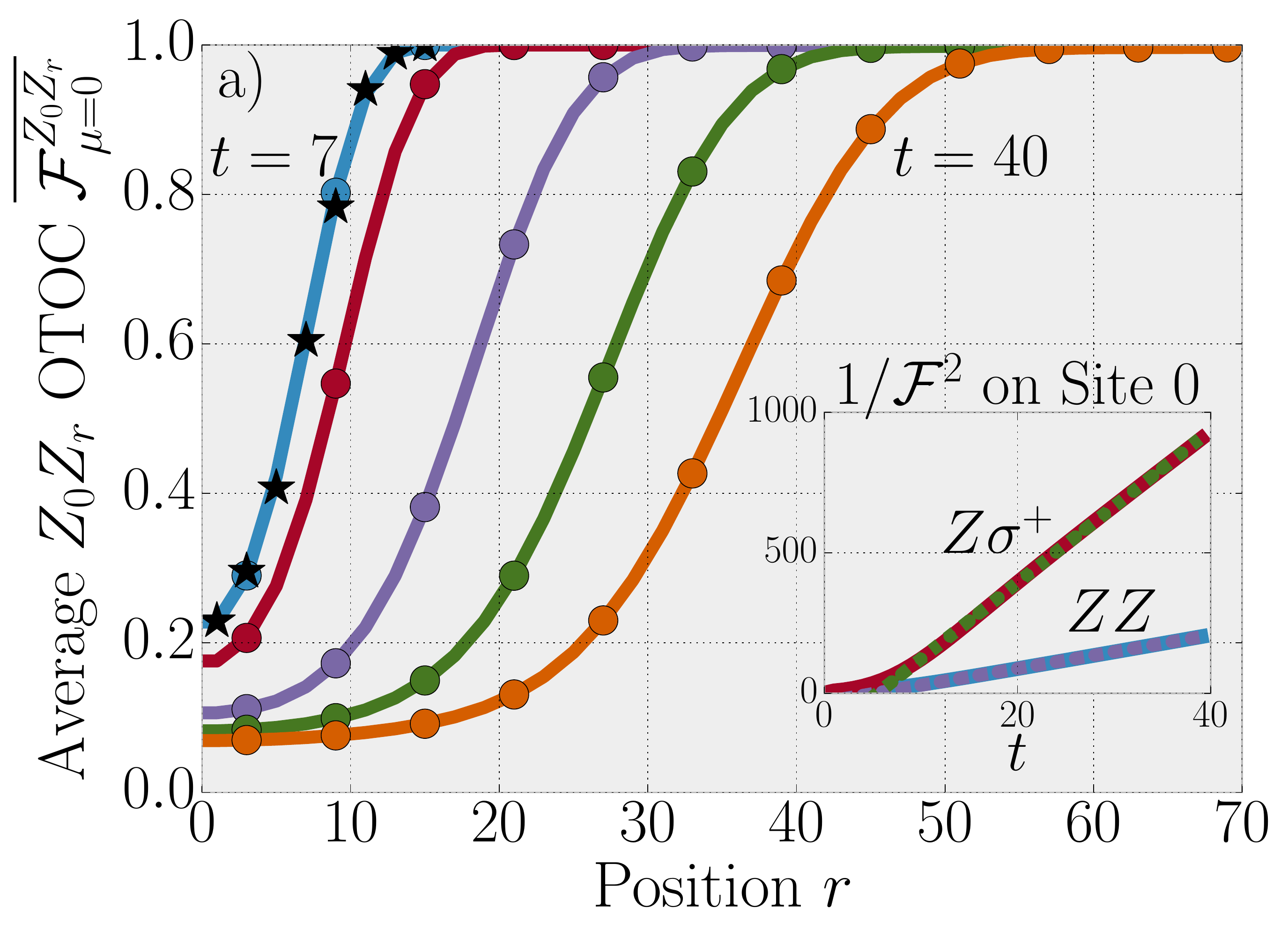}
	\includegraphics[width=0.49\columnwidth]{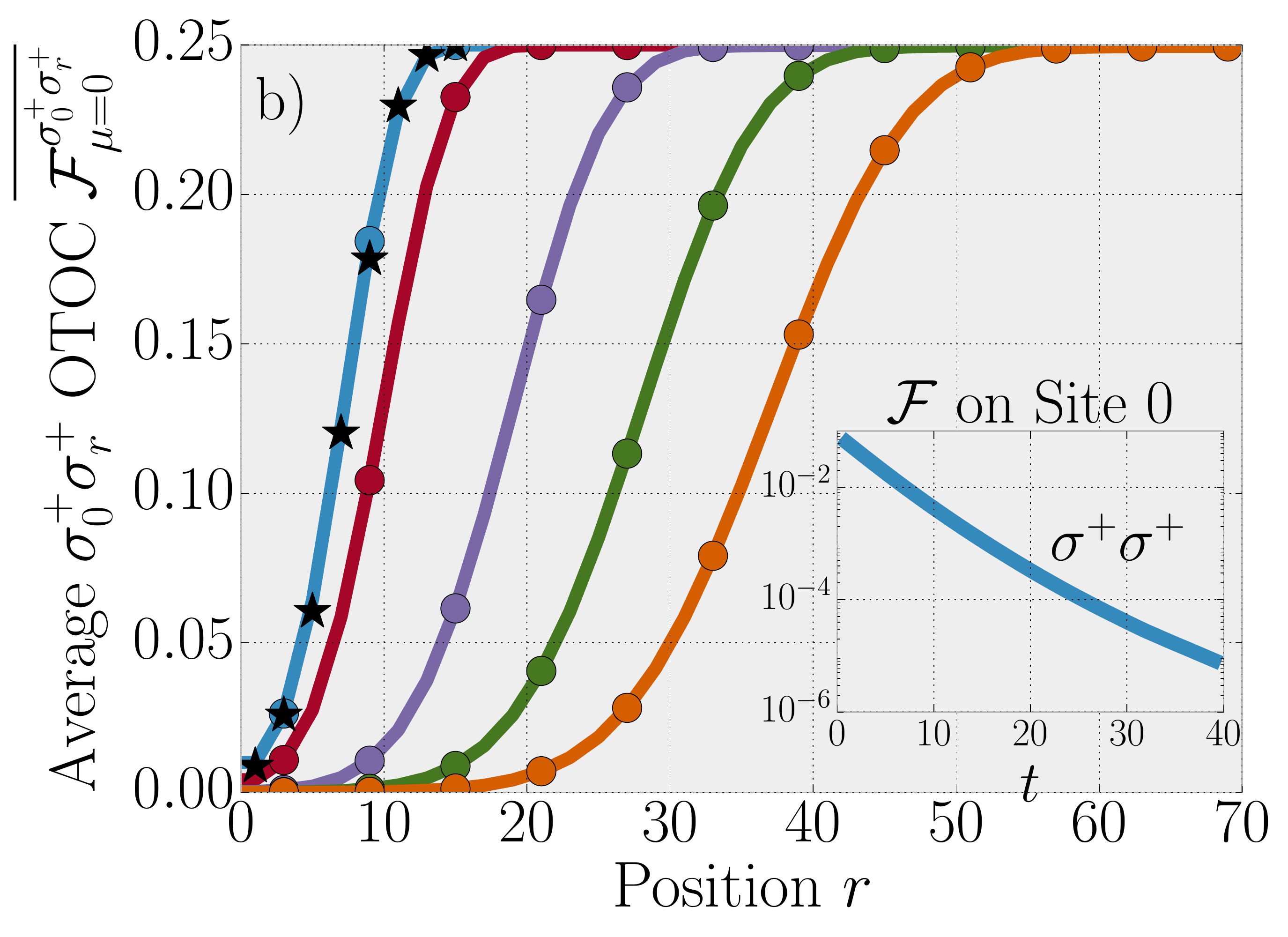}
	\includegraphics[width=0.49\columnwidth]{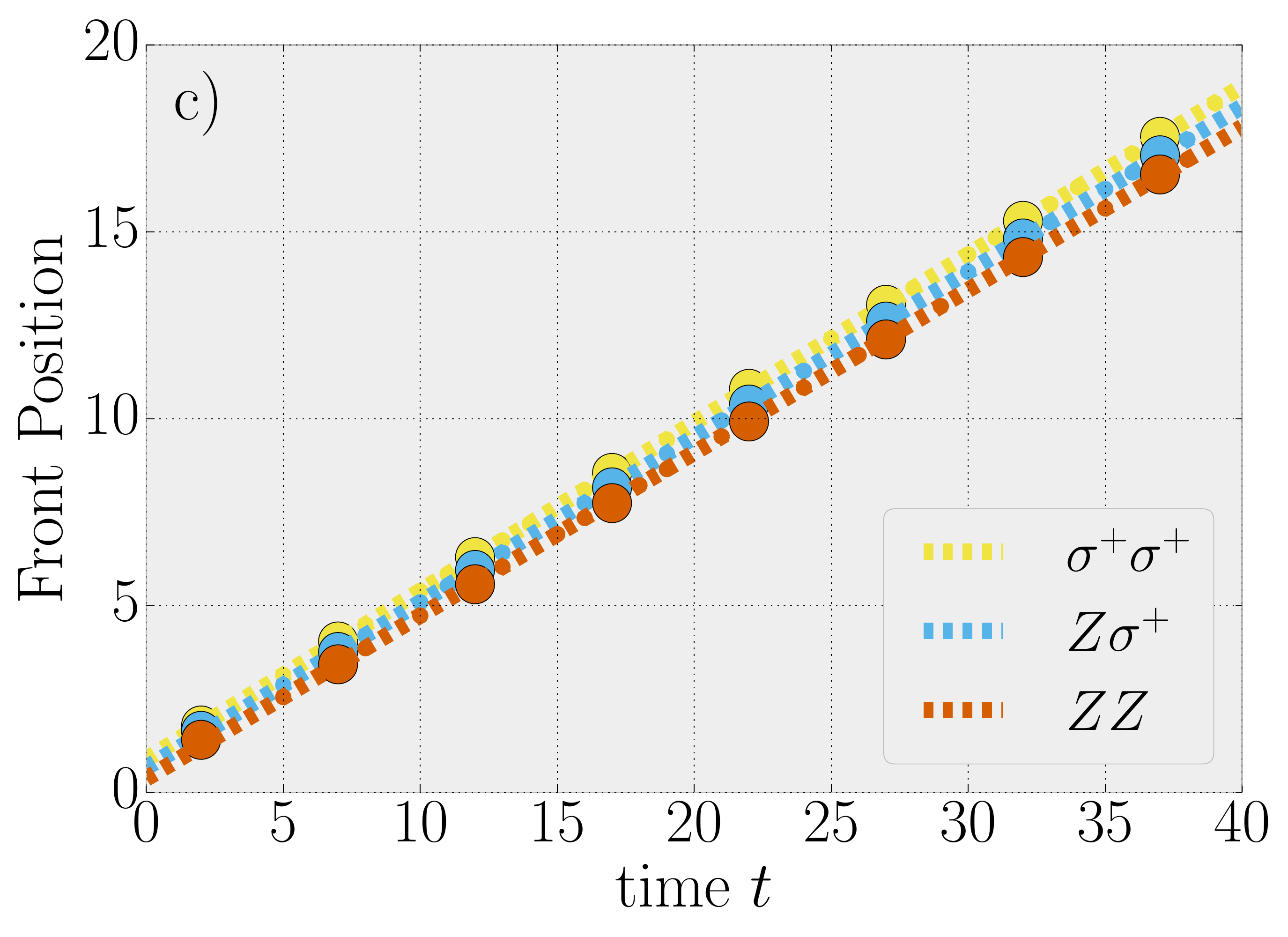}
	\includegraphics[width=0.49\columnwidth]{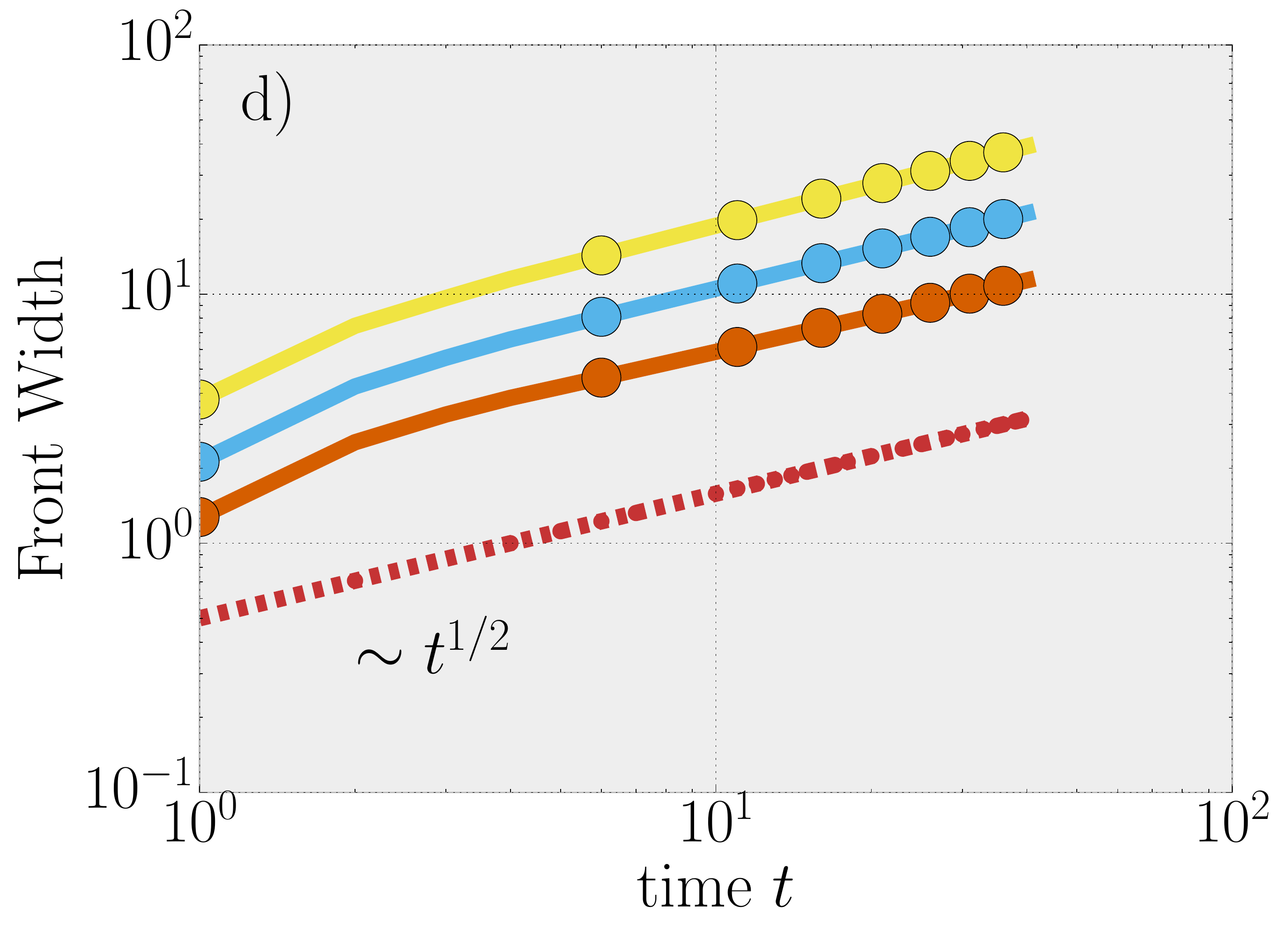}
	\caption{The average OTOC $\mathcal{F}$, defined in Eq.~\eqref{eq:def_oto_part}, at $\mu=0$, evaluated as a classical partition function. All OTOCs spread in a ballistically propagating front which itself diffusively broadens in time, and saturate to zero behind the front. The shape of the front is shown for a) the $\hat Z\hat Z$ and b) the $\hat \sigma^+\hat \sigma^+$ OTOCs for times (from left to right) $t=7,10,20,30,40$. The black stars represent data obtained by performing the unitary time evolution with TEBD and averaging over 100 realizations. The insets show the value of $\mathcal{F}$ for different operators at site $0$ as a function of time. For OTOCs involving $\hat Z$ we find that $1/\mathcal{F}^2$ grows linearly in time, indicating a saturation $\mathcal{F} \propto 1 / \sqrt{t}$ at long times, while the $\hat \sigma^+\hat \sigma^+$ OTOC saturates exponentially fast. The two lower figures show the c) position and d) width of the front as a function of time. The front moves ballistically with the three types of OTOCs having similar front velocities $v_\text{B}\approx 0.45$ in units of the circuit light cone velocity, while they all broaden diffusively. The position and the width are extracted from a curve that smoothly interpolates between the data points: the front position is defined by the point where the OTOC decays to half of its original value, while the width is computed as the inverse of the maximal derivative of this curve near the front. }
	\label{fig:otoc_mu0}
\end{figure}

In this section, we focus on the case where $\mu=0$; the finite $\mu$ case is treated in separately in \secref{Sec:FiniteMu}. The results, in Fig.~\ref{fig:otoc_mu0}, demonstrate that OTOCs exhibit ballistic behavior much like that which has been analytically described for random circuits without conserved quantities~\cite{Nahum17,RvK17}. In particular, there exists a velocity scale~\cite{VBNote} $v_{B}$ such that the OTOC $\mathcal{F}^{V_0W_r}_{\mu=0}(t)$ is of $\mathcal{O}(1)$ at $\left|r\right|>v_{B}t$, decreases near the so-called ``butterfly front'' $|r| \approx v_{B}t$, and to saturates to 0 for $v_{B}t \gg \left|r\right|$, as shown in Fig~\ref{fig:otoc_mu0}. In line with previous work, our numerics indicate that the regime over which the OTOCs obtain an $\mathcal{O}(1)$ value (the ``width of the front'') broadens diffusively ($\sim\sqrt{t}$) in time (see in particular the last panel of Fig~\ref{fig:otoc_mu0}). 

Our results indicate that OTOCs saturate to zero behind the front (this is peculiar to the $\mu=0$ case considered here: as we argue in~\secref{Sec:Saturation} at finite $\mu$ certain OTOCs have finite saturation values). Our main new finding is that OTOCs $\mathcal{F}^{VW}(t)$ for which at least one of $\hat V,\hat W$ is the conserved density $\hat Z_r$ decay at long times as $\mathcal{F} \propto 1 / \sqrt{t}$. We refer to this slow, power-law relaxation as a ``hydrodynamic tail'', due to its analogy with similar slow dacay of time-ordered correlation functions in classical and quantum hydrodynamics~\cite{KADANOFF1963,TailsReview05,Huse06,Rosch13}. The $\hat \sigma^+_0 \hat \sigma^+_0$ OTOC on the other hand decays exponentially with time, and using \eqnref{eq:relatepm}, the same is true for the $\hat \sigma^+_0 \hat \sigma^-_0$ OTOC. These results, are in contrast with previous results on random circuits without symmetries, where all OTOCs showed an exponential decay. As we explain in~\secref{Sec:mu0} the hydrodynamic tails we observe here are a natural consequence of the diffusive charge transposrt discussed in~\secref{Sec:Diffusion}.

\begin{figure}[t]
	\includegraphics[width=0.6\columnwidth]{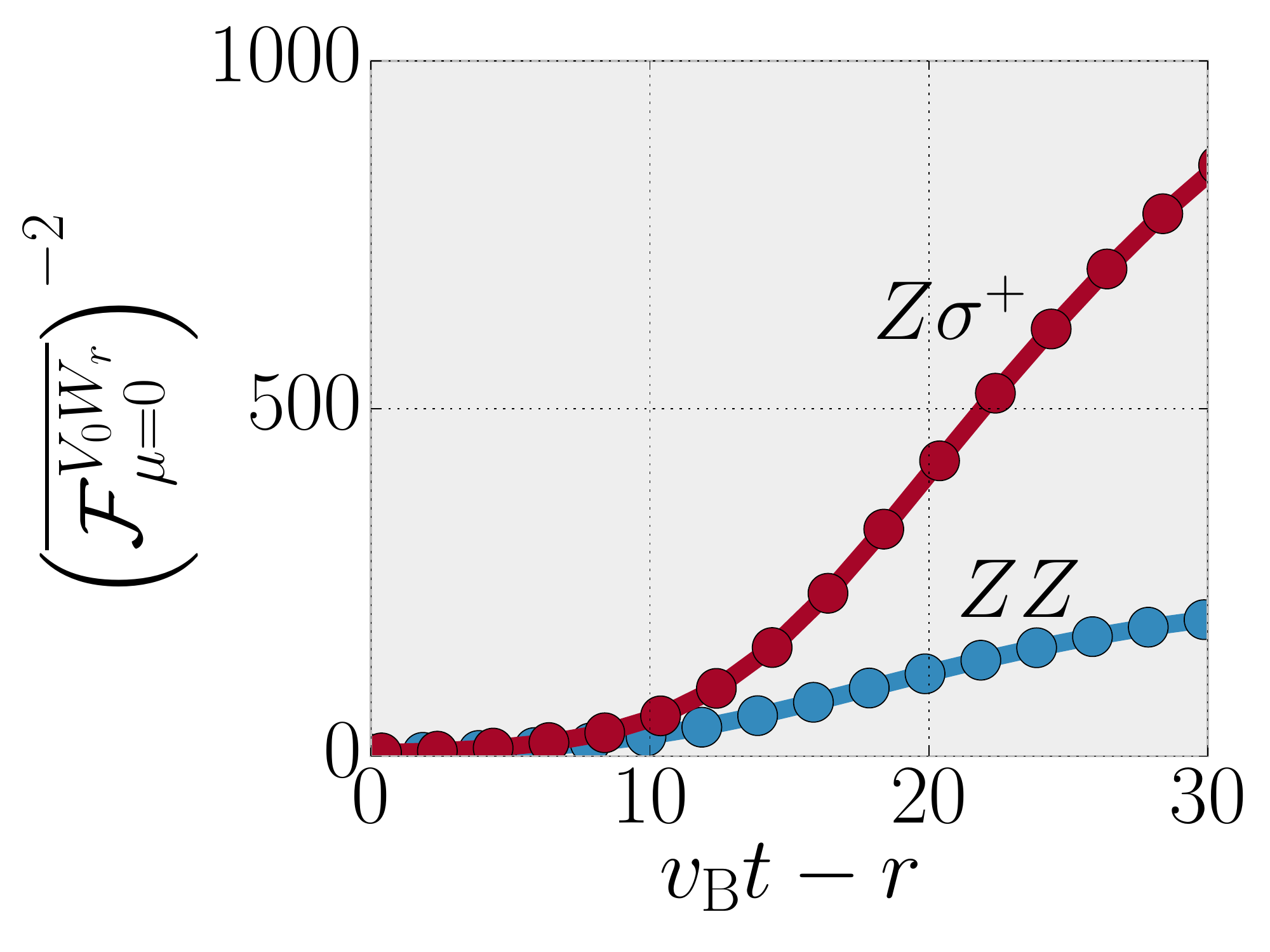}
	\caption{Space-time structure of the wave front in Fig.~\ref{fig:otoc_mu0} at time $t=40$ for OTOCs involving the conserved density $\hat Z_0$. We plot $\mathcal{F}^{-2}$ at $\mu = 0$ for $\hat V = \hat Z_0$ and $\hat W = \hat Z_r, \hat \sigma^+_r$ as a function of the distance from the front $v_\text{B}t - r$ and find a linearly growing regime in both quantities, indicating a decay of the form $\mathcal{F} \propto 1 / \sqrt{v_\text{B}t - r}$.}
	\label{fig:otoc_tails}
\end{figure}

Going beyond the decay of the OTOC at a given position, we can ask the question of what is the shape of the front at a fixed time. As we noted above, there is a diffusive broadening around the middle of the front which is expected from previous studies of random circuits without symmetries, where the OTOC near the front was well approximated by an error function~\cite{RvK17}. While this form is in agreement with the behavior of the  $\hat \sigma^+ \hat \sigma^+$ OTOC in Fig.~\ref{fig:otoc_mu0}, note in the same figure that the $\hat Z \hat Z$ OTOC has a position dependence that differs significantly from this error-function profile. Well behind the butterfly front, the position dependence of the OTOC is well described by $\mathcal{F} \propto 1 / \sqrt{v_\text{B}t - r}$, i.e. it exhibits power law decay as a function of the distance from the position of the front. This is shown in Fig.~\ref{fig:otoc_tails}. These results indicate that the simple biased diffusion description of OTOCs, developed previously for random circuits without symmetries, has to be supplemented by considerations of the diffusion of the local conserved quantities. We will elaborate on this further in~\secref{Sec:coarsegrainedanalyis}, where, by considering a modified random circuit with a large parameter, we derive such a ``hydrodynamic'' equation of motion for the OTOCs. The solution of this equation shows the same $1 / \sqrt{v_\text{B}t - r}$ behavior observed in our numerics (see Fig.~\ref{fig:ZZsolution} in particular).

\subsection{Hydrodynamic tails in deterministic systems}\label{Sec:Floquet}

As we argued in the introduction, we expect random unitary circuits, like the one studied in this paper, to capture universal properties of non-noisy, ergodic quantum systems in the strongly interacting, high-temperature regime. This is true for the diffusion of conserved quantities, discussed in~\secref{Sec:Diffusion}, and recent numerical work suggests that it is also the case for the diffusive broadening of the OTOC wavefront~\cite{RvK17,Leviatan2017}. We expect that the hydrodynamic tails that we observed in the previous section -- and which, as we argue below, are a direct consequence of charge diffusion -- should therefore also be present in non-noisy systems, provided that they exhibit diffusive transport of conserved quantities. Such power law decay of OTOCs in a weakly disordered Hamiltonian system has already been observed numerically in \cite{MBLOTOC1}. Here we show that the same phenomena appears also in a system without any randomness. We do this by considering a periodically driven (Floquet) spin-chain, where the total spin z component is conserved, and find that OTOCs involving $\hat Z$ show similar diffusive decay to the one seen in the random circuit model, while OTOCs between non-conserved operators decay exponentially.

We consider a one-dimensional chain of spin 1/2 degrees of freedom. A single driving sequence consists of four parts, with the so-called Floquet unitary given by
\begin{align}\label{eq:Floquet_def}
U_\text{F} &= e^{-i\tau H_4} e^{-i\tau H_3} e^{-i\tau H_2} e^{-i\tau H_1} \nonumber \\
H_1 &= J_{z}^{(1)} \sum_r \hat Z_r \hat Z_{r+1} \nonumber \\
H_3 &= J_{z}^{(2)} \sum_r \hat Z_r \hat Z_{r+2} \nonumber \\
H_2 &= H_4 =  J_{xy} \sum_r \left(\hat X_r \hat X_{r+1} + \hat Y_r \hat Y_{r+1}\right),
\end{align}
where $\hat X_r \equiv \hat \sigma^+_r + \hat \sigma^-_r$ and $\hat Y_r \equiv -i (\hat \sigma^+_r - \hat \sigma^-_r)$ are Pauli spin operators on site $r$, and we take periodic boundary conditions $r \equiv r+L$. Every part of the drive individually conserves the spin z component, $[H_a,\sum_r \hat Z_r] = 0$ for $a=1,2,3,4$. We take the period time to be $T \equiv 4\tau = 1$ and the couplings to be all order 1, namely $J_z^{(1)} = (\sqrt{3} + 5) / 6$, $J_z^{(2)} = \sqrt{5} / 2$ and $J_{xy} = (2 \sqrt{3} + 3) / 7$. 

We compute the OTOC~\eqref{eq:def_oto_part} at $\mu=0$ (infinite temperature) in this system using exact diagonalization methods, up to system size $L=24$. We do this be approximating the trace in the infinite temperature average by an expectation value in a random state, $\ket{\Psi}$, drawn from the Haar measure over the whole Hilbert space. The OTOC is then calculated as the overlap $\mathcal{F}^{VW}_{\mu=0}(t) \approx \text{Re}\,\braket{\Psi_1|\Psi_2}$ where $\ket{\Psi_1} \equiv \hat W U_\text{F}^{-t} \hat V U_\text{F}^{t} \ket{\Psi}$ and $\ket{\Psi_2} \equiv U_\text{F}^{-t} \hat V U_\text{F}^{t} \hat W \ket{\Psi}$. This overlap approximates the infinite temperature average up to an error which is exponentially small in the system size~\cite{Luitz17} and we indeed find numerically that the deviation of the OTOC between different random states is negligible at the system sizes we consider. 

\begin{figure}
	\includegraphics[width=0.85\columnwidth]{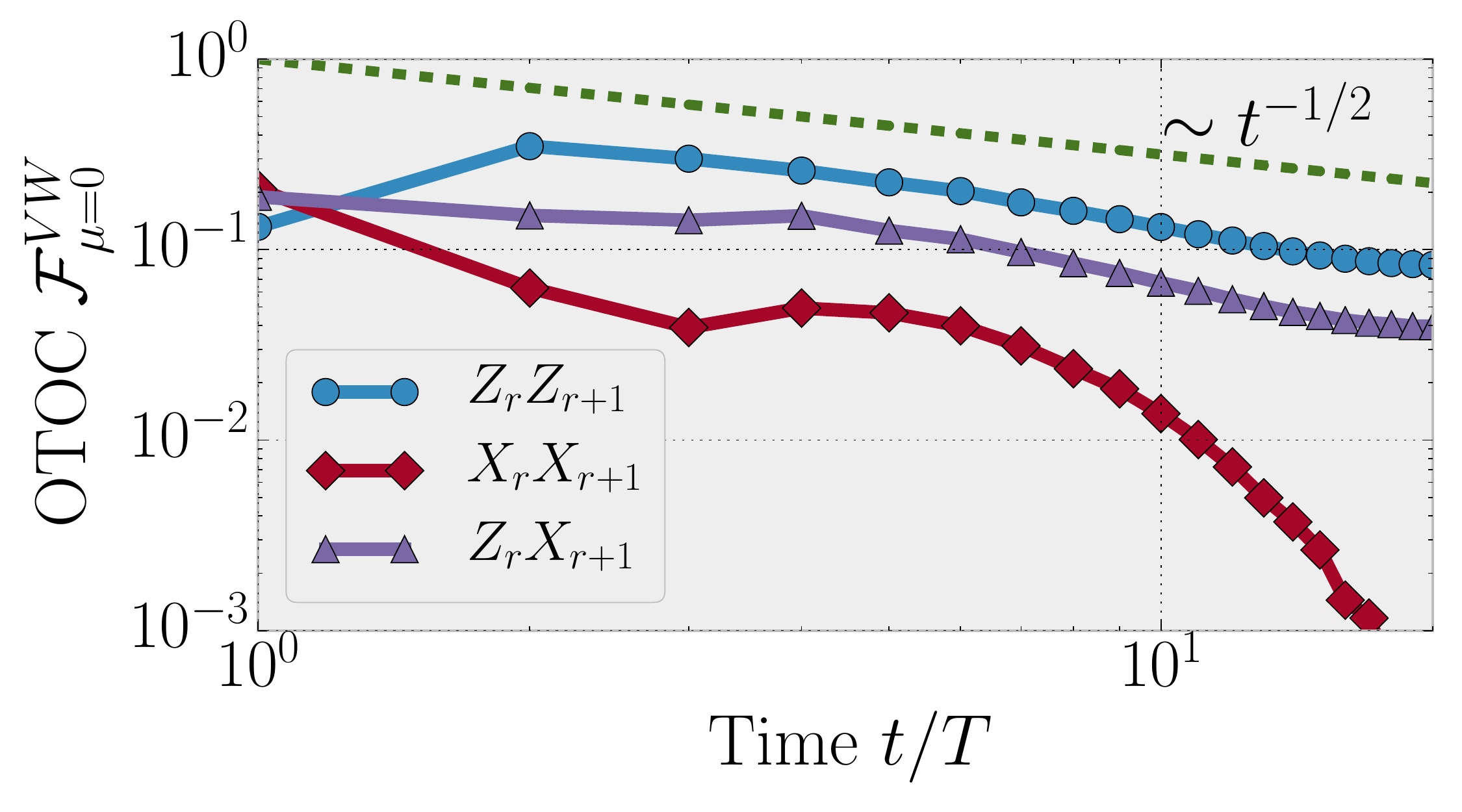}
	\includegraphics[width=0.49\columnwidth]{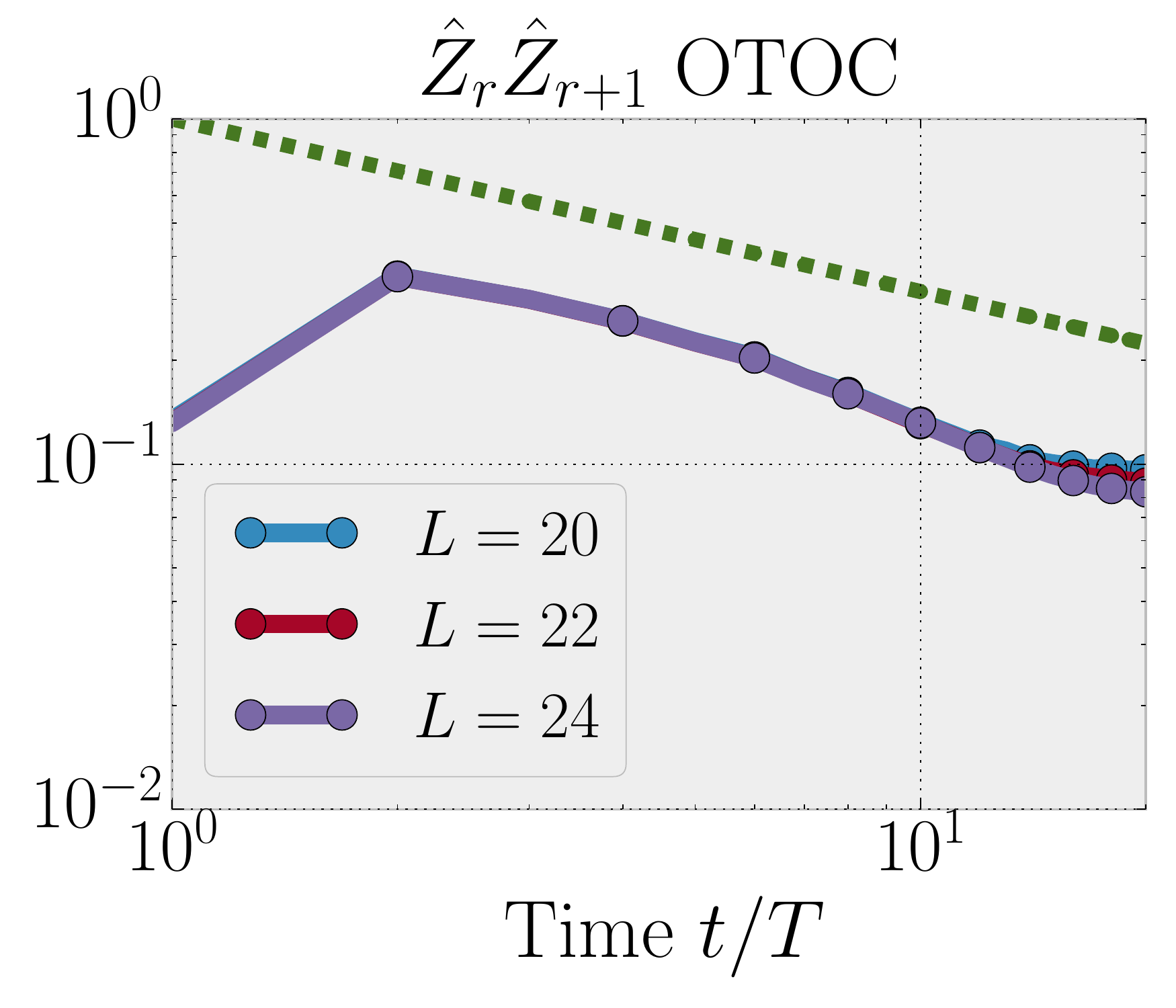}
	\includegraphics[width=0.49\columnwidth]{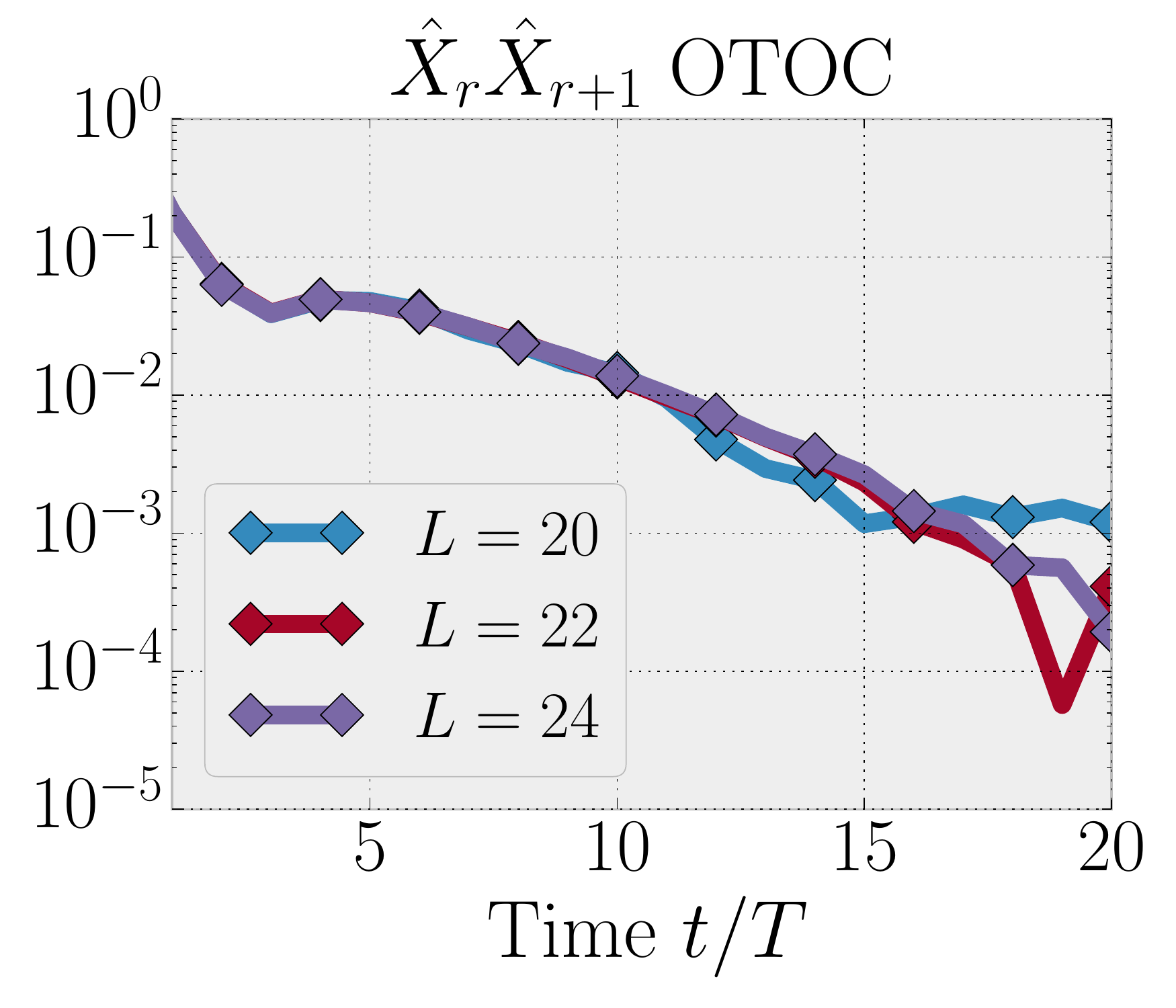}
	\caption{OTOC for different operators on nearest neighbor sites at zero chemical potential in the non-random Floquet model~\eqnref{eq:Floquet_def} which conserves $\sum_r \hat Z_r$. Upper panel: OTOCs on nearest neighbor sites for system size $L=24$ sites. Both the $\hat Z \hat Z$ and the $\hat Z \hat X$ OTOC decay as a power law, approximately as $\sim t^{/1/2}$,  while the $\hat X \hat X$ OTOC decays faster than a power law, as shown by the log-log plot in the upper panel. The lower two panels show a comparison between system sizes $L=20,22,24$ for the $\hat Z \hat Z$ and $\hat X \hat X$ OTOCs, respectively. The latter is shown in a lin-log plot and is well approximated by an exponential decay.}
	\label{fig:otoc_floquet}
\end{figure}

In Fig.~\ref{fig:otoc_floquet} we show the results for the $\hat Z \hat Z$, $\hat X \hat X$ and $\hat Z \hat X$ OTOCs. The local operator $\hat X$ has no overlap with the conserved quantity and is therefore expected to behave similarly to $\hat \sigma^+$ discussed above for the random circuit (indeed, it is a simple linear combination of $\hat \sigma^+$ and $\hat \sigma^-$). We take the OTOC between two operators on nearest neighbor sites such that they all have the same initial value 1. We find numerically that while the decay of the $\hat Z \hat Z$ and $\hat Z \hat X$ OTOCs is well described by the diffusive $\propto 1 / \sqrt{t}$ behavior, the $\hat X \hat X$ OTOC decays more quickly, approximately as an exponential. This is all in in agreement with the results found for the random circuit model, and we conjecture that it is the generic behavior for one-dimensional systems with diffusive transport properties. Indeed, we will argue in the following section that these hydrodynamic tails, exhibited by OTOCs involving the conserved density, are a natural consequence of diffusion.

\section{Explaining the presence and structure of the tails}\label{Sec:mu0}

In~\secref{Sec:Diffusion} we showed that the presence of a local conservation law leads to a diffusive, rather than exponential, relaxation of the associated charge. We now discuss how the presence of the same conserved quantity alters the behavior of OTOCs, leading to the hydrodynamic tails observed numerically in the previous section. Our discussion will focus on the random circuit model, but since most of our arguments are based on the diffusion of the conserved quantity, they should apply, with some slight changes, to other systems with diffusive transport, such as the Floquet system introduced in~\secref{Sec:Floquet} or the Hamiltonian system described in Ref. \onlinecite{MBLOTOC1}.

First in~\secref{ss:Optails} and  \secref{ss:supoptails} we argue that OTOCs $\mathcal{F}^{VW}(t)$ (with traceless local operators $\hat V, \hat W$) decay slowly (as a diffusive power law) precisely when at least one of $\hat V,\hat W$ has nonvanishing overlap with the conserved charge, i.e., $\text{tr}(\hat Q \hat V) \neq 0$. In the case that neither $\hat V,\hat W$ have such an overlap, the decay is expected to be exponentially quick, identical to the behavior observed in random circuits without symmetries.~\cite{Nahum17,RvK17}. Our arguments in~\secref{ss:Optails} and  \secref{ss:supoptails} are not fully controlled analytically. However, they (particularly those in~\secref{ss:supoptails}) have the merit of being completely analogous to those well established by previous work on hydrodynamical tails in regular observables, both in quantum and classical dynamics~\cite{TailsReview05,Huse06,Rosch13}. 

In~\secref{Sec:coarsegrainedanalyis} we provide a more controlled analytical argument which yields the detailed space-time structure of the front in those cases where it is present. For this analysis, we use a ``coarse grained'' model with a different circuit geometry, and a built-in large parameter which simplifies the calculations. Our conjecture is that such a model should describe the long-time dynamics of the circuit in~\secref{Sec:Random-Local-Unitary}. Indeed, we find in this coarse-grained circuit an OTOC that has the space-time profile $\propto 1/\sqrt{v_B t - x}$, in agreement with the the numerical data presented for the original circuit in~\secref{Sec:otoc_partfunc_results}. 

\subsection{Operator spreading explanation for the presence of tails}\label{ss:Optails}
The presence of hydrodynamic tails in out-of-time-ordered correlators is, in our opinion, most neatly explained in the language of superoperators. However, this requires introducing the appropriate formalism, which we delay until the next section. Before doing that, we will explain the presence of tails for a particular OTOC, in the case $\hat V=\hat Z$, in the (perhaps) more familiar language of operator spreading. 

Consider an OTOC between two local Pauli operators $\hat \sigma_{0}^{\alpha=\text{x,y,z}},\hat \sigma_{r}^{\beta=\text{x,y,z}}$ on sites $0$ and $r$. At time $t$, $\hat \sigma_{0}^{\alpha}$ evolves into a superposition of operators
\begin{equation}\label{eq:opspreading}
\hat \sigma_{0}^{\alpha}(t)=\sum_{\boldsymbol{\nu}}\hat \sigma^{\boldsymbol{\nu}}c_{\boldsymbol{\nu}}(t),
\end{equation}
where $\hat \sigma^{\boldsymbol{\nu}}$ denotes a \emph{Pauli string} of the form $\bigotimes_{s=1}^{L} \hat \sigma_s^{\nu_s}$, with $\nu_s = 0,x,y,z$. The out-of-time-ordered part of the OTOC at zero chemical potential then takes the form
\begin{equation}\label{eq:otocnosymm}
\mathcal{F}^{\alpha\beta}_{\mu=0}=\sum_{\boldsymbol{\nu}} S(\nu_{r},\beta) |c_{\boldsymbol{\nu}}(t)|^{2},
\end{equation}
where $S(\nu_{r},\beta)=\pm 1$ depending on whether $\hat \sigma^{\boldsymbol{\nu}}$ commutes or anti-commutes with $\hat \sigma_r^\beta$. A similar expression can be derived for the case of general $q$~\cite{RvK17}.

In the case \emph{without} symmetries it was found that the distribution of $|c_{\boldsymbol{\nu}}(t)|^{2}$ is strongly dominated by Pauli strings $\boldsymbol{\nu}$ which fill most of the spatial region $[-v_{B}t,+v_{B}t]$, while the total weight contained in strings of a fixed length decays exponentially, an observation that follows simply from the fact that there are exponentially more long operators than short ones~\cite{RvK17}. Since the operator norm is conserved, the average weight on a single string is $|c_{\boldsymbol{\nu}}|^{2}\sim q^{-4v_{B}t}$. Assuming $|c_{\boldsymbol{\nu}}|^{2}$
is largely independent of $\nu_{r}$ for typical strings when $|r| \ll v_{B}t$, the sum
in Eq.~\eqref{eq:otocnosymm} would average to $0$, as the positive and negative contributions cancel. In practice not all strings have the same weight, but
we expect such fluctuation to be suppressed (by the law of large numbers)
as $\mathcal{O}(\sqrt{1/q^{-4v_{B}t}})$. Accounting for these fluctuations,
and exponentially small contributions from Pauli strings well behind
the front, we were able to prove~\cite{RvK17} exponential decay of the OTOC $\mathcal{F}^{\alpha\beta}_{\mu=0} \sim \mathcal{O}(e^{-a t})$. 

\begin{figure}
	\includegraphics[width=0.85\columnwidth]{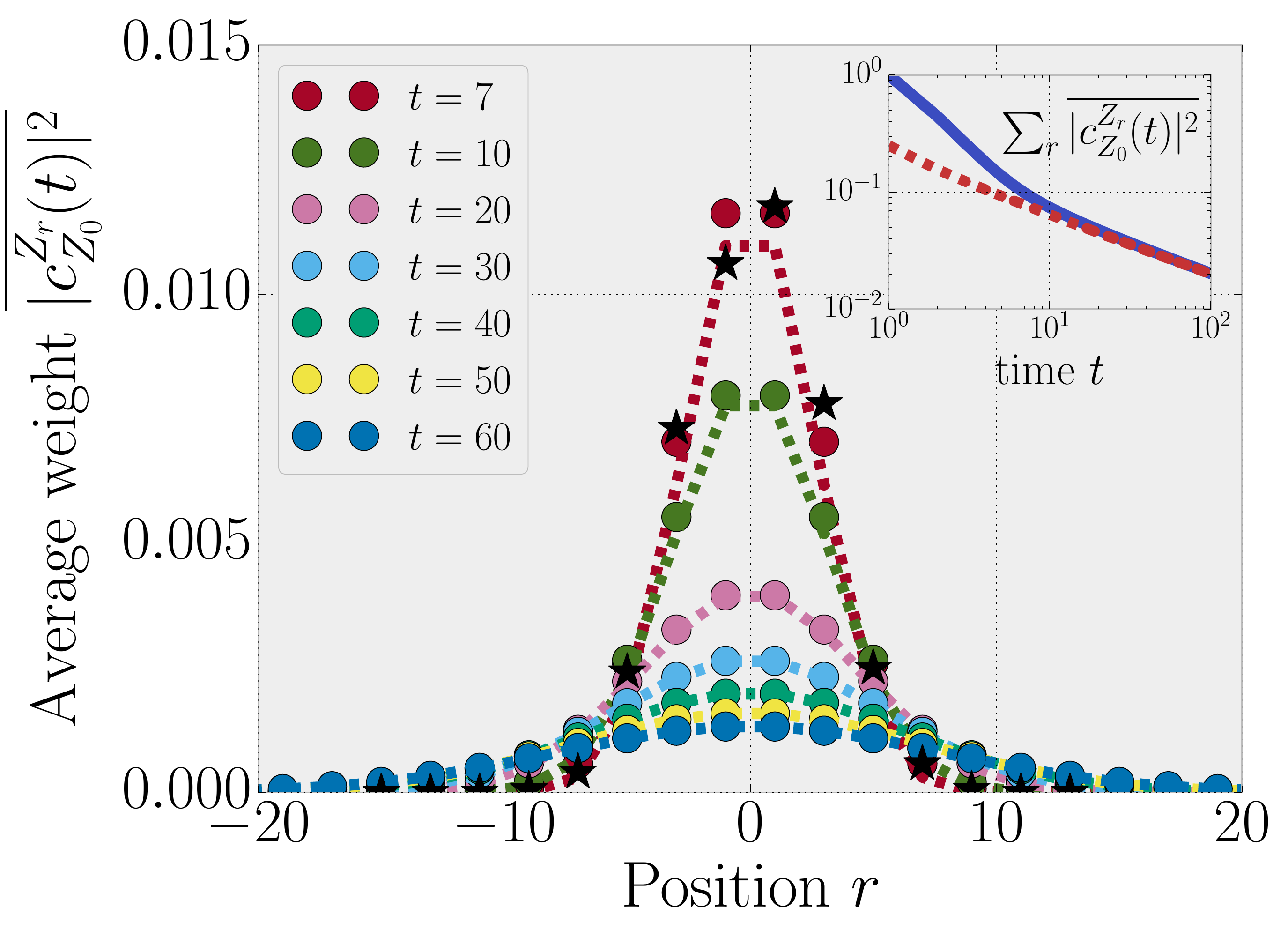}
	\caption{Average weights of single-site $\hat Z$ operators, $\overline{|c_{Z_{0}}^{Z_{r}}(t)|^2}$ (dots), compared to their lower bound given by the squares of the average coefficients $\left(\overline{|c_{Z_{0}}^{Z_{r}}(t)|}\right)^2$ (dashed lines). The weights approach the lower bound at times $t \approx 10$. The weight are computed using the classical partition function formalism of~\ref{Sec:partfunc}, and for comparison we show TEBD data averaged over 100 circuit realizations at time $t=7$ (black stars). Inset: the total weight contained in single-site $\hat Z$ operators decays in time at the rate set by the diffusion of the coefficients $c_{Z_{0}}^{Z_{r}}(t)$, as $t^{-1/2}$ at long times.}
	\label{fig:z_weights}
\end{figure}

The presence of conserved charges constrains some of the operator spread coefficients and alters the above argument significantly. In particular, expressions of the form $\text{tr}(f(\hat{Q}) \hat \sigma_{0}^{\alpha}(t))$ are independent of time, due to charge conservation, for any function $f$. One immediate consequence for the operator $\hat Z_0(t)$ is that its operator spread coefficients on single-site $\hat Z_r$ operators, defined as $c_{Z_{0}}^{Z_{r}}(t) \equiv q^{-L} \text{tr}(\hat Z_r \hat Z_0(t))$, satisfy $\sum_{r}c_{Z_{0}}^{Z_{r}}(t)=1$ at all times, where $r$ ranges over all sites in the forward light cone of $\hat Z_{0}$. Indeed, as we have shown in Sec.~\ref{Sec:Diffusion}, the coefficients decay on average as $t^{-1/2}$, rather than exponentially as they would without conservation laws. Using $\overline{|c_{\boldsymbol{\nu}}|^2} \geq \overline{|c_{\boldsymbol{\nu}}|}^2$ and \eqnref{eq:diffusiononlattice}, and summing over all sites $r$, implies that the total weight on single-site $\hat Z$ operators is lower bounded by a value that decays as $\sim t^{-1/2}$. We observe numerically that while this weight initially decays faster (approximately as $t^{-0.8}$), it approaches this lower bound at times $\approx 10$ (see Fig.~\ref{fig:z_weights}). Based on these numerical results, we expect that at longer times $\sum_{r}\overline{|c_{Z_{0}}^{Z_{r}}|^2}(t) \sim 1/\sqrt{t}$. In contrast, recall that the same weight is expected to decay exponentially quickly for circuits without a conserved charge. 

The above argument shows that if we pick $\alpha=z$ in \eqnref{eq:otocnosymm} there is an anomalously large, slowly (diffusively) decaying positive contribution to the OTOC, coming from the $|c_{Z_{0}}^{Z_{l}}(t)|^2$ coefficients discussed above. This suggests that OTOCs involving $\hat Z$ relax to their long-time value as $\sim 1/\sqrt{t}$ at leading order in $t$, in agreement with the numerical evidence in Fig.~\ref{fig:otoc_mu0}. We expect similar behavior for the relaxation of OTOCs in Hamiltonian systems for operators that have a non-vanishing overlap with the local energy density, an effect already observed numerically in Ref. \onlinecite{MBLOTOC1}. 


\subsection{Diffusive tails in a superoperator formalism}\label{ss:supoptails}
An alternative, possibly more general way of understanding the effects of charge conservation on operator spreading is possible using the language of superoperators.

\subsubsection{Superoperator formalism}\label{Sec:supop_language}
First we show that the objects defined in the formalism of~\secref{Sec:partfunc} have a natural interpretation as superoperators that act on the operators of the original spin chain. This provides a general language for describing OTOCs which we will use for the ``hydrodynamic'' interpretation of our results on OTOC dynamics in the rest of this section as well as in~\secref{Sec:Saturation} where we show that the long-time limit of OTOCs can be understood by generalizing the notion of thermalization to the space of operators (rather than states). We expect these concepts to be useful in the future study of scrambling. 

As described in \secref{Sec:partfunc}, the quantities $\ket{\mathcal{P}_V}$ and $\ket{\mathcal{D}_W}$, appearing in Eq.~\eqref{eq:otoc_as_partfunc}, have four indices, i.e. they live in a Hilbert space which comprises four copies of the original system. We can naturally interpret this as the space of superoperators acting on the spin chain. This identification goes as follows. Let us first note that an operator $\hat O$ of the original system can be though of as a state $|\hat{O}\rangle \in \mathcal{H} \otimes \mathcal{H}$ in two copies of the original Hilbert space, such that $\braket{\alpha\beta|\hat{O}} \equiv \braket{\alpha|\hat{O}|\beta} = O_{\alpha\beta}$. Operators thus naturally correspond to states in a doubled space. Iterating this procedure once more we arrive at superoperators $\mathcal{A}$ that map one operator to another, i.e. $\mathcal{A} \ket{\hat{O}} = \ket{\hat{O}'}$ (throughout the text, we use capital calligraphic letters to denote superoperators~\cite{CalligraphicNote}). These can then be reinterpreted as states in \emph{four} copies of the original Hilbert space, defined as $\braket{\alpha\beta\gamma\delta|\mathcal{A}} \equiv \braket{\beta\alpha|\mathcal{A}|\gamma\delta}$.

A natural basis of superoperators is of the form $\ket{\hat A}\bra{\hat B}$. Another way to turn operators into superoperators is via left and right multiplication, defined as $\mathcal{L}_A \ket{\hat B} \equiv \ket{\hat A \hat B}$ and $\mathcal{R}_A \ket{\hat B} \equiv \ket{\hat B \hat A}$. The two states appearing in Eq.~\eqref{eq:otoc_as_partfunc} can then be interpreted as follows. $\mathcal{P}_V = \ket{\hat V}\bra{\hat V}$ is the ``density matrix'' corresponding to the state $\ket{\hat V}$, while $\mathcal{D}_W = \mathcal{L}_{W^\dagger} \mathcal{R}_{W^{\phantom{\dagger}}}$ corresponds to multiplying from left and right with $\hat W^\dagger$ and $\hat W$, respectively. The OTOC then has the interpretation of the time evolved expectation value of a superoperator,
\begin{equation*}
\mathcal{F}^{VW}_{\mu=0}(t) \propto \braket{\mathcal{P}_V | \mathcal{D}_W(t)} = \braket{\hat V(t)|\mathcal{D}_W|\hat V (t)}. 
\end{equation*}
As we show in~\secref{Sec:FiniteMu}, the OTOC at $\mu \neq 0$ can be written in a similar form, with $\hat V$ replaced by a slightly modified operator. As we argue in~\secref{Sec:Saturation}, the long-time limit of the ``state'' $\ket{\hat V(t)} \bra{\hat V(t)}$ can be understood as relaxation to a state analogous to a thermal equilibrium. We note here that the states appearing in Eq.~\eqref{eq:onegate_avg_4layers} also have simple interpretations in the superoperator language as $\mathcal{I}^+_{Q_1Q_2} = \ket{\hat P_{Q_1}} \bra{\hat P_{Q_2}}$ and $\mathcal{I}^-_{Q_1Q_2} = \mathcal{L}_{P_{Q_1}} \mathcal{R}_{P_{Q_2}}$, where $\hat P_Q$ are the projectors defined in Eq.~\eqref{eq:onestep_opavg}, acting on the two-site Hilbert space.

\subsubsection{OTOC tails are associated with diffusion of $\mathcal{L}_{Q}$ and $\mathcal{R}_{Q}$}\label{ss:OTOCtailsLR}
We now address the issue of tails, and when they appear in OTOCs, in the language of superoperators. Consider, from the preceeding section, the superoperators corresponding to left and right multiplication by $\hat{Q}$, namely $\mathcal{L}_{Q}$ and $\mathcal{R}_{Q}$. These super-operators are conserved as a function of time
\begin{equation*}
\mathcal{L}_{Q},\mathcal{R}_{Q}\rightarrow\mathcal{L}_{U^{\dagger}QU},\mathcal{R}_{U^{\dagger}QU}=\mathcal{L}_{Q},\mathcal{R}_{Q}.
\end{equation*}
Both superoperators are local densities, in that they can be written
as sums of local superoperators, e.g., $\mathcal{R}_{Q}=\sum_{r}\mathcal{R}_{Q_{r}}$. Note that the relation $\overline{\mathcal{L}_{Q_{r}}(t)},\overline{\mathcal{R}_{Q_{r}}(t)}=\mathcal{L}_{\overline{Q_{r}(t)}},\mathcal{R}_{\overline{Q_{r}(t)}}$, together with the diffusion of the local charge $\hat{Q}_{r}$, derived in Eq.~\eqref{eq:diffusiononlattice}, imply that $\mathcal{L}_{Q_{r}},\mathcal{R}_{Q_{r}}$ diffuse on average as well. Thus, the presence of a diffusing conserved
quantity in the original many body problem leads to the presence of
two new diffusing conserved quantities in operator space. In the continuum approximation we can write this as
\begin{align}\label{eq:supopdiff}
\partial_t \mathcal{L}_{Q_r(t)} &= D \partial^2_r \mathcal{L}_{Q_r(t)}; \nonumber\\
\partial_t \mathcal{R}_{Q_r(t)} &= D \partial^2_r \mathcal{R}_{Q_r(t)}. 
\end{align}
We can make use of the conservation of $\mathcal{L}_{Q},\mathcal{R}_{Q}$ to shed new light on the diffusive relaxation of certain OTOCs, discussed previously in~\secref{Sec:otoc_partfunc_results}. There, we noted that the $\hat Z\hat Z$  and $\hat Z\hat \sigma^{+}$  OTOCs have a power law relaxation. We can account for both of these tails in the following way. These two OTOCs can be written in the form $\langle \hat V(t) \mid \mathcal{D}_{Z_{r}} \mid \hat V(t) \rangle$ where $\hat V=\hat Z, \hat \sigma^+$. Note that the superoperator in this expression can be rewritten
\begin{equation*}
\mathcal{D}_{Z_{r}} = 1-2(\mathcal{L}_{Q_r(t)} -\mathcal{R}_{Q_r(t)})^2 
\end{equation*}
This expression is quadratic in $\mathcal{L}_{Q_{r}}-  \mathcal{R}_{Q_{r}}$, which is a conserved density. In ergodic theories, conserved densities, \textit{and their variances}, are expected to show $1/\sqrt{t}$ relaxation in 1D. The variances in conserved densities show this diffusive behavior even states with initially spatially homogeneous distributions of the conserved quantity~\cite{Rosch13}. In the superoperator language, it is thus natural to conclude at $\hat Z$-type OTOCs relax as $1/\sqrt{t}$.~\cite{SupopDiffusionNote} The remaining OTOCs $\hat \sigma^+,\hat \sigma^\pm$ do not involve $\hat Z$, and also do not show diffusive decay. We attribute this to the fact that the corresponding OTOC operator $\mathcal{D}_{\hat \sigma^+_r}$ is orthogonal to any algebraic combination of the only two available local conserved densities, $\mathcal{L}_{Q}$ and $\mathcal{R}_{Q}$.

\subsection{OTOC Hydrodynamics from coarse-graining}\label{Sec:coarsegrainedanalyis}
Since an exact analytical calculation of the partition function~\eqref{eq:otoc_as_partfunc} is out of reach, we instead consider a modified version of the random circuit which we expect to reproduce the behavior of the original model at long time- and length scales. We use this simplified model to shed light on the hydrodynamic nature of the OTOC dynamics and give an analytical estimate for the shape of the wavefront, reproducing the numerical results of~\ref{Sec:otoc_partfunc_results}.

\begin{figure}
\includegraphics[width=0.8\columnwidth]{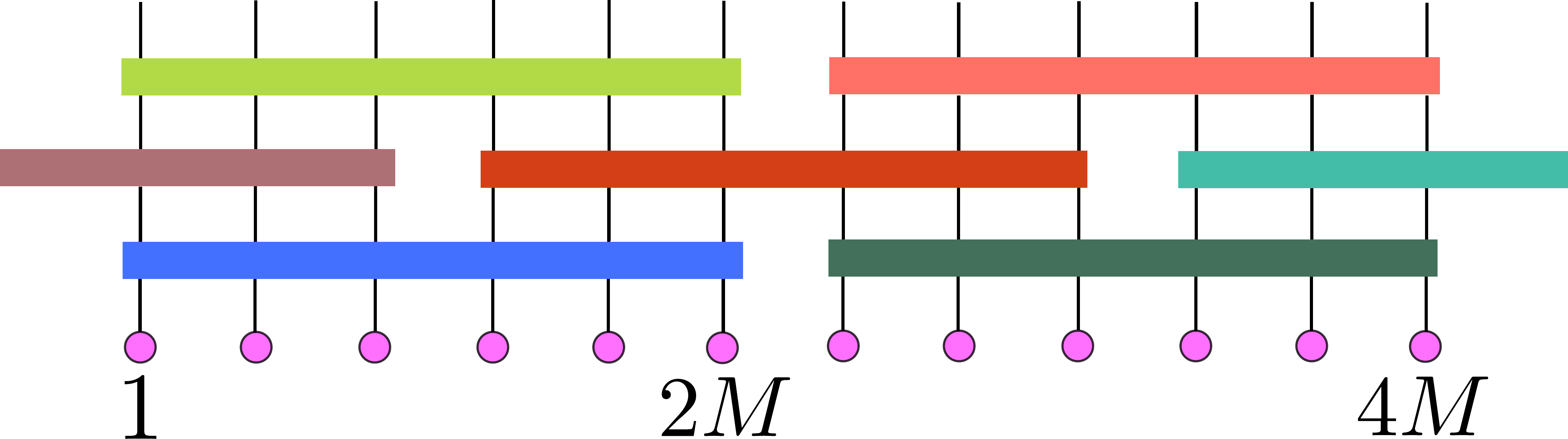}

\caption{Structure of the first few layers of the coarse-grained random circuit. Each gate acts on $2M$ consecutive sites (the on-site Hilbert space is $q=2$) and subsequent layers are shifted by $M$ sites. Every gate is an independently chosen random unitary, block-diagonal with respect to the total charge on the $2M$ sites and Haar-random in each block.}\label{fig:coarsegrained_circuit}
\end{figure}

To arrive at this approximate description, we consider a ``coarse-grained'' version of the circuit, defined by increasing the range of the random unitary gates such that each one acts on $2M$ consecutive sites, as illustrated in Fig.~\ref{fig:coarsegrained_circuit}. We label physical sites by $r$ and super-sites (consisting of $M$ copies of the $q=2$ Hilbert space) by $x$. Time evolution is then described by a network of these longer range random symmetric unitaries, with a geometry similar to the original $M=1$ case illustrated in Fig.~\ref{fig:circuit}. We find that in the limit $M\gg 1$ the dynamics simplifies considerably, allowing for a closed approximate expression for the OTOC, which we detail below. We consider evolving the operator $\hat Z_{r}$ here and leave the $\hat \sigma_r^+$ case for later work.

As noted above the $\hat Z_r$ OTOC superoperator can be decomposed as 
\begin{equation}\label{eq:decompose_Zotoc}
\mathcal{D}_{Z_{r}}\sim\mathcal{D}_{1\!\!1}-2\mathcal{L}_{Q_{r}}-2\mathcal{R}_{Q_{r}}+4\mathcal{L}_{Q_{r}}\mathcal{R}_{Q_{r}},
\end{equation}
where $\mathcal{D}_{1\!\!1} \equiv \mathcal{L}_{1\!\!1} \mathcal{R}_{1\!\!1}$ is a superoperator that acts on an operator by multiplying it from both sides with the identity, which is equivalent to the ``super-identity'', obeying $\mathcal{D}_{1\!\!1} \ket{\hat O} = \ket{\hat O}$ for any operator $\hat O$. The decomposition~\eqref{eq:decompose_Zotoc} already suggests that the superoperator $\mathcal{D}_{Z_{r}}$ should have a diffusive component, since, as we showed previously, $\mathcal{L}_{Q_{r}}$ and $\mathcal{R}_{Q_{r}}$ diffuse on average. The main technical aim of this section is to understand the average behavior of the non-linear term $\mathcal{L}_{Q_{r}}\mathcal{R}_{Q_{r}}$. Let us first evolve $\mathcal{D}_{Z_{r}}$ with one unitary gate on sites $r,r+1,\ldots,r+2M-1$, corresponding to `super-sites' $x,x+1$. A straightforward application of~\eqnref{eq:onegate_avg_4layers} yields a sum of two terms 
\begin{equation}\label{eq:Zotoc_onestep}
\mathcal{D}_{Z_{r}}(\Delta \tau) = \sum_{Q}\frac{b_{Q}}{d_{Q}}\mathcal{I}_{Q_{1}Q_{2}}^{x,x+1}
+  \frac{1}{M^{2}}\mathcal{D}_{\frac{1}{2}\left(\zeta_{x}+\zeta_{x+1}\right)},
\end{equation}
where $\hat \zeta_{x}\equiv\sum_{r\in x}\hat Z_{r}$ is the total charge on supersite $x$, while $\mathcal{I}_{Q_{1}Q_{2}} \equiv \mathcal{I}_{Q_{1}Q_{2}}^+ \equiv \ket{\hat P_{Q_{1}}} \bra{ \hat P_{Q_{2}}}$ and $\mathcal{D}_{\frac{1}{2}\left(\zeta_{x}+\zeta_{x+1}\right)} \equiv \mathcal{L}_{\frac{1}{2}\left(\zeta_{x}+\zeta_{x+1}\right)}\mathcal{R}_{\frac{1}{2}\left(\zeta_{x}+\zeta_{x+1}\right)}$ are the superoperators (acting on $x,x+1$) introduced in~\secref{Sec:partfunc}, and $b_{Q}\equiv1-(1-\frac{Q}{M})^{2}$. In this equation we neglected terms that are suppressed by at least $1/M^2$. As we argue in App.~\ref{App:SupOpDeriv}, the first term in Eq.~\eqref{eq:Zotoc_onestep} grows ballistically upon applying further layers of unitaries. For the second term, on the other hand, we find that superoperators of the form  $\mathcal{L}_{Q_{x}}\mathcal{R}_{Q_{y}}$ diffuse, unless super-sites $x$ and $y$ are acted upon by the same gate in the circuit, in which case extra contact terms arise. Summing up these different contributions, and applying some further approximations which we detail in App.~\ref{App:SupOpDeriv}, we arrive at the following form of the OTOC operator at time $t$:
\begin{multline}\label{eq:Zotocapprox}
\mathcal{D}_{Z_{r\in x}}\left(t\right) \approx \frac{1-2M}{2M}\,\mathcal{P}^{A_{x}(t)} +\frac{1}{M^{2}}\mathcal{D}_{\zeta_{x}(t)} - \\
-\frac{1}{2M}\sum_{t'<t}\sum_{y\in t'+2\mathbb{Z}}\left(K_{x,y+1}-K_{x,y}\right)^{2}(t')\mathcal{P}^{A_{y}(t-t')}.
\end{multline} 
Here $A_x(t) = [x-t,x+t]$ is a ballistically spreading region around $x$ and $\mathcal{P}^{A}\equiv \mathcal{P}_{1\!\!1_A} / \text{tr}(1\!\!1_A)$ is a projection unto $1\!\!1_A$, the identity operator acting on this region. $K_{x,y}$ denotes the diffusion kernel defined by the right hand side of Eq.~\eqref{eq:diffusiononlattice}. Note that the time evolution of the OTOC involves summing up contributions from diffusion processes starting at different times $t'$. This is a consequence of the aforementioned contact terms, wherein the diffusively spreading densities $\mathcal{R}_{Q_r}$, $\mathcal{L}_{Q_r}$ can be converted into ballistically propagating $\mathcal{P}_{1\!\!1}$ superoperators. In~\appref{App:SupOpDeriv} we derive a more general version of~\eqnref{eq:Zotocapprox} which also involves corrections from finite $\mu$.

\begin{figure}
	\includegraphics[width=0.7\columnwidth]{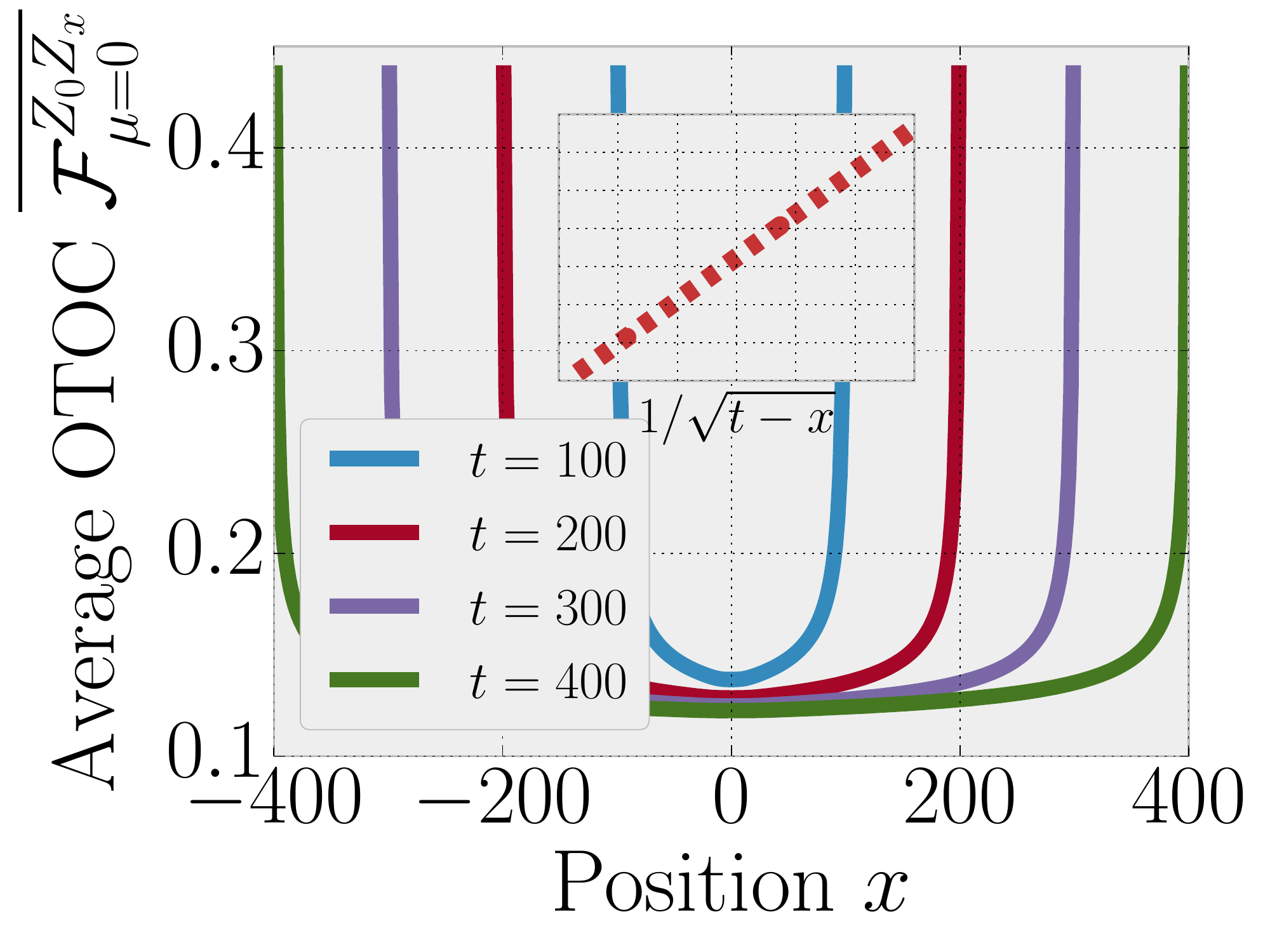}
	\caption{Shape of the OTOC wave front for an initial operator that overlaps with the conserved quantity. Left: front shape arising from the solution~\eqnref{eq:ZZsolution} of the coarse-grained circuit, substituting $M=1$. In evaluating the formula we used the continuum form of the diffusion kernel $K_{x,y}(t) = \exp\left[\frac{(x-y)^2}{4t}\right] / \sqrt{4\pi t}$. While the numerical values (for example the saturation value behind the front) have $\mathcal{O}(1/M)$ corrections, the main features of the shape of the OTOC behind the front should be captured by this solution. Notably, we find that the hydrodynamic tail behind the main front (which travels at speed $v_\text{B} = 1$ in this case) is proportional to $\sim 1 / \sqrt{t-x}$, as illustrated by the inset.}
	\label{fig:ZZsolution}
\end{figure}

Applying the approximate solution~\eqref{eq:Zotocapprox} for the $\hat Z\hat Z$ OTOC, we get
\begin{equation}\label{eq:ZZsolution}
\mathcal{F}^{Z_0Z_x}_{\mu=0} \approx 1 - \delta(x\in A_0(t)) \left[ \frac{2M-1}{2M} - \mathcal{Y}(t,x) \right],
\end{equation}
where $\mathcal{Y}(t,x)$ stands for the double sum appearing in Eq.~\eqref{eq:Zotocapprox} evaluated at the operator $\hat Z_x$, which reads
\begin{multline*}
\mathcal{Y}(t,x) = \frac{1}{2M-1} \sum_{t' < t}\sum_{y} \left( K_{0,y+1}(t')
-K_{0,y}(t') \right)^2 \times \\ \times \delta(x\in A_y(t-t')). 
\end{multline*}
The formula~\eqnref{eq:ZZsolution} is plotted in Fig.~\ref{fig:ZZsolution}. We find that it exhibits a tail behind the front, where the OTOC decays as $(t-x)^{-1/2}$, reproducing the shape found numerically in~\secref{Sec:otoc_partfunc_results}. These hydrodynamic tails were also studied in more detail by Khemani et. al.~\cite{KhemaniOTOC} finding a similar scaling near the front. At the origin, the function $\mathcal{Y}$ relaxes as $at^{-1} + bt^{-1/2}$, consistent with our earlier discussion of power law tails in~\secref{Sec:mu0}. 

Evaluating~\eqnref{eq:Zotocapprox} on the local operator $\hat \sigma_x^+$ gives the result 
\begin{multline*}
\mathcal{F}^{Z_0\sigma^+_x}_{\mu=0} \approx \delta(r\in A_0(t)) \left[ -\frac{1}{2} \mathcal{Y}(t,x) - \frac{K_{0x}^2(t)}{M^2}  \right] + \\ 
+\frac{1}{2} \delta(r\notin A_0(t)).
\end{multline*}
The main contribution that determines the shape of the OTOC front, is given by the same function, $\mathcal{Y}(t,x)$, as for the $\hat Z\hat Z$ OTOC. This implies that the shape of the tail behind the front is the same as the one seen in Fig.~\ref{fig:ZZsolution}. This is also in agreement with the results of Ref. \onlinecite{KhemaniOTOC}.
%

The equation and the formalism in this section marries two notions of hydrodynamics. As shown by our previous work in Ref. \onlinecite{RvK17}, the ballistic spreading of OTOCs can be understood to follow from a biased diffusion equation that describes the spatial growth of an initially local operator. The hydrodynamic nature of this equation is related to an emergent conservation law, that of the norm of the evolving operator, which follows from the unitarity of time evoltuion. A second, more conventional, notion of hydrodynamics arises, as detailed in Sections~\ref{Sec:Diffusion} and~\ref{Sec:mu0} due to the presence of the conserved charge $\hat Q$, which leads to the two locally conserved, diffusing, superoperator densities $\mathcal{L}_{Q_r}$ and $\mathcal{R}_{Q_r}$. Our approximate coarse-grained description couples these two types of quantities: \eqnref{eq:Zotocapprox} includes ballistically spreading terms (namely $\mathcal{P}$), as well as the conserved densities $\mathcal{L}_{Q_r},\mathcal{R}_{Q_r}$, and couplings between these terms. The couplings lead to a conversion of the conserved densities into ballistically propagating components and all such terms, created at different times $t'<t$, need to be summed up to get the OTOC at time $t$. This process results in the OTOC decaying as $\sim 1 / \sqrt{t-x}$ behind the front, as shown in Fig.~\ref{fig:ZZsolution}.

\section{Finite chemical potential}\label{Sec:FiniteMu}
In this section we extend our discussion from the properties of OTOCs in the infinite temperature state $\mu=0$, discussed so far, to the case of a finite chemical potential $\mu$. The chemical potential controlc the equilibrium entropy density of the systems, and in this sense plays a similar role to temperature in Hamiltonian systems. Our results are as follows. i) We show how the finite chemical potential affects the long-time saturation value of the OTOC and use the superoperator formalism developed above in~\secref{Sec:supop_language} to interpret this as a form of thermalization on the space of operators. ii) We confirm our prediction for the saturation values in the random circuit model and show that the hydrodynamic tails observed at $\mu=0$ are also present at small but finite chemical potential. iii) In the $\mu \to \infty$ limit the OTOC spreads out diffusively as a function of space and time, as opposed to having a ballistic light-cone seen in previous sections. In this somewhat special limit, the OTOC can exhibit a double plateau structure: it initially relaxes to a value different from the one predicted in i), only decaying to its final saturation value on time scales $\mathcal{O}(L^2)$; we refer to this initial relaxation as a ``prethermal plateau''. iv) Through a perturbative expansion, we show that the $\mu \to \infty$ results, including the diffusive space-time behavior and double plateau structure, can survive at finite $\mu$ to times $t\sim\mathcal{O}(e^{2\mu})$.

We begin our discussion by rewriting the OTOC, originally defined in Eq.~\eqref{eq:def_oto_part}, as
\begin{equation}\label{eq:otoc_newreg}
\langle \hat V^\dagger(t) \hat W^\dagger \hat V(t) \hat W \rangle_\mu = e^{\frac{\mu}{2}(\lambda_V + \lambda_W)} \frac{\text{tr}\left( \tilde V^\dagger(t) \hat W^\dagger \tilde V(t) \hat W \right)}{\text{tr}\left( e^{-\mu\hat Q} \right)} ,
\end{equation}
where we defined $\tilde V \equiv e^{-\frac{\mu}{4}\hat Q} \hat V e^{-\frac{\mu}{4}\hat Q}$ (this is similar to the regularized version of the OTOC introduced in Ref. \onlinecite{Maldacena2016}). Thus we see that the effect of finite chemical potential can be incorporated entirely into modifying the boundary conditions of the partition function defined in Eq.~\eqref{eq:otoc_as_partfunc}. These new boundary conditions penalize boundary states (of the four-layer system) with large total charges. We can therefore easily generalize the calculation of the classical partiton function, originally introduced in~\secref{Sec:partfunc}, to the case with finite $\mu$. In the following section we will also use the form~\eqref{eq:otoc_newreg} of the OTOC to show that its saturation value can be understood by assuming that the ``state'' $\ket{\tilde V}$ thermalizes at long times.

\subsection{Long-time saturation of OTOCs}\label{Sec:Saturation}

Before examining the time evolution of OTOCs at finite $\mu$, we derive some analytical results on their expected long-time behavior. For the purposes of this section we return to the definition of the OTOC in terms of the squared commutator~\eqref{eq:otoc_def}. We will show that the saturation value that the OTOC approaches as $t\to\infty$ depends non-trivially on both the chemical potential and the type of operators considered (i.e., their charges $\lambda_{V,W}$). Notably, we will show that the out-of-time-ordered part has a non-zero saturation value for $\mu>0$ if either $\lambda_V = 0$ or $\lambda_W = 0$. 

\begin{figure}
	\includegraphics[width=0.7\columnwidth]{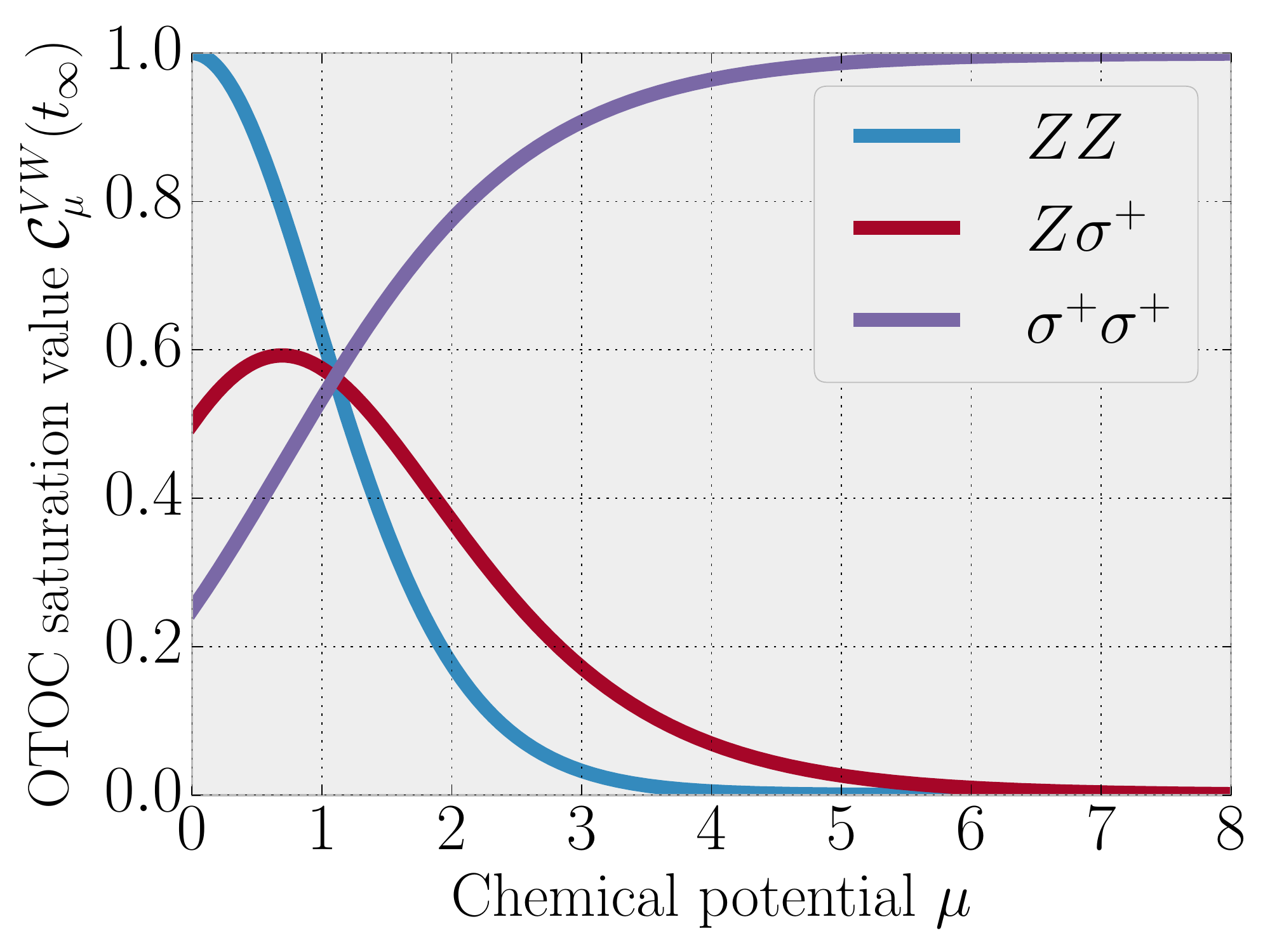}
	\caption{Long-time saturation values of different OTOCs $\mathcal{C}^{WV}$, predicted by Eq.~\eqref{eq:otoc_sat}, as a function of the chemical potential $\mu$.}
	\label{fig:otoc_sat}
\end{figure}

As a starting point of this calculation, we will assume that over vast time scales, well in excess of
the system size, our local random unitary dynamics for the OTOC becomes indistinguishable
from non-local dynamics with the same conserved quantity $\hat Q$. An analogous statement is known to hold for a random circuit without symmetries~\cite{Harrow2009,Brandao2016}, which approximate the first two moments of the Haar-distribution at long times, and therefore it is natural to assume that the same would happen in our case for each symmetry sector. Thus
we will estimate the long time value of Eq.~\eqref{eq:otoc_def} by taking
$\hat V(t)=U^{\dagger}(t)\hat VU(t)$ where $U$ is a unitary that conserves $\hat Q$,
but which is otherwise completely Haar random, without any notion of locality. Averaging the OTOC over such unitaries is expected to yield the saturation values $\overline{\mathcal{C}^{VW}_\mu}\left(t_{\infty}\right)$. In the limit of large system size, provided $\hat V,\hat W$ are operators with subextensive charge (which automatically holds in the case of interest where $\hat V,\hat W$ are local) this approach yields:
\begin{align}\label{eq:otoc_sat}
\overline{\mathcal{C}^{VW}_\mu}\left(t_{\infty}\right)= \frac{1}{2}e^{\mu\lambda_{V}}\braket{\hat W_{\perp}^{\dagger}\hat W_{\perp}^{\phantom{\dagger}}}_\mu \braket{\hat V_{\perp}^{\phantom{\dagger}}\hat V_{\perp}^{\dagger}}_\mu + \nonumber \\
+\frac{1}{2}e^{\mu\lambda_{W}} \braket{\hat W_{\perp}^{\phantom{\dagger}}\hat W_{\perp}^{\dagger}}_\mu \braket{\hat V_{\perp}^{\dagger}\hat V_{\perp}^{\phantom{\dagger}}}_\mu +\mathcal{O}(1/L),
\end{align}
where $\hat W_{\parallel}\equiv\sum_{Q}\frac{\hat P_{Q}}{d_{Q}}\text{tr}\left(\hat P_{Q}W\right)$ is the part of $\hat W$ that is diagonal in charge and $\hat W_{\perp}\equiv \hat W-\hat W_{\parallel}$ is the off-diagonal part. The behavior of Eq.~\eqref{eq:otoc_sat} as a function $\mu$ for different OTOCs of interest is shown in Fig.~\ref{fig:otoc_sat}. Note that Eq.~\eqref{eq:otoc_sat} indicates that for $\mu\neq 0$, if one of the operators involved in the OTOC has non-zero overlap with $\hat Q$, then the out-of-time-ordered part does not saturate to zero, i.e., $\mathcal{F}^{VW}_{\mu\neq 0}(t_\infty) \neq 0$. This fact might also be of relevance for Hamiltonian systems if the operators considered overlap with the local energy density.

We can gain some further insight into the meaning of Eq.~\eqref{eq:otoc_sat} by relating it to the superoperator formalism developed in~\secref{Sec:supop_language}. As we show in App.~\ref{App:Opequilibration}, the saturation values, and indeed the long-time value of any superoperator, can be understood as a form of thermalization, wherein the initial state (in operator space), $\ket{\tilde V (0)} \bra{\tilde V (0)}$, for the operator $\tilde V \equiv e^{-\frac{\mu}{4}\hat Q} \hat V e^{-\frac{\mu}{4}\hat Q}$ introduced in Eq.~\eqref{eq:otoc_newreg}, becomes at long times locally indistinguishable from the ``thermal'' state
\begin{multline}\label{eq:operatorethmaintext}
\overline{\ket{\tilde V (t_\infty)} \bra{\tilde V (t_\infty)}} = \text{tr}\left(\tilde V_{\parallel}^{\dagger}\tilde V_{\parallel}^{\phantom{\dagger}}\right)\frac{\ket{e^{-\frac{\mu}{2}\hat Q}} \bra{e^{-\frac{\mu}{2}\hat Q}}}{Z_{\mu}}+ \\ 
+\text{tr}\left(\tilde V_{\perp}^{\dagger}\tilde V_{\perp}^{\phantom{\dagger}}\right)\frac{e^{-\mu\left(\mathcal{L}_{Q}+\mathcal{R}_{Q}\right)}}{Z_{\mu}^{2}},
\end{multline}
where $Z_{\mu}=\text{tr}\left(e^{-\mu \hat Q}\right)$. The latter part of this expression in~\eqnref{eq:operatorethmaintext} is none other than the Gibbs ensemble with respect to the conserved quantities $\mathcal{L}_{Q},\mathcal{R}_{Q}$ defined in~\secref{ss:OTOCtailsLR}. This result suggests (in a manner we detail in App.~\ref{App:Opequilibration}) that when considering objects like OTOCs or operator weights, the usual notion of thermalization should be supplemented by considering that of equilibration in operator space, as defined above.

The above result relies on averaging over all possible charge-conserving time evolutions without restrictions of locality, which is a valid approximation at time scales long compared to the system size. One might expect that this saturation value is in fact approached on a much shorter, $L$-independent time scale. This is indeed the case for example at $\mu=0$ where the OTOCs relax to the above predicted long-time values either exponentially or as a power law, as we showed above. We observe a similar behavior at sufficiently small $\mu$, as we show in~\secref{Sec:musmall}. In the limit of $\mu\gg 1$, however, we find that the saturation of certain OTOCs can take a time which grows exponentially with $\mu$ and in the limit $\mu\to\infty$, the long-time value of the $\hat \sigma^+\hat \sigma^+$ OTOC in an infinite system deviates from the above prediction by an $\mathcal{O}(1)$ value. For a finite system this means that the OTOC first saturates to a prethermal plateau and only approaches its final value on a time scale that grows as $\sim L^2$. We discuss this in~\secref{Sec:mubig}.

\subsection{Relaxation of OTOCs at $\mu \sim \mathcal{O}(1)$}\label{Sec:musmall}

We now confirm the predictions of the previous section regarding the long-time saturation values of OTOCs at finite $\mu$, by computing their time-evolution numerically in the random circuit model. We also show that the relaxation to these long-time values exhibits the same hydrodynamic tails (at least for small $\mu$) as the ones observed previously in~\secref{Sec:otoc_partfunc_results}. 

\begin{figure}[t]
	\centering
	\includegraphics[width=0.49\columnwidth]{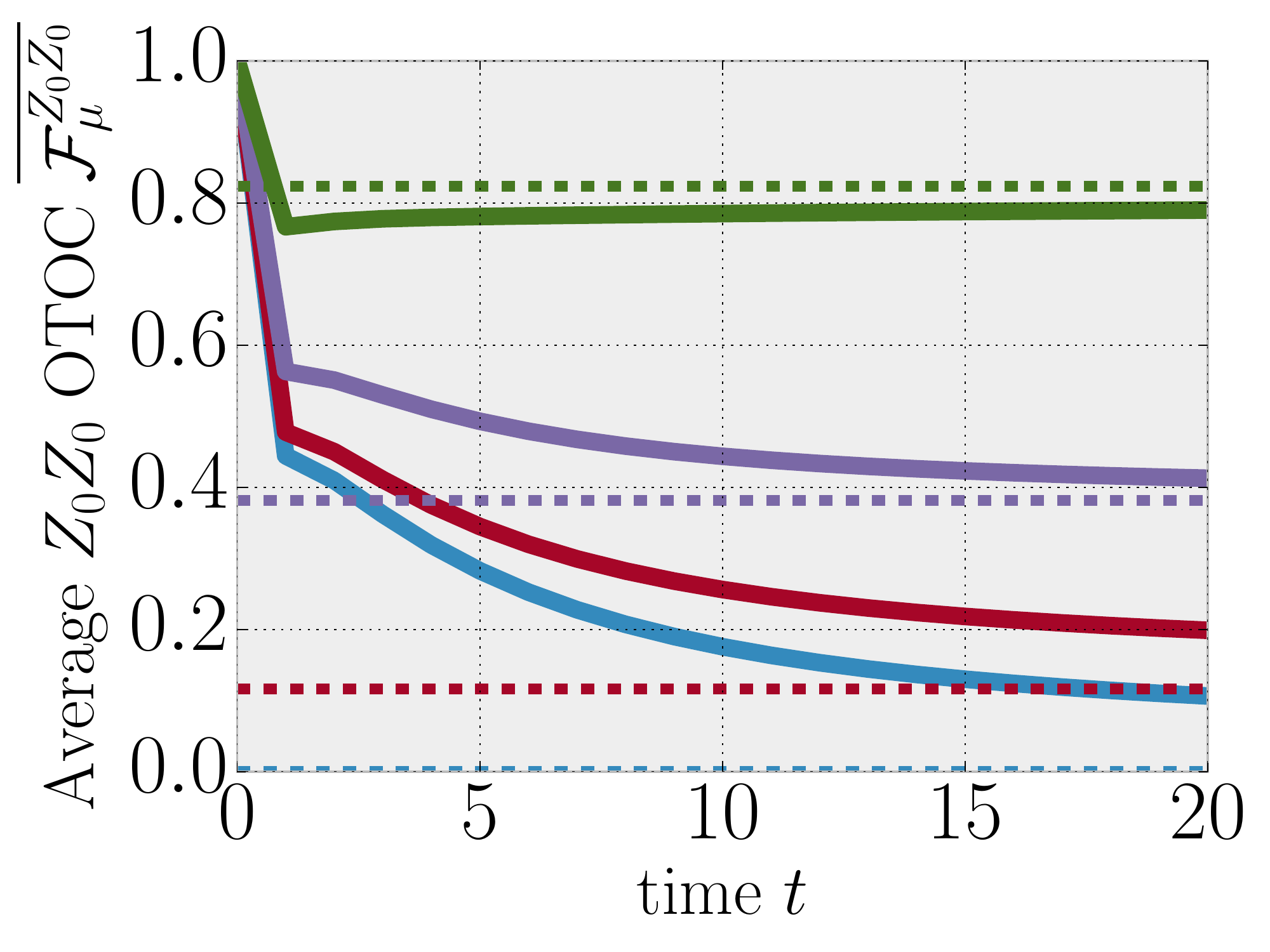}
	\includegraphics[width=0.49\columnwidth]{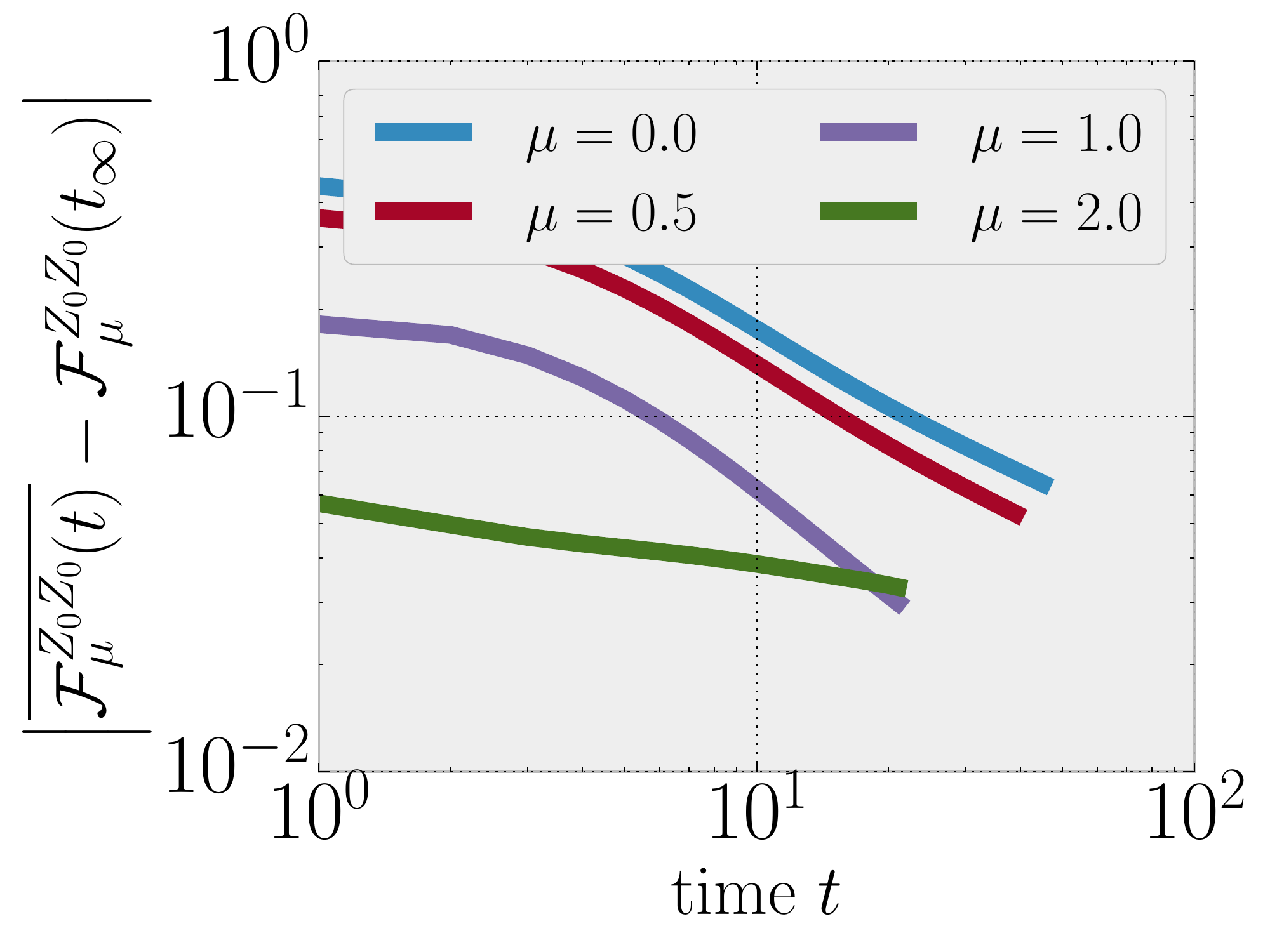}
	\caption{Time evolution of the $\hat Z \hat Z$ OTOC on site zero for chemical potentials $\mu = 0,1/2,1,2$ in the random circuit model. The left panel shows the OTOC approaching the saturation value predicted by Eq.~\eqref{eq:otoc_sat} (dashed horizontal lines). For $\mu = 2$ the OTOC tends to the long-time value from below. The right panel shows the distance from this equilibration value as a function of time. The lines for $\mu = 0$ and $\mu = 1/2$ are parallel, indicating that the latter also exhibits the diffusive $t^{-1/2}$ decay discussed in~\ref{Sec:mu0}.}
	\label{fig:OTOC_finite_mu}
\end{figure}

As discussed at the beginning of~\secref{Sec:FiniteMu}, the mapping of the average OTOC to the classical partition function problem remains intact in the presence of finite $\mu$, except for some additional Gibbs factors which can be incorporated into either (or both) of the boundary conditions. We can then evaluate the average OTOC using the same tensor network methods that we used at $\mu=0$. The results for the OTOC between operators $\hat Z_0, \hat Z_0$ on the same site are shown in Fig.~\ref{fig:OTOC_finite_mu}. We find that the OTOC indeed saturates to the value predicted by Eq.~\eqref{eq:otoc_sat}. Interestingly, for $\mu=2$, the OTOC first drops below this value and then approaches it from below. Moreover, by plotting the distance from saturation, $\left|\overline{\mathcal{F}^{Z_0Z_0}_{\mu}(t)} - \mathcal{F}^{Z_0Z_0}_{\mu}(t_\infty)\right|$, we find that the hydrodynamical tail of the form $\left|\overline{\mathcal{F}^{Z_0Z_0}_{\mu}(t)} - \mathcal{F}^{Z_0Z_0}_{\mu}(t_\infty)\right| \propto t^{-1/2}$ is present also for finite small $\mu$. Namely, we observe that the $\mu = 1/2$ OTOC decays in exactly the same manner as the one at $\mu = 0$ , which has diffusive relaxation as discissed at length in the previous sections. We find deviations from the $t^{-1/2}$ behavior for larger chemical potentials, although these might correspond to some intermediate-time behavior. As we discuss in the following section, the finite $\mu$ behavior of OTOCs at short times can be very different from the one decribed so far in the $\mu = 0$ case.

\subsection{$\mu \gg 1$ and OTOC diffusion}\label{Sec:mubig}

We next turn our attention to the behavior of OTOCs at low fillings, or large chemical potentials, and argue perturbatively that there is an additional structure arising in this limit, wherein the ballistic OTOC front does not appear at times that are short compared to $e^{2\mu}$. Moreover, certain OTOCs ($\mathcal{F}^{\sigma^+\sigma^+}$ in particular) can initially relax to a value different from the one predicted in the previous section, only approaching their long-time limit at $t \gg e^{2\mu}$.

\begin{figure}[h!]
 \centering
  	\includegraphics[width=0.75\columnwidth]{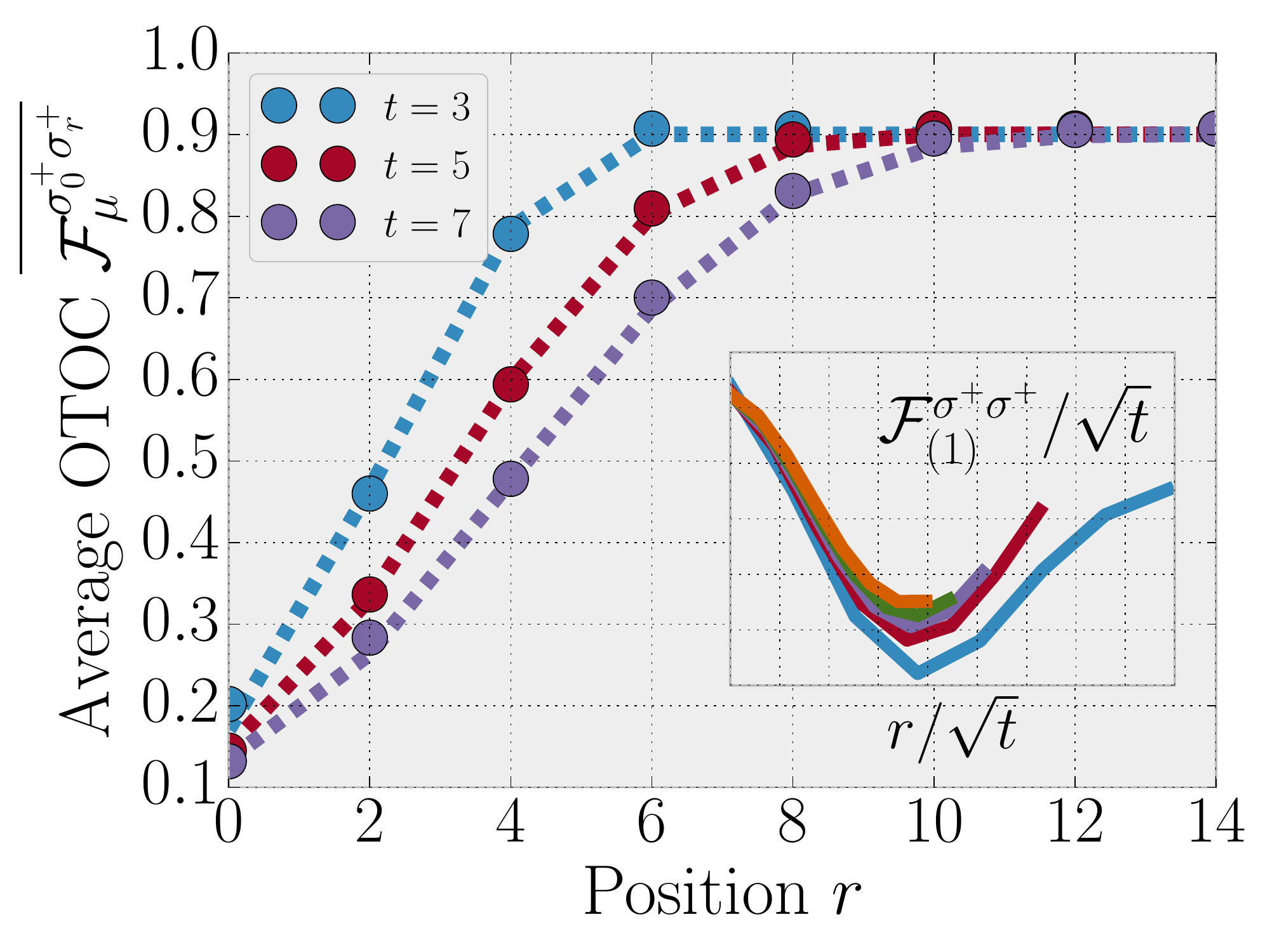}
\caption{Comparison of the perturbative expansion to TEBD results at short-times for the average $\hat \sigma^+ \hat \sigma^+$ OTOC at chemical potential $\mu=3$. Dots: TEBD results averaged over 100 circuits; dashed lines: perturbative expansion at $\mathcal{O}(e^{-\mu})$. The perturbative result agrees very well with the TEBD numerics up to the times considered. Inset: the $\mathcal{O}(e^{-\mu})$ correction to the OTOC $\mathcal{F}^{\sigma^+_0\sigma^+}_{(1)}$. We observe an approximate collapse of the data when $\mathcal{F}^{\sigma^+_0 \sigma^+_r}_{(1)} / \sqrt{t}$ is plotted as a function of $r/\sqrt{t}$, indicating that the OTOC is still diffusive in nature.}
 \label{fig:OTOC_pp_compare}
 \end{figure}

As discussed in~\secref{ss:Optails}, in the infinite temperature ensemble, i.e., at $\mu=0$, OTOCs are closely related to the problem of operator spreading, sampling over all coefficients appearing in Eq.~\eqref{eq:opspreading} with equal measures (see Eq.~\eqref{eq:otocnosymm}). This explains the ballistic spreading of OTOCs, which in this language is a simple consequence of the fact that there are exponentially more long Pauli strings than short ones. However, when $\mu$ is increased the OTOC will be more and more dominated by states with a few charges. Here, we set out to explain how this affects their space-time structure and saturation behavior in the limit $\mu\gg 1$. In this limit we can expand the OTOC in powers of $e^{-\mu}$ and find that the terms in this perturbative expansion describe a diffusively, rather than ballistically, spreading OTOC. This diffusive behavior is exhibited by the three lowest orders of the expansion, and we conjecture that it survives up to a time scale $t \sim \mathcal{O}(e^{2\mu})$, at which point the perturbation theory breaks down. While the method is only well controlled in the $\mu\gg 1$ limit, there is excellent qualitative agreement between the results of this section and those from TEBD even when $\mu\approx 3$ (see \figref{fig:OTOC_pp_compare}). Thus, we believe the results in this section could be very useful in developing a qualitative description of the early time behavior of OTOCs in low temperature strongly coupled systems (in particular systems not permitting a quasiparticle description). 

The starting point of the perturbative description is given by expanding the boundary conditions in orders of $e^{-\mu/4}$ as 
\begin{align}\label{eq:expand_omega}
e^{-\frac{\mu}{4}\hat Q} = \prod_{r=1}^{L} \left( \hat P_r + e^{-\frac{\mu}{4}} \hat Q_r \right) & & \hat P_r = 1\!\!1 - \hat Q_r.
\end{align}
When expanding this product, the different terms correspond to different number of particles as defined in~\secref{Sec:partfunc} (see Fig.~\ref{fig:linetypes} in particular). Since the total number of such particles is conserved during evolution with the circuit, they have to be the same in both boundary conditions. Gathering all terms with the same power of $e^{-\mu}$ we find that the average value of the OTOC can be expanded as a power series of the form
\begin{equation}\label{eq:otoc_pert}
\overline{\mathcal{F}^{VW}_{\mu}}(t) \approx \sum_N e^{-N\mu} \mathcal{F} ^{VW}_{(N)}(t),
\end{equation}
where the $\mathcal{O}(e^{-N\mu})$ term corresponds to initial and final conditions with $2N + \lambda_V + \lambda_W$ particles. $\mathcal{F} ^{VW}_{(N)}(t)$ can then be evaluated by considering the same partition function as defined in~\secref{Sec:partfunc} but with the boundary conditions restricted by the total number of particles, which can therefore be more efficiently calculated, even using exact diagonalization techniques.

\begin{figure}[t]
 \centering
  	\includegraphics[width=0.49\columnwidth]{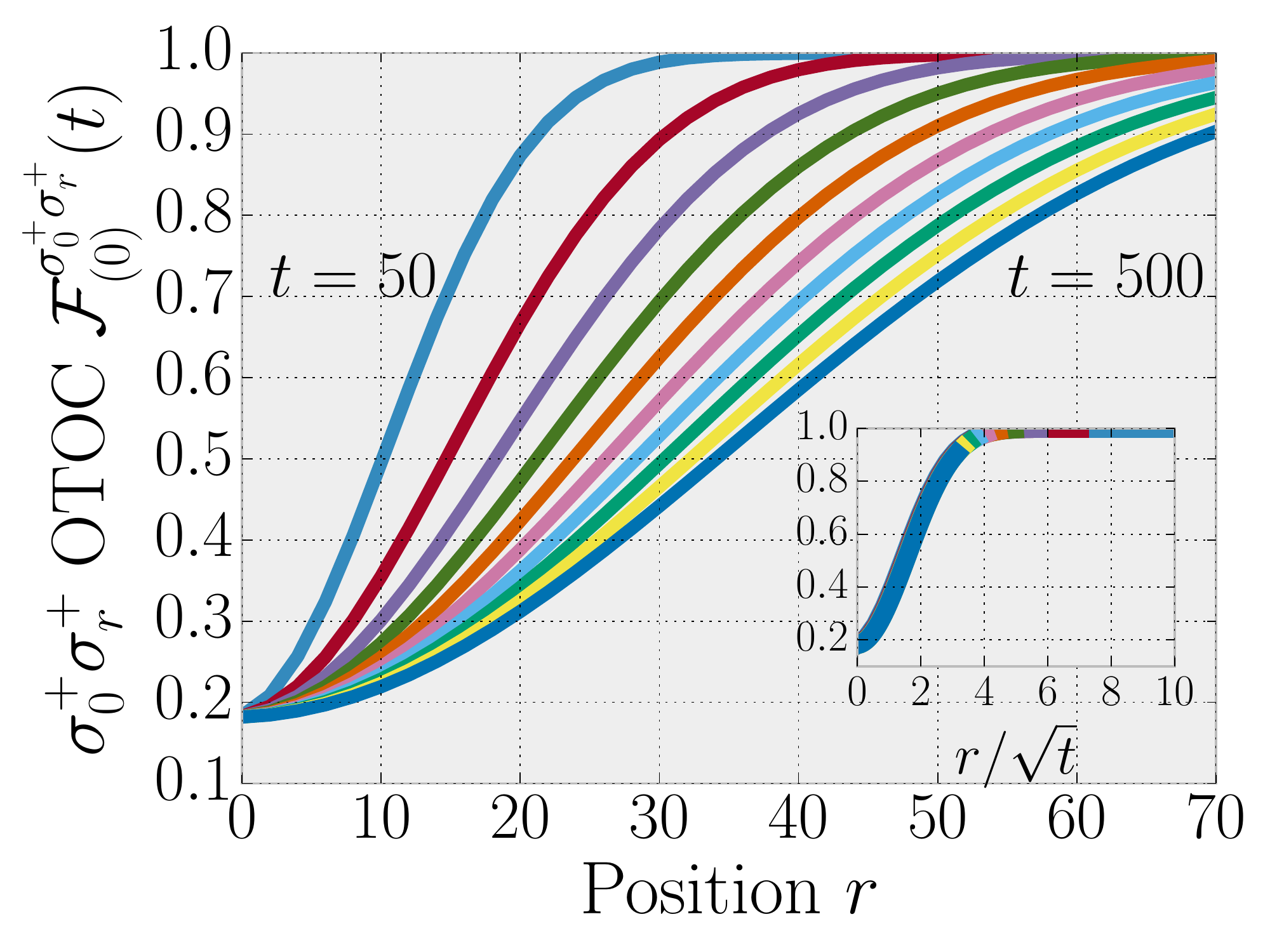}
  	\includegraphics[width=0.49\columnwidth]{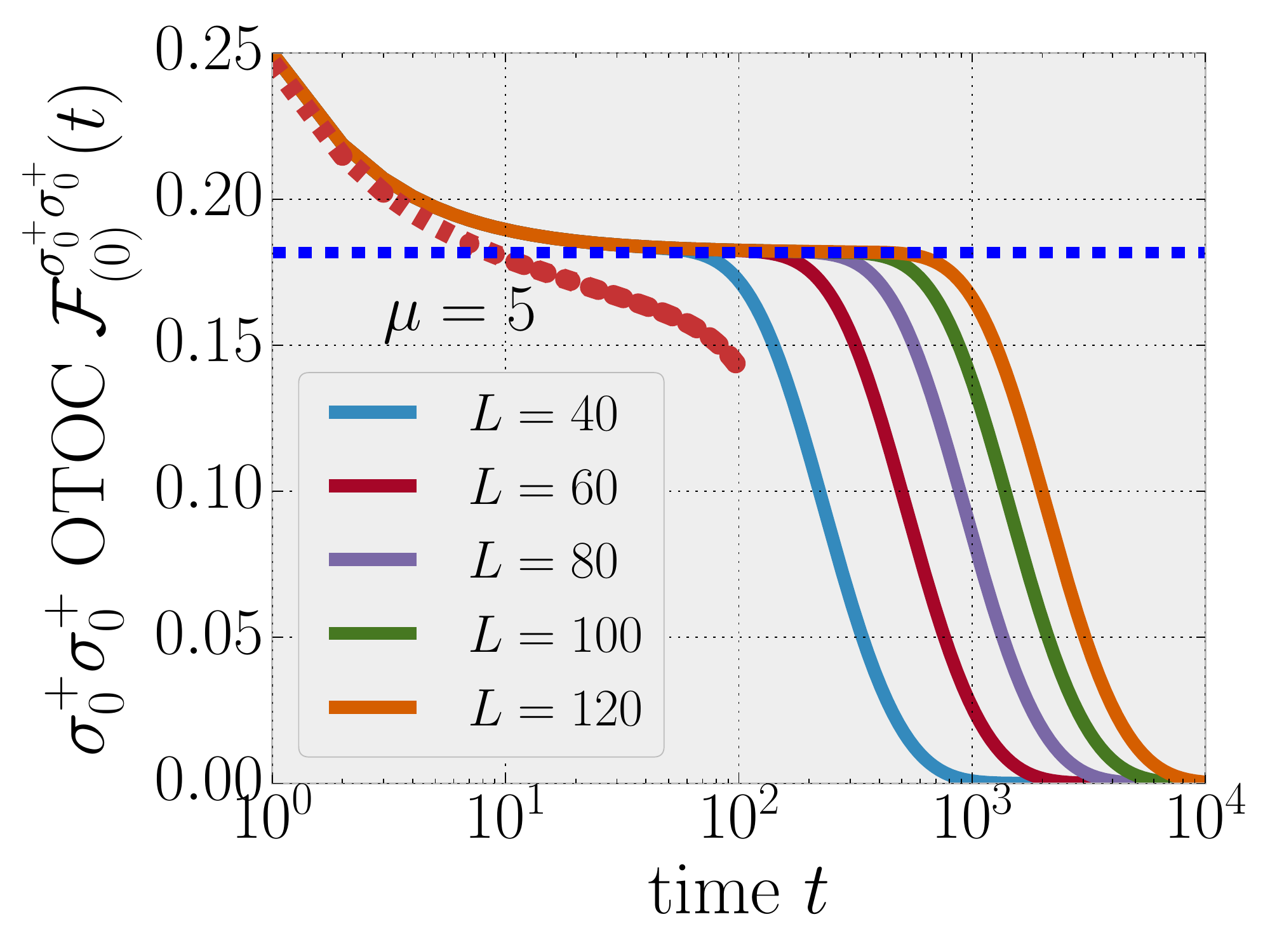}
\caption{The OTOC $\mathcal{F}$ between $\hat \sigma_0^+$ and $\hat \sigma_r^+$ at infinite chemical potential. Left: The OTOC as a function of initial distance $r$ for times $t=50,100,\ldots,500$ in an infinite system. The OTOC spreads diffusively and saturates to a ``prethermal'' plateau behind the front. The inset shows the data collapse when the position is rescaled as $r\to r/\sqrt{t}$ Right: The OTOC first saturates to the value $\frac{1}{2} - \frac{1}{\pi}$ (dashed horizontal line) as $1/\sqrt{t}$. Then at a later timescale $t\sim L^2$ it decays to zero. At late times its value decreases as $\exp(-\pi^2 t/L^2)$. The red dashed line shows the next order prediction at $\mu=5$, which indicates that the plateau survives up to a time scale that diverges with $\mu$.}
 \label{fig:OTOC_pp_exact}
 \end{figure}

Here we detail the behavior of the $\hat \sigma^+\hat \sigma^+$ OTOC which has a non-trivial behavior even at zeroth order, leaving the discussion of other OTOCs to~\appref{App:Qp_lowT}. The zeroth order contribution can be computed from a random walk problem involving a pair of particles that annihilate upon meeting each other, which has an exact solution as we show in App.~\ref{App:2pwalk} One particularly interesting property of the solution is the absence of a ballistically propagating front. Indeed, as suggested by the formulation of the problem as a two-particle random walk, the spreading of the OTOC front is entirely diffusive, depending only on the combination $r/\sqrt{t}$, where $r$ is the spatial separation of the operators in the OTOC. This is demonstrated in the left panel of Fig.~\ref{fig:OTOC_pp_exact}. This is in contrast to the behavior seen at $\mu=0$ in \figref{fig:otoc_mu0} (and general expectations of ballistic propagation), and as we argue below is a property of the perturbative expansion that is in general valid up to a $\mu$-dependent time scale. 

Furthermore, if we consider the same OTOC at a fixed distance $r$ as a function of time, we find a double plateau structure: it first saturates to the value $\frac{1}{2} - \frac{1}{\pi}$ on an $\mathcal{O}(r^2/D)$ timescale, where $D$ is the charge diffusion constant defined by~\eqnref{eq:diffusion}, and only goes to zero, as predicted by Eq.~\eqref{eq:otoc_sat}, when the particles reach around the whole system, at times $\mathcal{O}(L^2/D)$. This is illustrated on the right panel of Fig.~\ref{fig:OTOC_pp_exact}. As we show in App.~\ref{App:2pwalk}, this latter result, the non-commutativity of the $L\to\infty$ and the $t\to\infty$ limits, can be understood from the fact that in an infinite system two random walkers have a finite probability of avoiding each other forever, while they have to meet eventually if the system is finite. Moreover, the mapping from the OTOC to the above random walk problem also holds if we consider similar random circuits in higher dimensions, in which case the probability for crossing paths is smaller, and the deviation from the thermal expectation value in the thermodynamic limit is even larger.

Computing the next term, $\mathcal{F}^{\sigma^+\sigma^+}_{(1)}$, which is of order $\mathcal{O}(e^{-\mu})$, we find that it increases as $\sqrt{t}$ up to an $\mathcal{O}(L)$ value, as shown in the inset of Fig.~\ref{fig:OTOC_pp_compare}. Similarly, we find that the ratio $\mathcal{F}^{\sigma^+\sigma^+}_{(2)} / \mathcal{F}^{\sigma^+\sigma^+}_{(1)}$ of the second and first order terms (not shown here) also increases as $\sqrt{t}$. This suggests that the perturbative expansion is valid up to a time scale of order $t \sim e^{2\mu}$, at which point all terms become of comparable size. Moreover, while the second order contribution does lead to a speed-up of the spreading of the OTOC compared to the $\mu=\infty$ result shown in Fig.~\ref{fig:OTOC_pp_exact}, it is still diffusive as also illustrated by the same inset. This suggests that the diffusive behavior persists up to the aforementioned $\mathcal{O}(e^{2\mu})$ time scale. Therefore, the $\hat \sigma^+ \hat \sigma^+$ OTOC will saturate to the prethermal plateau seen in Fig.~\ref{fig:OTOC_pp_exact}, if $\mu$ is sufficiently large, indicating that the scrambling time~\cite{Maldacena2016,RobertsDesign} associated to saturation of the OTOC can be exponentially large in $\mu$. As shown by Fig.~\ref{fig:OTOC_pp_compare}, the expansion up to $\mathcal{O}(e^{\mu})$ agrees well with numerical TEBD results even for $\mu=3$ at short times $t\leq 7$.

\begin{figure}
	\centering
	\includegraphics[width=1.\columnwidth]{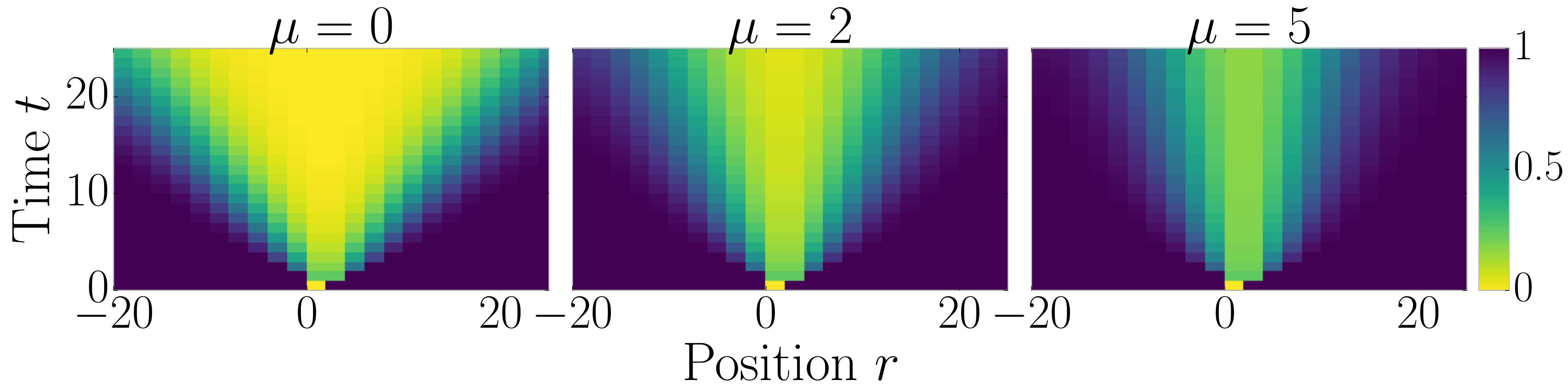}
	\caption{Space-time spreading of the $\hat \sigma^+ \hat \sigma^+$ OTOC $\mathcal{F}^{\sigma_0^+\sigma_r^+}_{\mu}$ for different the chemical potentials $\mu=0,2,5$. The ballistic light cone observed for $\mu = 0$ slows down and gives way to a diffusively spreading OTOC at large $\mu$.}
	\label{fig:lightcones_mu}
\end{figure}

To confirm the results from the perturbative expansion we also computed the $\hat \sigma^+ \hat \sigma^+$ otoc exactly at different finite $\mu$, using the exact partition function~\eqref{eq:otoc_newreg}. We observe a pronounced slow-down of the OTOC spreading even for $\mu = 2$ up to times $t\approx 20$, as shown in Fig.~\ref{fig:lightcones_mu}.
 
We find similar behavior for the out-of-time-ordered part of the OTOC between operators $\hat{Q}_0$ and $\hat{Q}_r$ (note that the OTOC $\mathcal{F}^{Q_0Q_r}_{\mu}$ is related to $\mathcal{F}^{Z_0Z_r}_{\mu}$ via Eq.~\eqref{eq:QtoZ}). While both the first and second order contributions decay in time as a power law and than saturate to an $L$-dependent value, their ratio, $\mathcal{F}^{Q_0Q_r}_{(2)} / \mathcal{F}^{Q_0Q_r}_{(1)}$, increases in time as $\sqrt{t}$ until it saturates to a value which is linear in system size. The OTOC between $\hat{Q}_0$ and $\hat \sigma^+_r$ on the other hand shows a more complicated, non-monotonic behavior. The data for these two cases is presented in App.~\ref{App:Qp_lowT}.

In summary, we find that a variety of intriguing phenomena can occur in OTOCs at early times when the available space for the dynamic is restricted by a finite, large chemical potential. The most robust of these seems to be the initial diffusive spreading of the OTOC at early times. Whether this initial behavior has some bearing on the shape of the OTOC front at later times is an interesting question for further study. 

\section{Conclusions}\label{Sec:Conclusion}

In this work we investigated the dynamics of a U$(1)$ symmetric local unitary circuit, which we propose as a toy model for ergodic many-body systems at long length and time scales for the purposes of calculating transport and OTOCs. We proved that the conserved charge in this system obeys an exact diffusion equation on average, in agreement with the expectation that conserved quantities diffuse in generic, locally interacting lattice spin systems at high temperatures~\cite{BLOEMBERGEN1949,DEGENNES1958,KADANOFF1963}. We provided both analytical arguments and numerical evidence that this leads to the appearance of hydrodynamic tails in out-of-time-ordered correlators of operators that overlap with the total conserved quantity. We also provided numerical evidence that the same hydrodynamic tails appear in a clean periodically driven spin-chain. Furthermore, we argued that these hydrodynamic tails manifest themselves in a particular shape of the OTOC wave front.

In the course of explaining the hydrodynamics of OTOCs we developed a general formalism, involving superoperators, to describe the spatial spreading of operators and the evolution of OTOCs in a unified framework. We believe that this formalism will prove useful in other settings as well. In particular we noted the appearance of conserved superoperators $\mathcal{L}_Q,\mathcal{R}_Q$, whose diffusive behavior is connected to the power law relaxation of the OTOC. Since the diffusion of $\mathcal{L}_Q,\mathcal{R}_Q$ is a direct consequence of the diffusion of $Q$ itself, these arguments, and their conclusions regarding hydrodynamic tails in OTOCs, should generalize to other systems that exhibit diffusive transport, including those with energy conservation. A corollary of this new formalism is an interpretation of long-time saturation of OTOCs in terms of a generalized notion of thermalization for operators rather than states

In the last part of the paper we developed a perturbatve expansion capturing the short-time behavior of OTOCs at low filling, and found that the ballistic behavior usually associated with OTOCs can only develop at time scales that are exponentially large compared to the chemical potential $\mu$, while they initially have a diffusive space-time structure instead. Moreover, we found that in this regime a peculiar double plateau structure appears for the $\hat \sigma^+ \hat \sigma^+$ OTOC, wherein it initially saturates to a prethermal plateau and only decays to zero on a similar, $\mathcal{O}(e^{2\mu})$ time scale.

It would be an interesting direction for future reseach to probe the detailed space-time structure of OTOCs in Hamiltonian systems, in search for the particular front shape we predicted in this paper. Similarly, it is an important open question whether the same behavior can be extracted from field theoretic calculations of OTOCs~\cite{Swingle17,Aleiner16,Stanford15,Stanford2016}. Another possible direction is to extend our results to higher dimensions, possibly by considering random circuit models similar to the one introduced here.

\emph{Related Work:} Shortly before the completion of this manuscript we became aware of closely related work by Khemani et. al.~\cite{KhemaniOTOC}, which appeared in the same ArXiv posting. While they take a slightly different approach, our results appear to agree where they overlap.

\acknowledgements
We are grateful to Michael Knap, Johannes Hauschild and especially Shivaji Sondhi for many useful discussions. TR and FP acknowledge the support of  Research Unit FOR 1807 through grants no. PO 1370/2-1, and the Nanosystems Initiative Munich (NIM) by the German Excellence Initiative. FP also acknowledges support from the European Research Council (ERC) under the European Union's Horizon 2020 research and innovation programme (grant agreement no. 771537). CvK is supported by a Birmingham Fellowship. 


\input{OSWS_resubmit.bbl}

\onecolumngrid
\appendix

\section{Average effect of a single gate}\label{App:Identities}

In this Appendix we derive Eq.~\eqref{eq:onegate_avg_4layers} that describes the average effect of a single 2-site gate on four copies of the Hilbert space (i.e., on the time evolution of superoperators), relevant for calculation of the OTOC. Let us start by examining the simpler problem of the average time evolution of an operator, already discussed in Eq.~\eqref{eq:onestep_opavg}, and re-derive the result in a slightly different language. An operator $\hat O$ evolves under the effect of the unitary $U$ as 
\begin{equation}
(U^\dagger \hat O U)_{\alpha\beta} = U^*_{\gamma\alpha} O_{\gamma\delta} U^{\phantom{*}}_{\delta\beta} = O_{\gamma\delta} (U^* U)_{(\gamma\delta)(\alpha\beta)},
\end{equation}
i.e. we can think of it as being evolved by the superoperator $U^* \otimes U$. Now let us imagine that $U$ is a 2-site unitary, with the block-diagonal structure $U = \sum_Q U_Q$, where $U_Q$ acts on states with total charge $Q$. We can then use the fact that the blocks are independent Haar-random matrices to evaluate the average. For an $n\times n$ random unitary matrix $u$, the properties of the Haar-distribution imply that $\overline{u} = 0$ and $\overline{u^* \otimes u} = \frac{1}{n} \ket{\hat{1\!\!1}}\bra{\hat{1\!\!1}}$, where $\ket{\hat{1\!\!1}}\bra{\hat{1\!\!1}} \equiv \mathcal{P}_{1\!\!1}$ is a superoperator projecting (up to a normalization constant) on the identity, using the notation of~\secref{Sec:supop_language}. We conclude that 
\begin{equation}\label{eq:2layer_avg}
\overline{U^* \otimes U} = \sum_{Q_1,Q_2} \overline{U^*_{Q_1} \otimes U^{\phantom{*}}_{Q_2}}  = \sum_{Q} \overline{U^*_{Q} \otimes U^{\phantom{*}}_{Q}}  = \sum_Q \frac{1}{d_Q} \ket{\hat P_Q} \bra{\hat P_Q},
\end{equation}
where $\hat P_Q$ are projectors acting on the two-site Hilbert space. The above expression acts on an operator as $\hat O \to \sum_Q \frac{1}{d_Q} \text{tr}(\hat{P}_Q \hat O) \hat{P_Q}$. 

For evaluating the OTOC we need to know how to average the time evolution on four, rather than two copies of the Hilbert space. For this four-layer case we have three distinct ways of pairing up the unitaries which gives
\begin{equation}\label{eq:4layer_avg_all}
\overline{U^* \otimes U \otimes U^* \otimes U} = \sum_{Q_1 \neq Q_2} \left( \overline{U^{*}_{Q_1} \otimes U^{\phantom{*}}_{Q_1} \otimes U^{*}_{Q_2} \otimes U^{\phantom{*}}_{Q_2}} + \overline{U^{*}_{Q_1} \otimes U^{\phantom{*}}_{Q_2} \otimes U^{*}_{Q_2} \otimes U^{\phantom{*}}_{Q_1}} \right ) + \sum_{Q} \overline{U^{*}_{Q} \otimes U^{\phantom{*}}_{Q} \otimes U^{*}_{Q} \otimes U^{\phantom{*}}_{Q}}.
\end{equation}
For the $Q_1 \neq Q_2$ terms, where each block only appears at most twice, we can use the result of Eq.~\eqref{eq:2layer_avg}. In the simplest case this gives
\begin{equation}\label{eq:4layer_avg_t1}
\overline{U^{*}_{Q_1} \otimes U^{\phantom{*}}_{Q_1} \otimes U_{Q_2}^{*} \otimes U^{\phantom{*}}_{Q_2}} = \frac{1}{d_{Q_1}} (\ket{\hat P_{Q_1}}\bra{\hat P_{Q_1}}) \otimes \frac{1}{d_{Q_2}} (\ket{\hat P_{Q_2}}\bra{\hat P_{Q_2}}) \equiv \frac{1}{d_{Q_1} d_{Q_2}} \ket{\mathcal{I}_{Q_1Q_2}^{+}}\bra{\mathcal{I}_{Q_1Q_2}^{+}},
\end{equation}
where we defined $\ket{\mathcal{I}_{Q_1Q_2}^{+}} \equiv \sum_{\alpha \in \mathcal{H}_{Q_1}} \sum_{\beta \in \mathcal{H}_{Q_2}} \ket{\alpha\alpha\beta\beta}$. The second term in Eq.~\eqref{eq:4layer_avg_all} corresponds to swapping the second and fourth copies and thus gives
\begin{equation}\label{eq:4layer_avg_t2}
\overline{U^{*}_{Q_1} \otimes U^{\phantom{*}}_{Q_2} \otimes U^{*}_{Q_2} \otimes U^*_{\phantom{*}}} = \frac{1}{d_{Q_1} d_{Q_2}} \ket{\mathcal{I}_{Q_1Q_2}^{-}} \bra{\mathcal{I}_{Q_1Q_2}^{-}},
\end{equation}
where $\ket{\mathcal{I}_{Q_1Q_2}^{-}} \equiv \sum_{\alpha \in \mathcal{H}_{Q_1}} \sum_{\beta \in \mathcal{H}_{Q_2}} \ket{\alpha\beta\beta\alpha}$. 

For the last term we need to apply the Haar identity for the fourth moment of $U$. The result is given by~\cite{Nahum17,RvK17}
\begin{equation}\label{eq:4layer_avg_t3}
\overline{U^{*}_{Q} \otimes U^{\phantom{*}}_{Q} \otimes U^{*}_{Q} \otimes U^{\phantom{*}}_{Q}} = \frac{1}{d_Q^2 - 1} \left[ \left( \ket{\mathcal{I}_{QQ}^{+}} \bra{\mathcal{I}_{QQ}^{+}} + \ket{\mathcal{I}_{QQ}^{-}} \bra{\mathcal{I}_{QQ}^{-}} \right) - \frac{1}{d_Q} \left( \ket{\mathcal{I}_{QQ}^{+}} \bra{\mathcal{I}_{QQ}^{-}} + \ket{\mathcal{I}_{QQ}^{-}} \bra{\mathcal{I}_{QQ}^{+}} \right) \right].
\end{equation}
Combining Eqs.~\eqref{eq:4layer_avg_t1}-\eqref{eq:4layer_avg_t3} we get the full result for the average of a single gate in the four-layer system given in Eq.~\eqref{eq:onegate_avg_4layers}.

A significant difficulty of this charge-conserving circuit, compared to the one without symmetries, is that the states that appear when averaging over a two-site gate do not factorize into independent states on the two sites. If we want to write them in terms of such single-site states (living on four copies of a single site) they become
\begin{align}\label{eq:split_legs}
\ket{\mathcal{I}_{Q_1Q_2}^{+}}  = \sum_{\alpha\beta\gamma\delta} \ket{\alpha\alpha\beta\beta}_1 \ket{\gamma\gamma\delta\delta}_2 \, \delta_{\alpha + \gamma = Q_1} \delta_{\beta + \delta = Q_2} & & \ket{\mathcal{I}_{Q_1Q_2}^{-}}  = \sum_{\alpha\beta\gamma\delta} \ket{\alpha\beta\beta\alpha}_1 \ket{\gamma\delta\delta\gamma}_2 \, \delta_{\alpha + \gamma = Q_1} \delta_{\beta + \delta = Q_2}.
\end{align}
Let us focus on the case of only two states, $\ket{0}$ and $\ket{1}$, on each site site. A possible basis of operators on a single sites is then given by $\hat Q = \ket{1}\bra{1}$, $\hat P \equiv 1\!\!1 - \hat Q = \ket{0}\bra{0}$, $\hat \sigma^+ = \ket{1}\bra{0}$ and $\hat \sigma^- = \ket{0}\bra{1}$. We can then write Eq.~\eqref{eq:split_legs} in terms of the following six local states:
\begin{align}
\ket{0} &\equiv \ket{0000} = \ket{\hat P}\bra{\hat P} = \mathcal{L}_{P} \mathcal{R}_{P} & \ket{1} &\equiv \ket{1111} = \ket{\hat Q}\bra{\hat Q} = \mathcal{L}_{Q} \mathcal{R}_{Q} \nonumber \\
\ket{A} &\equiv \ket{1100} = \ket{\hat Q} \bra{\hat P} = \mathcal{L}_{\sigma^+} \mathcal{R}_{\sigma^-} & \ket{B} &\equiv \ket{0011} = \ket{\hat P} \ket{\hat Q} = \mathcal{L}_{\sigma^-} \mathcal{R}_{\sigma^+} \nonumber \\
\ket{C} &\equiv \ket{1001} = \ket{\hat \sigma^+} \bra{\hat \sigma^+} = \mathcal{L}_{Q} \mathcal{R}_{P} &\ket{D} &\equiv \ket{0110} = \ket{\hat \sigma^-} \bra{\hat \sigma^-} = \mathcal{L}_{P} \mathcal{R}_{Q},
\end{align}
where we have included their interpretation as superoperators, using the definitions of~\secref{Sec:supop_language}. The states appearing in the tensor corresponding to a single two-site gate can then be written in terms of the above states on each site as
\begin{align}
\ket{\mathcal{I}_{00}^{+}} &= \ket{\mathcal{I}_{00}^{-}} = \ket{0} \ket{0} &
\ket{\mathcal{I}_{22}^{+}} &= \ket{\mathcal{I}_{22}^{-}} = \ket{1} \ket{1} \nonumber \\
\ket{\mathcal{I}_{02}^{+}} &= \ket{A} \ket{A} &
\ket{\mathcal{I}_{02}^{-}} &= \ket{C} \ket{C} \nonumber \\
\ket{\mathcal{I}_{20}^{+}} &= \ket{B} \ket{B} &
\ket{\mathcal{I}_{20}^{-}} &= \ket{D} \ket{D} \nonumber \\
\ket{\mathcal{I}_{01}^{+}} &= \ket{0} \ket{A} + \ket{A} \ket{0} &
\ket{\mathcal{I}_{01}^{-}} &= \ket{0} \ket{C} + \ket{C} \ket{0} \nonumber \\
\ket{\mathcal{I}_{10}^{+}} &= \ket{0} \ket{B} + \ket{B} \ket{0} &
\ket{\mathcal{I}_{10}^{-}} &= \ket{0} \ket{D} + \ket{D} \ket{0} \nonumber \\
\ket{\mathcal{I}_{21}^{+}} &= \ket{1} \ket{A} + \ket{A} \ket{1} &
\ket{\mathcal{I}_{21}^{-}} &= \ket{1} \ket{C} + \ket{C} \ket{1} \nonumber \\
\ket{\mathcal{I}_{12}^{+}} &= \ket{1} \ket{B} + \ket{B} \ket{1} &
\ket{\mathcal{I}_{12}^{-}} &= \ket{1} \ket{D} + \ket{D} \ket{1} \nonumber \\
\ket{\mathcal{I}_{11}^{+}} &= \ket{0} \ket{1} + \ket{1} \ket{0} + \ket{A} \ket{B} +  \ket{B} \ket{A}&
\ket{\mathcal{I}_{11}^{-}} &= \ket{0} \ket{1} + \ket{1} \ket{0} + \ket{C} \ket{D} +  \ket{D} \ket{C},
\end{align}
where the two states on the right hand side correspond to the two neighboring sites on which the gate acts. Based on these we can compute all the matrix elements of the form $\braket{IJ| \overline{U^* \otimes U \otimes U^* \otimes U} |KL}$ for $I,J,K,L = 0,A,B,C,D,1$ which give us the transition coefficients illustrated in Fig.~\ref{fig:linetypes}. Applying these for each gate in the circuit, and contracting with the appropriate boundary conditions defined by the operators $\hat V$, $\hat W$ and the chemical potential $\mu$, give the 2D partition function one needs to evaluate to compute the OTOC $\mathcal{F}^{VW}_{\mu}$.

We end this appendix by noting that the above formula for the average effect of the 2-site gate can be written in somewhat more compact form by introducing the states $\ket{\mathcal{J}_{Q_1Q_2}} \equiv \ket{\mathcal{I}_{Q_1Q_2}^-} - \frac{\delta_{Q_1Q_2}}{d_{Q_1}} \ket{\mathcal{I}_{Q_1Q_2}^+}$ and renaming $\ket{\mathcal{I}_{Q_1Q_2}^+} \to \ket{\mathcal{I}_{Q_1Q_2}}$. Using this notation Eq.~\eqref{eq:onegate_avg_4layers} becomes
\begin{equation}\label{eq:onegate_Curtbasis}
\overline{U^* \otimes U \otimes U^* \otimes U} = \sum_{Q_1,Q_2} \frac{1}{d_{Q_1}d_{Q_2}} \ket{\mathcal{I}_{Q_1Q_2}}\bra{\mathcal{I}_{Q_1Q_2}} +  \sum_{Q_1,Q_2} \frac{1}{d_{Q_1}d_{Q_2} - \delta_{Q_1Q_2}} \ket{\mathcal{J}_{Q_1Q_2}}\bra{\mathcal{J}_{Q_1Q_2}}.
\end{equation}
We will use this version of the formula in the following appendix to derive Eq.~\eqref{eq:Zotocapprox}.

\section{Derivation of Eq.~\eqref{eq:Zotocapprox}}\label{App:SupOpDeriv}

In this appendix we detail the derivation that leads us to the conjectured long-time form ot the OTOC presented in Eq.~\eqref{eq:Zotocapprox}. Here we consider a more general version of the equation of motion in Eq.~\eqref{eq:Zotoc_onestep}, where we also take into account effects of finite chemical potential. We do this my considering a modified superoperator 
\begin{equation}\label{eq:thermalOTOCoperator}
\mathcal{D}_{\mu,V} \equiv \mathcal{L}_{\omega_\mu V \omega_\mu}\mathcal{R}_{\omega_\mu V^\dagger \omega_\mu},
\end{equation}
where we introduced the notation $\hat \omega_\mu \equiv e^{-\frac{\mu}{4} \hat Q}$. Similarly, it is useful to define a generalized version of the projection superoperators as 
\begin{equation}
\mathcal{P}_{\mu}^{x} \equiv \frac{\ket{e^{-\mu\hat Q_x}}\bra{e^{-\mu\hat Q_x}}}{\text{tr} (e^{-\mu \hat Q_x})},
\end{equation}
where $\hat Q_x$ is the total charge within a single `supersite' $x$, consisting of $M$ sites of the original lattice, as shown in Fig.~\ref{fig:coarsegrained_circuit}.

The superoperator defined in Eq.~\eqref{eq:thermalOTOCoperator} becomes, on average, after a single application of a unitary gate on $2M$ sites,
\begin{align*}
\mathcal{D}_{\mu,Z_{0}}\left(\Delta\tau\right) & =\sum_{Q}\frac{1}{d_{Q}}\mathcal{I}_{QQ}^{x,x+1}e^{-\mu Q}b_{Q} & :\blacklozenge\\
+ & \frac{1}{M^{2}}\mathcal{L}_{\omega_{\mu}\frac{1}{2}\left(\zeta_{x}+\zeta_{x+1}\right)\omega_{\mu}}\mathcal{R}_{\omega_{\mu}\frac{1}{2}\left(\zeta_{x}+\zeta_{x+1}\right)\omega_{\mu}}. & :\bigstar,
\end{align*}
where $b_{Q}\equiv1-(1-\frac{Q}{M})^{2}$. In the main text we argued that the $\blacklozenge$ term is mainly responsible for ballistic spreading, while the $\bigstar$ term is more complicated and involves diffusion of conserved superoperator densities. We detail these arguments below. The resulting solution for the OTOC superoperator is
\begin{equation}
\mathcal{D}_{\mu,Z_{x}}\left(t\right) \approx \alpha_\mu\,\mathcal{P}_{\mu}^{A_{x}(t)} +\frac{1}{M^{2}}\mathcal{D}_{\mu,\zeta_{x}(t)}
+\frac{\alpha_\mu}{2M-1}\sum_{t'<t}\sum_{y\in t'+2\mathbb{Z}}\left(K_{x,y+1}-K_{x,y}\right)^{2}(t')\mathcal{P}_{\mu}^{A_{y}(t-t')},
\end{equation} 
where $\alpha_\mu \equiv \frac{1-2M}{2M} \cosh^{-2}(\mu/2)$ and $\mathcal{P}_\mu^A \equiv \bigotimes_{x\in A} \mathcal{P}_\mu^x$ for the ballistically growing region $A_x(t) = [x-t,x+t]$.

\subsection{Ballistic expansion of $\blacklozenge$\label{sec:Ballistic-expansion-of}}

The operator $\blacklozenge$, defined above, occupies two supersites, $x,x+1$.
Considered as a superoperator on the $2^{2M}$ dimensional Hilbert
space on these two sites, the individual terms give typical expectation
values on local operators of size $d_{Q}e^{-\mu Q}$. Such summands
are, for large $M$, dominated by $\overline{Q}$ in a small window
around $2M/(1+e^{\mu})$. The most significant term is therefore
\begin{align*}
\left(\ket{\hat P_{\overline{Q}}} \bra{\hat P_{\overline{Q}}}\right)^{x,x+1}=\sum_{e_{L}+e_{R}=\overline{Q}}\sum_{f_{L}+f_{R}=\overline{Q}} \ket{\hat P_{e_{L}}^{x} \hat P_{e_{R}}^{x+1}} \bra{\hat P_{f_{L}}^{x} \hat P_{f_{R}}^{x+1}},
\end{align*}
where $e_{L,R}$ and $f_{L,R}$ are local charges on the two sites and $\hat P_e^x$ is a projection unto that charge on site $x$. This term is similarly dominated by those terms with $e_{L,R}=f_{L,R}\approx\overline{Q}/2$.
Taking the approximations together gives
\begin{equation}\label{eq:effbellterm}
\blacklozenge\approx\frac{\alpha_\mu}{d_{\overline{Q}}} e^{-\mu\overline{Q}}
\ket{\hat P_{\overline{Q}/2}^{x} \hat P_{\overline{Q}/2}^{x+1}} \bra{\hat P_{\overline{Q}/2}^{x} \hat P_{\overline{Q}/2}^{x+1}},
\end{equation}
where  $\alpha_\mu \equiv \frac{1-2M}{2M} \cosh^{-2}(\mu/2)$.

We will probe the dynamics of~\eqnref{eq:effbellterm} under unitary
dynamics on $x+1,x+2$. At this point it is useful to remember that the OTOC operator
was originally defined on the whole Hilbert space as $\mathcal{L,R}$ superoperators being pre-
and post-multiplied by thermal factors as in \eqnref{eq:thermalOTOCoperator}. Taking into account these additional factors coming
from site $x+2$ 
\begin{align*}
\blacklozenge & =\frac{e^{-\mu\overline{Q}}}{d_{\overline{Q}}} \ket{P_{\overline{Q}/2}^{x}P_{\overline{Q}/2}^{x+1}}  \bra{P_{\overline{Q}/2}^{x}P_{\overline{Q}/2}^{x+1}}\mathcal{L}_{e^{-\mu Q_{x+2}/2}}\mathcal{R}_{e^{-\mu Q_{x+2}/2}}\\
 & =\frac{e^{-\mu\overline{Q}}}{d_{\overline{Q}}}\mathcal{P}_{\overline{Q}/2}^{x}\mathcal{P}_{\overline{Q}/2}^{x+1}\mathcal{L}_{e^{-\mu Q_{x+2}/2}}\mathcal{R}_{e^{-\mu Q_{x+2}/2}},
\end{align*}
where $\mathcal{P}_{Q}^{x} \equiv \ket{\hat P_Q^x}\bra{\hat P_Q^x}$ is a local projector on the space of operators.

A straightforward but tedious application of~\eqnref{eq:onegate_Curtbasis} gives two contributions to the expression for $\blacklozenge(\Delta\tau)=\blacklozenge(\Delta\tau)_{1}+\blacklozenge(\Delta\tau)_{2}$, corresponding to the first and second terms in Eq.~\eqref{eq:onegate_Curtbasis}, respectively. We detail the calculation of both of these separately below. The first we can write,
making similar large $M$ approximations as above, as
\begin{align*}
\blacklozenge(\Delta\tau)_{1} & \approx e^{-3\mu\overline{Q}}\frac{\alpha_\mu}{\left(\mathcal{Z}_{\mu}^{1}\right)^{3}} \ket{\hat P_{\overline{Q}/2}^{x} \hat P_{\overline{Q}/2}^{x+1} \hat P_{\overline{Q}/2}^{x+2}} \bra{\hat P_{\overline{Q}/2}^{x} \hat P_{\overline{Q}/2}^{x+1} \hat P_{\overline{Q}/2}^{x+2}}
\approx\mathcal{P}^{x,x+1,x+2}_\mu,
\end{align*}
where we have defined $\mathcal{Z}_{\mu}^{1}$ as the partition
function of a singe supersite, which is peaked at charge $\overline{Q}/2$. The
second term, $\blacklozenge(\Delta\tau)_{2}$, is obtained by applying the second term in Eq.~\eqref{eq:onegate_Curtbasis}; it is sub-leading
by a factor at least $\mathcal{O}(1/d_{Q})$, which is typically exponentially
small in $Q$. 

On net, considering the full Hilbert space, we can iterate the above
procedure to argue that
\begin{align*}
\blacklozenge\left(t\right)=\alpha_\mu\bigotimes_{x\in A(t)}\mathcal{P}_\mu^x \bigotimes_{x\notin A(t)}\mathcal{L}_{e^{-\mu Q_{x}/2}}\mathcal{R}_{e^{-\mu Q_{x}/2}},
\end{align*}
where $A(t)$ is a region that ballistically spreads out from initial
site $1$ at a velocity of $2M$. We anticipate that there are $\mathcal{O}(1/M)$
errors involved in this approximation associated with neglecting fluctuations
in the charge arguments of the projectors $\mathcal{P}$.
We leave a more thorough accounting of these errors to other works. 

\subsubsection{Computing $\blacklozenge(\Delta\tau)_{1}$}

Here we apply the first line
of Eq.~\eqref{eq:onegate_Curtbasis} to $\blacklozenge$. We act on supersites
$x+1,x+2$. For the sake of the calculation it is useful to define the local superoperators $\mathcal{I}^{x}_{ef} \equiv \ket{\hat P_e}\bra{\hat P_f}$, acting on a single supersite. Using these we can write
\begin{align*}
\braket{\mathcal{I}_{Q_{1}Q_{2}}^{x+1,x+2}|\mathcal{P}_{\overline{Q}/2}^{x+1}\mathcal{L}_{e^{-\mu Q_{x+2}/2}}\mathcal{R}_{e^{-\mu Q_{x+2}/2}}}  & =\sum_{e_{L}+e_{R}=Q_{1}}\sum_{f_{L}+f_{R}=Q_{2}}\braket{ \mathcal{I}_{e_{L}f_{L}}^{x+1}\mathcal{I}_{e_{R}f_{R}}^{x+2}|\mathcal{P}_{\overline{Q}/2}^{x+1}\mathcal{L}_{e^{-\mu Q_{x+2}/2}}\mathcal{R}_{e^{-\mu Q_{x+2}/2}}} \\
 & =\sum_{e_{L}+e_{R}=Q_{1}}\sum_{f_{L}+f_{R}=Q_{2}}\braket{ \mathcal{I}_{e_{L}f_{L}}^{x+1}|\mathcal{P}_{\overline{Q}/2}^{x+1}} \braket{ \mathcal{I}_{e_{R}f_{R}}^{x+2}|\mathcal{L}_{e^{-\mu Q_{x+2}/2}}\mathcal{R}_{e^{-\mu Q_{x+2}/2}}} \\
 & =\sum_{e_{L}+e_{R}=Q_{1}}\sum_{f_{L}+f_{R}=Q_{2}}\chi_{\overline{Q}/2}\chi_{\overline{Q}/2}\delta_{e_{L}\overline{Q}/2}\delta_{f_{L}\overline{Q}/2}e^{-\mu e_{R}}\delta_{e_{R}f_{R}}\chi_{e_{R}}\\
 & =\delta_{Q_{1}Q_{2}}\left(\chi_{\overline{Q}/2}\right)^{2}e^{-\mu(Q_{1}-\overline{Q}/2)}\chi_{(Q_{1}-\overline{Q}/2)},
\end{align*}
where $\chi_{Q}$ is the size of the 1-supersite Hilbert space with
charge $Q$. Now we put this back into the first line of the evolution
equation to get
\begin{align*}
\blacklozenge(\Delta\tau)_{1}= & \sum_{Q_{1},Q_{2}}\frac{1}{d_{Q_{1}}d_{Q_{2}}}\mathcal{I}_{Q_{1}Q_{2}}^{x+1,x+2}\times\frac{\alpha_\mu e^{-\mu\overline{Q}}}{d_{\overline{Q}}}\mathcal{P}_{\overline{Q}/2}^{x}\times\delta_{Q_{1}Q_{2}}\left(\chi_{\overline{Q}/2}\right)^{2}e^{-\mu(Q_{1}-\overline{Q}/2)}\chi_{(Q_{1}-\overline{Q}/2)}\\
= & \left(\chi_{\overline{Q}/2}\right)^{2}\alpha_\mu e^{-\mu\overline{Q}}\sum_{Q_{1}}\frac{\mathcal{P}_{\overline{Q}/2}^{x}}{d_{\overline{Q}}}\frac{\mathcal{I}_{Q_{1}Q_{1}}^{x+1,x+2}}{d_{Q_{1}}^{2}}\times e^{-\mu(Q_{1}-\overline{Q}/2)}\chi_{Q_{1}-\overline{Q}/2}.
\end{align*}
Note that when we take expectation values of this quantity, we should
find a value of size $\chi^{3}$ where $\chi$ is the typical value
of $\chi_{Q}$ in the thermal ensemble, which is exponentially large
in $M$ for large system size. As a function of $Q_{1}$ the norm
of the terms is peaked around $Q_{1}=\overline{Q}$ giving
\begin{align*}
\blacklozenge(\Delta\tau)_{1} & \approx\left(d_{\overline{Q}/2}^{1}\right)^{3}\alpha_\mu e^{-3\mu\overline{Q}/2}\frac{\mathcal{P}_{\overline{Q}/2}^{x}\mathcal{P}_{\overline{Q}/2}^{x+1}\mathcal{P}_{\overline{Q}/2}^{x+2}}{d_{\overline{Q}}^{3}}\\
 & \approx\alpha_\mu\frac{1}{\left(Z_{\mu}^{1}\right)^{3}} \ket{e^{-\mu\left(\hat Q_{x}+\hat Q_{x+1}+\hat Q_{x+1}\right)}} \bra{ e^{-\mu\left(\hat Q_{x}+\hat Q_{x+1}+\hat Q_{x+1}\right)}}\\
 & =\alpha_\mu\mathcal{P}_{x,x+1,x+2}\left(\mu\right),
\end{align*}
where $\mathcal{Z}_{\mu}^{1}$ is the $1$-site partition function.

\subsubsection{Computing $\blacklozenge(\Delta\tau)_{2}$}

We now apply the second line of~\eqnref{eq:onegate_Curtbasis} to $\blacklozenge$
i.e., to calculate $\blacklozenge(\Delta\tau)_{2}$. The main object of
interest is
\begin{align*}
 \braket{ \mathcal{L}_{P_{Q_{1}}}\mathcal{R}_{P_{Q_{2}}}|\mathcal{P}_{\overline{Q}/2}^{x+1}\mathcal{L}_{e^{-\mu Q_{x+2}/2}}\mathcal{R}_{e^{-\mu Q_{x+2}/2}}}
 & =\sum_{e_{L}+e_{R}=Q_{1}}\sum_{f_{L}+f_{R}=Q_{2}}\braket{ \mathcal{L}_{P_{e_{L}}^{x+1}}\mathcal{R}_{P_{f_{L}}^{x+1}}|\mathcal{P}_{\overline{Q}/2}^{x+1}} \braket{ \mathcal{L}_{P_{e_{R}}^{x+2}}\mathcal{R}_{P_{f_{R}}^{x+2}}|\mathcal{L}_{e^{-\mu Q_{x+2}/2}}\mathcal{R}_{e^{-\mu Q_{x+2}/2}}} \\
 & =\sum_{e_{L}+e_{R}=Q_{1}}\sum_{f_{L}+f_{R}=Q_{2}}\delta_{e_{L}\overline{Q}/2}\delta_{f_{L}\overline{Q}/2}\chi_{\overline{Q}/2}\times e^{-\mu\left(f_{R}+e_{R}\right)/2}\chi_{e_{R}}\chi_{f_{R}}\\
 & =\chi_{\overline{Q}/2}\times e^{-\mu\left(Q_{1}+Q_{2}-\overline{Q}\right)/2}\chi_{Q_{1}-\overline{Q}/2}^{1}\chi_{Q_{2}-\overline{Q}/2}^{1}.
\end{align*}
We also need to calculate 
\begin{align*}
 \frac{\delta_{Q_{1}Q_{2}}}{d_{Q_{1}}}\braket{\mathcal{I}_{Q_{1}Q_{1}}^{x+1,x+2}|\mathcal{P}_{\overline{Q}/2}^{x+1}\mathcal{L}_{e^{-\mu Q_{x+2}/2}}\mathcal{R}_{e^{-\mu Q_{x+2}/2}}} 
 \approx\frac{\delta_{Q_{1}Q_{2}}}{d_{Q_{1}}}(\chi_{\overline{Q}/2})^{2}\times\chi_{Q_{1}-\overline{Q}/2}e^{-\mu(Q_{1}-\overline{Q}/2)}.
\end{align*}
Now by putting everything together we arrive at
\begin{align*}
\blacklozenge(\Delta\tau)_2 \approx \sum_{Q_{1},Q_{2}}\frac{1}{d_{Q_{1}}d_{Q_{2}}-\delta_{Q_{1}Q_{2}}}\frac{\alpha_\mu e^{-\mu\overline{Q}}}{d_{\overline{Q}}}\mathcal{P}_{\overline{Q}/2}^{x}\left(\mathcal{L}_{P_{Q_{1}}}^{x+1,x+2}\mathcal{R}_{P_{Q_{2}}}^{x+1,x+2}-\frac{\delta_{Q_{1}Q_{2}}\mathcal{I}_{Q_{1}Q_{1}}^{x+1,x+2}}{d_{Q_{1}}}\right)\\
\times\left(\chi_{\overline{Q}/2}\times e^{-\mu\left(Q_{1}+Q_{2}-\overline{Q}\right)/2}\chi_{Q_{1}-\overline{Q}/2}\chi_{Q_{2}-\overline{Q}/2}-\frac{\delta_{Q_{1}Q_{2}}}{d_{Q_{1}}}(\chi_{\overline{Q}/2})^{2}\times\chi_{Q_{1}-\overline{Q}/2}e^{-\mu(Q_{1}-\overline{Q}/2)}\right).\\
\end{align*}
Note that expectation values here will take values of order $O(\chi)$ on local product operators. For the $Q$ of interest, this is a factor of $O(\chi^{2})$ smaller than $\blacklozenge(\Delta\tau)_{1}$. So we ignore $\blacklozenge(\Delta\tau)_{2}$. 

\subsection{Evolution of $\bigstar$}\label{sec:Evolution-ofstar}

\subsubsection{Evolve $\frac{1}{M^{2}}\mathcal{L}_{\zeta_{x}}\mathcal{R}_{\zeta_{x}}$ on
sites $x,x+1$}

We now investigate the evolution of the $\frac{1}{M^{2}}\mathcal{L}_{\zeta_{x}}\mathcal{R}_{\zeta_{x}}$ term under a unitary gate on $x,x+1$. Label the two lines of the OTOC evolution in~\eqnref{eq:onegate_Curtbasis} as $\blacktriangle$ and $\blacksquare$, respectively. Consider $\blacktriangle$ first:
\begin{align*}
\blacktriangle= \frac{1}{M^{2}}\braket{ \mathcal{I}_{Q_{1}Q_{2}}^{x,x+1}|\mathcal{L}_{\zeta_{x}}\mathcal{R}_{\zeta_{x}}} =\delta_{Q_{1}Q_{2}}\sum_{e_{L}}\left(1-\frac{2e_{L}}{M}\right)^{2}\chi_{e_{L}}\chi_{Q_{1}-e_{L}},
\end{align*}
where we have used
\begin{align*}
\text{tr}_{x}\left(\hat \zeta_{x}\hat P_{f_{L}}^{x}\hat \zeta_{x}\hat P_{e_{L}}^{x}\right) & =\text{tr}_{x}\left(\left(1-\frac{2\hat{Q}_{x}}{M}\right)\hat P_{f_{L}}^{x}\left(1-\frac{2\hat{Q}_{x}}{M}\right)\hat P_{e_{L}}^{x}\right)
= M^{2}\left(1-\frac{2e_{L}}{M}\right)^{2}\chi_{e_{L}}\delta_{e_{L}f_{L}}.
\end{align*}
Hence 
\begin{align*}
\blacktriangle=\sum_{Q_{1},Q_{2}}\frac{1}{d_{Q_{1}}^{2}}\mathcal{I}_{Q_{1}Q_{1}}^{x+1,x+2}\sum_{e_{L}}\left(1-\frac{2e_{L}}{M}\right)^{2}\chi_{e_{L}}\chi_{Q_{1}-e_{L}}.
\end{align*}

Now we estimate $\blacksquare$ as
\begin{align*}
 \frac{1}{M^{2}}\braket{ \mathcal{L}_{P_{Q_{1}}^{x,x+1}}\mathcal{R}_{P_{Q_{2}}^{x,x+1}}-\frac{\delta_{Q_{1}Q_{2}}\mathcal{I}_{Q_{1}Q_{1}}^{x,x+1}}{d_{Q_{1}}}|\mathcal{L}_{\zeta_{x}}\mathcal{R}_{\zeta_{x}}}
\approx d_{Q_{1}}d_{Q_{2}}\left(1-\frac{Q_{1}}{M}\right)\left(1-\frac{Q_{2}}{M}\right).
\end{align*}
We can drop the second term in the last line because it is a factor
of $O(d_{Q}^{2})$ smaller than the first --- this translates into
being exponentially smaller in $M$ as our final expressions for OTOCs
are dominated by $Q$ for which $d_{Q}^{2}$ is exponentially large
in $M$ at finite chemical potential. This leads to 
\begin{eqnarray*}
\blacksquare & \approx & \sum_{Q_{1},Q_{2}}\frac{1}{d_{Q_{1}}d_{Q_{2}}-\delta_{Q_{1}Q_{2}}}\left(\mathcal{L}_{P_{Q_{1}}^{x,x+1}}\mathcal{R}_{P_{Q_{2}}^{x,x+1}}-\frac{\delta_{Q_{1}Q_{2}}\mathcal{I}_{Q_{1}Q_{1}}^{x,x+1}}{d_{Q_{1}}}\right)\times d_{Q_{1}}d_{Q_{2}}\left(1-\frac{Q_{1}}{M}\right)\left(1-\frac{Q_{2}}{M}\right).\\
\end{eqnarray*}

Combining the two terms $\blacksquare+\blacktriangle$ and dropping further terms
of relative size $O(1/d_{Q}^{2})$ gives 
\begin{align*}
 & \sum_{Q_{1},Q_{2}}\frac{1}{d_{Q_{1}}d_{Q_{2}}}\mathcal{L}_{P_{Q_{1}}^{x,x+1}}\mathcal{R}_{P_{Q_{2}}^{x,x+1}}\left(1-\frac{Q_{1}}{M}\right)\left(1-\frac{Q_{2}}{M}\right)\\
 & +\sum_{Q_{1}}\frac{1}{d_{Q_{1}}^{2}}\mathcal{I}_{Q_{1}Q_{1}}^{x,x+1}\left(\sum_{e_{L}}\left(1-\frac{2e_{L}}{M}\right)^{2}\chi_{e_{L}}\chi_{Q_{1}-e_{L}}-d_{Q_{1}}\left(1-\frac{Q_{1}}{M}\right)^{2}\right).
\end{align*}
The former is readily expressed as $\frac{1}{M^{2}}\mathcal{L}_{\frac{\zeta_{x}+\zeta_{x+1}}{2}}\mathcal{R}_{\frac{\zeta_{x}+\zeta_{x+1}}{2}}$.
The latter term requires more work. Note first that we can exactly
evaluate
\begin{align*}
\sum_{e_{L}}\left(1-\frac{2e_{L}}{M}\right)^{2}d_{e_{L}}^{1}d_{Q_{1}-e_{L}}^{1}-d_{Q_{1}}\left(1-\frac{Q_{1}}{M}\right)^{2}=\frac{Q_{1}}{M^{2}}\left(1-\frac{Q_{1}}{2M}\right)\left(\frac{1}{1-\frac{1}{2M}}\right)d_{Q_{1}},
\end{align*}
so that in total we get
\begin{align*}
 & \frac{1}{M^{2}}\mathcal{L}_{\frac{\zeta_{x}+\zeta_{x+1}}{2}}\mathcal{R}_{\frac{\zeta_{x}+\zeta_{x+1}}{2}}+\frac{1}{M}\left(\frac{1}{1-\frac{1}{2M}}\right)\sum_{Q_{1}}\frac{1}{d_{Q_{1}}}\mathcal{I}_{Q_{1}Q_{1}}^{x,x+1}\frac{Q_{1}}{M}\left(1-\frac{Q_{1}}{2M}\right)\\
= & \frac{1}{M^{2}}\mathcal{L}_{\frac{\zeta_{x}+\zeta_{x+1}}{2}}\mathcal{R}_{\frac{\zeta_{x}+\zeta_{x+1}}{2}}+\frac{1}{2M-1}\sum_{Q_{1}}\frac{1}{d_{Q_{1}}}\mathcal{I}_{Q_{1}Q_{1}}^{x,x+1}b_{Q}.
\end{align*}

\subsection{Evolve $\frac{1}{M^{2}}\mathcal{L}_{\zeta_{x+1}}\mathcal{R}_{\zeta_{x}},\frac{1}{M^{2}}\mathcal{L}_{\zeta_{x}}\mathcal{R}_{\zeta_{x+1}}$
on sites $x,x+1$}

The result of such an evolution can be obtained from that of $\frac{1}{M^{2}}\mathcal{L}_{\zeta_{x}}\mathcal{R}_{\zeta_{x}}(\Delta\tau)$
in the previous section by noting $\frac{1}{M^{2}}\mathcal{L}_{\zeta_{x+1}}\mathcal{R}_{\zeta_{x}}=\frac{1}{M^{2}}\mathcal{L}_{\zeta_{x+1}+\zeta_{x}}\mathcal{R}_{\zeta_{x}}-\frac{1}{M^{2}}\mathcal{L}_{\zeta_{x}}\mathcal{R}_{\zeta_{x}}$
and that $\zeta_{x+1}+\zeta_{x}$ is conserved on $x,x+1$ for the
gate considered. As a result,
\begin{align*}
\frac{1}{M^{2}}\mathcal{L}_{\zeta_{x+1}}\mathcal{R}_{\zeta_{x}}(\Delta\tau) & =\frac{1}{M^{2}}\mathcal{L}_{\zeta_{x+1}+\zeta_{x}}\mathcal{R}_{\frac{1}{2}\left(\zeta_{x}+\zeta_{x+1}\right)}-\frac{1}{M^{2}}\mathcal{L}_{\zeta_{x}}\mathcal{R}_{\zeta_{x}}(\Delta\tau)\\
 & =\frac{1}{M^{2}}\mathcal{L}_{\frac{1}{2}\left(\zeta_{x+1}+\zeta_{x}\right)}\mathcal{R}_{\frac{1}{2}\left(\zeta_{x}+\zeta_{x+1}\right)}-\frac{1}{2M-1}\sum_{Q_{1}}\frac{1}{d_{Q_{1}}}\mathcal{P}_{Q_{1}}^{x,x+1}b_{Q}.
\end{align*}
The result is the same for $\frac{1}{M^{2}}\mathcal{L}_{\zeta_{x}}\mathcal{R}_{\zeta_{x+1}}(\Delta\tau)$. 

\subsubsection{Summing up contact terms}

Let us start by evolving the purely diffusive term at time $t$ by
one time step. Using our results earlier in this section, we obtain
a sum of contact terms in addition to the expected purely diffusive
term:
\begin{align*}
\frac{1}{M^{2}}\mathcal{L}_{\zeta_{x}(t)}\mathcal{R}_{\zeta_{x}(t)} & =\frac{1}{M^{2}}\sum_{yy'}K_{xy'}K_{xy}\mathcal{L}_{\zeta_{y}}\mathcal{R}_{\zeta_{y'}}\\
\rightarrow & \frac{1}{M^{2}}\mathcal{L}_{\zeta_{x}(t+1)}\mathcal{R}_{\zeta_{x}(t+1)}+\frac{1}{2M-1}\sum_{y:y=t\mod2}\left(K_{x,y}^{2}(t)+K_{x,y+1}^{2}(t)\right)\sum_{Q_{1}}\frac{1}{d_{Q_{1}}}\mathcal{P}_{Q_{1}}^{y,y+1}b_{Q_{1}}\\
 & -\frac{1}{2M-1}\sum_{y:y=t\mod2}2K_{x,y}K_{x,y+1}(t)\sum_{Q_{1}}\frac{1}{d_{Q_{1}}}\mathcal{P}_{Q_{1}}^{y,y+1}b_{Q}\\
\rightarrow & \frac{1}{M^{2}}\mathcal{L}_{\zeta_{x}(t+1)}\mathcal{R}_{\zeta_{x}(t+1)}+\frac{1}{2M-1}\sum_{y:y=t\mod2}\left(K_{x,y+1}-K_{x,y}\right)^{2}(t)\sum_{Q_{1}}\frac{1}{d_{Q_{1}}}\mathcal{P}_{Q_{1}}^{y,y+1}b_{Q_{1}},
\end{align*}
where $\zeta_{x}(t)=\sum_{y}K_{xy}\zeta_{y}$ and $K_{xy}$ is
the diffusion kernel of Eq.~\eqref{eq:diffusiononlattice}.

\begin{table}
\begin{tabular}{|c|c|c|c|c|}
\hline 
$v_{r}$ & $\left\langle v_{r}\right\rangle $ & $\left\langle v_{r}^{\dagger}v_{x}\right\rangle $ & $\frac{1}{M^{2}}\left\langle v_{r}^{\dagger}\zeta_{x}(t)v_{r}\zeta_{x}(t)\right\rangle $ & \tabularnewline
\hline 
\hline 
$1$ & 1 & $1$ & $\left(f^{2}+\frac{\sum_{y}K_{xy}^{2}(t)}{M}\left(1-f^{2}\right)\right)$ & \tabularnewline
\hline 
$Z$ & $f$ & $1$ & $\left(f^{2}+\frac{\sum_{y}K_{xy}^{2}(t)}{M}\left(1-f^{2}\right)\right)$ & \tabularnewline
\hline 
$\sigma^{\pm}$ & $0$ & $\frac{1\mp f}{2}$ & $\frac{1}{2}(1\pm f)\left[\frac{1}{M^{2}}\left\langle \zeta_{x}(t)\zeta_{x}(t)\right\rangle -K_{x[r]}\frac{2}{M}f^{2}-K_{x\left[r\right]}^{2}\frac{2}{M^{2}}\left(1-f^{2}\right)\right]$ & \tabularnewline
\hline 
 &  &  &  & \tabularnewline
\hline 
\end{tabular}

\caption{Useful expectation values for manipulating OTOC.}
\end{table}

\section{Equilibration in operator space}\label{App:Opequilibration}

In this appendix we use the superoperator formalism, developed in Sec.~\ref{Sec:supop_language}, to show that the expectation values of local superoperators, e.g. OTOCs, are at long times determined by a Gibbs ensemble on operator space, which reproduces the results established in Sec.~\ref{Sec:Saturation}. Consider a spin system with $L$ sites, each with on-site Hilbert space dimension $q=2$ (for concreteness). We want to time evolve the `density matrix' corresponding to a pure state in the space of operators, $\mathcal{P}_{V}(t)= \ket{\hat V(t)} \bra{\hat V(t)}$. It is convenient to consider initial operators which include a Gibbs factor, e.g., take an operator of form $\hat V = \hat \omega_{\mu}\hat O_0 \hat \omega_{\mu}$
where $\hat O_{0}=\hat Z_{0},\hat \sigma_{0}^{\pm}$ is a local Pauli matrix on site 0 and $\hat \omega_{\mu} \equiv e^{-\frac{\mu}{4} \hat Q}$. This
is useful for our purposes because the out-of-time-order part of the OTOC (the focus of our study) can be expressed as an expectation value
of a local superoperator with respect to such a $\mathcal{P}_V$ as
\begin{equation}
\braket{\mathcal{P}_V(t)|\mathcal{L}_{W_{r}^{\dagger}}\mathcal{R}_{W_{r}^{\phantom{\dagger}}}} = \braket{\hat V(t) |\mathcal{L}_{W_{r}^{\dagger}}\mathcal{R}_{W_{r}^{\phantom{\dagger}}}|\hat{V}(t)} =
\text{tr}\left(\hat \omega_{\mu}\hat O_{0}^{\dagger}\hat \omega_{\mu}\hat W_{r}^\dagger(t)\hat \omega_{\mu}\hat O_{0}\hat \omega_{\mu}\hat W_{r}^{\phantom{\dagger}}(t)\right).
\end{equation}

If we apply local two site U(1) random unitaries to such a spin system
for a very long time, we expect the system to scramble completely, such
that the time evolution is essentially a non-local random unitary operator
with conserved U(1) charge. Hence, at long times, we expect the average
density matrix to be that obtained by plugging $\mathcal{P}_V$
into Eq.~\eqref{eq:onegate_avg_4layers} for a unitary that acts on the whole chain. The result is a Haar averaged `density matrix' (on operator space) of the form 
\begin{align*}
\mathcal{P}_V\left(t_{\infty}\right) = \ket{\hat V_{\parallel}}\bra{\hat V_{\parallel}} +
\sum_{Q_{1}Q_{2}}\text{tr}\left(\hat P_{Q_{1}}\hat V_{\perp}^{\phantom{\dagger}}\hat P_{Q_{2}}\hat V_{\perp}^{\dagger}\right)\times\frac{\mathcal{L}_{P_{Q_{1}}}\mathcal{R}_{P_{Q_{2}}}-\frac{\delta_{Q_{1}Q_{2}}}{d_{Q_{1}}}\mathcal{P}_{Q_{1}}}{d_{Q_{1}}d_{Q_{2}}-\delta_{Q_{1}Q_{2}}},
\end{align*}
where we have separated $\hat V$ into orthogonal components $\hat V= \hat V_{\parallel}+ \hat V_{\perp}$,
with $\hat V_{\parallel}\equiv\sum_{Q}\frac{\hat P_{Q}\text{tr}\left(\hat P_{Q}\hat V\right)}{d_{Q}}$
and $\hat V_{\perp}=\hat V- \hat V_{\parallel}$, and used the notation $\mathcal{P}_{Q} \equiv \ket{\hat P_Q} \bra{\hat P_Q}$. In what follows, we consider the expectation value of a local superoperator --
for concreteness we will take a superoperator $\mathcal{L}_{W_{r}^{\phantom{\dagger}}}\mathcal{R}_{W_{r}^{\dagger}}$ where $\hat W_{r}$ is a local operator. When evaluated in the `state' $\mathcal{P}_V(t)$, this will have two separate contributions form the $\parallel,\perp$ components respectively. Let us deal first with the $\parallel$ component,
\begin{align*}
\text{tr}\left(\mathcal{P}_{V_{\parallel}}\left(t_{\infty}\right)\mathcal{L}_{W_{r}^{\phantom{\dagger}}}\mathcal{R}_{W_{r}^{\dagger}}\right) =  \text{tr}\left(\hat \omega_{2\mu}\hat O_{\parallel}^{\dagger}\hat W_{r}\hat \omega_{2\mu}\hat O_{\parallel}\hat W_{r}^{\dagger}\right) 
= \sum_{Q_{1}Q_{2}}e^{-\frac{\mu}{2}\left(Q_{1}+Q_{2}\right)}\text{tr}\left(\hat O_{\parallel}^{\dagger}\hat P_{Q_{1}}\hat W_{r}^{\phantom{\dagger}}\hat P_{Q_{2}}\hat O_{\parallel}\hat W_{r}^{\dagger}\right).
\end{align*}
It is readily verified by example that for two local observables, $\hat O$ and $\hat W_{r}$,
this sum is for large $L$ sharply peaked for $Q_{1,2}=\overline{Q}+\mathcal{O}(1)$
where $\overline{Q}=L/(1+e^{\mu})$. (The key observation here is that
local operators have $\mathcal{O}(1)$ charge under the adjoint action of $\hat Q$). This justifies replacing $\mathcal{P}_{V_{\parallel}}\left(t_{\infty}\right)$
with essentially any other distribution peaked in the same position.
A particularly simple choice is.
\begin{equation*}
\mathcal{P}_{V_{\parallel}}\left(t_{\infty}\right)\to\text{tr}\left(\hat V_{\parallel}^{\dagger}\hat V_{\parallel}^{\phantom{\dagger}}\right)\frac{\ket{e^{-\frac{\mu}{2}\hat Q}} \bra{e^{-\frac{\mu}{2}\hat Q}}}{Z_{\mu}},
\end{equation*}
where $Z_{\mu}=\text{tr}\left(e^{-\mu \hat Q}\right)$. 

The $\perp$ part of the density matrix takes form
\begin{equation*}
\mathcal{P}_{V_{\perp}}\left(t_{\infty}\right)=\sum_{Q_{1}Q_{2}}\text{tr}\left(\hat P_{Q_{1}}\hat V_{\perp}^{\phantom{\dagger}}\hat P_{Q_{2}}\hat V_{\perp}^{\dagger}\right)
\times\frac{\mathcal{L}_{P_{Q_{1}}}\mathcal{R}_{P_{Q_{2}}}-\frac{\delta_{Q_{1}Q_{2}}\mathcal{P}_{Q_{1}}}{d_{Q_{1}}}}{d_{Q_{1}}d_{Q_{2}}-\delta_{Q_{1}Q_{2}}}.
\end{equation*}
We again consider the expectation values of local superoperators (e.g.,
$\mathcal{L}_{W_{r}^{\phantom{\dagger}}}\mathcal{R}_{W_{r}^{\dagger}}$). Once again,
the sum is sharply peaked around $Q_{1,2}=\overline{Q}+\mathcal{O}(1)$ in the
large $L$ limit, i.e. 
\begin{align*}
\mathcal{P}_{V_{\perp}}\left(t_{\infty}\right)\sim \text{tr}\left(\hat V_{\perp}^{\dagger}\hat V_{\perp}^{\phantom{\dagger}}\right)\frac{\mathcal{L}_{P_{{\overline{Q}} + \lambda_{V}}}\mathcal{R}_{P_{{\overline{Q}}}}}{d_{\overline{Q} + \lambda_{V}}d_{\overline{Q}}}
\end{align*}
As before, this justifies replacing $\mathcal{P}_{V_{\perp}}\left(t_{\infty}\right)$
with a similar distribution peaked at the same charge,
\begin{align*}
\mathcal{P}_{V_{\perp}}\left(t_{\infty}\right) & \to \text{tr}\left(\hat V_{\perp}^{\dagger}\hat V_{\perp}^{\phantom{\dagger}}\right)\frac{e^{-\mu\left(\mathcal{L}_{Q}+\mathcal{R}_{Q}\right)}}{Z_{\mu}^{2}}.
\end{align*}
We reiterate that the above approximations are also only expected to hold weakly (i.e., when we calculate the expectation values of observables with an $\mathcal{O}(1)$ charge). 

In summary, our late time operator density matrix takes the form
\begin{equation}
\mathcal{P}_{V}\left(t_{\infty}\right)=\text{tr}\left(\hat V_{\parallel}^{\dagger}\hat V_{\parallel}^{\phantom{\dagger}}\right)\frac{\ket{e^{-\frac{\mu}{2}\hat Q}} \bra{e^{-\frac{\mu}{2}\hat Q}}}{Z_{\mu}}
+\text{tr}\left(\hat V_{\perp}^{\dagger}\hat V_{\perp}^{\phantom{\dagger}}\right)\frac{e^{-\mu\left(\mathcal{L}_{Q}+\mathcal{R}_{Q}\right)}}{Z_{\mu}^{2}}.\label{eq:finalform}
\end{equation}
This form, particularly the latter $\perp$ term, is nothing other
than a Gibbs ensemble for the superoperator conserved quantities
$\mathcal{L}_{Q},\mathcal{R}_{Q}$. In fact,
we could motivate~\eqnref{eq:finalform} using the
language standard to discussions of equilibration to the Gibbs ensemble.
Having identified $\mathcal{L}_{Q},\mathcal{R}_{Q}$
as the conserved local densities, we could have proposed an obvious ansatz of Gibbs form for the late time density matrix
\begin{equation}
\mathcal{P}_{V}^{\text{ansatz}}=\text{tr}\left(\hat V_{\parallel}^{\dagger}\hat V_{\parallel}^{\phantom{\dagger}}\right)\frac{\ket{e^{-\frac{1}{2}\eta_{\parallel}^{(1)}\hat Q}} \bra{e^{-\frac{1}{2}\eta_{\parallel}^{(2)}\hat Q}}}{Z_{\frac{1}{2}\eta_{\parallel}^{(1)} + \frac{1}{2}\eta_{\parallel}^{(2)}}}
+\text{tr}\left(\hat V_{\perp}^{\dagger}\hat V_{\perp}^{\phantom{\dagger}}\right)\frac{e^{-\eta_{\perp}^{(1)}\mathcal{L}_{Q}-\eta_{\perp}^{(2)}\mathcal{R}_{Q}}}{Z_{\eta_{\perp}^{(1)}}Z_{\eta_{\perp}^{(2)}}},\label{eq:ansatzforETH}
\end{equation}
and determined $\eta_{\perp,\parallel}^{(1,2)}$ via the conditions
depending on the initial state,
\begin{equation}
\frac{\braket{\hat V_{\perp}|\mathcal{L}_{Q}|\hat V_{\perp}}}{\braket{\hat V_{\perp}|\hat V_{\perp}}} =  \mathring{\text{tr}}\left(\frac{e^{-\eta_{\perp}^{(1)}\mathcal{L}_{Q}-\eta_{\perp}^{(2)}\mathcal{R}_{Q}}}{Z_{\eta_{\perp}^{(1)}}Z_{\eta_{\perp}^{(2)}}}\mathcal{L}_{Q}\right),\label{eq:genETHetaperp}
\end{equation}
\begin{equation}
\frac{\braket{\hat V_{\parallel}|\mathcal{L}_{Q}|\hat V_{\parallel}}}{\braket{\hat V_{\parallel}|\hat V_{\parallel}}} = \mathring{\text{tr}}\left(\frac{e^{-\eta_{\parallel}^{(1)}\mathcal{L}_{Q}-\eta_{\parallel}^{(2)}\mathcal{R}_{Q}}}{Z_{\frac{1}{2}\eta_{\parallel}^{(1)} + \frac{1}{2}\eta_{\parallel}^{(2)}}}\mathcal{L}_{Q}\right),\label{eq:genETHetaperp-1}
\end{equation}
and an otherwise identical pair of equations for $\mathcal{R}_{Q}$.
It is readily verified that for the choice of initial operator $\hat V=\hat \omega_{\mu}\hat O_{0}\hat \omega_{\mu}$, we get $\eta_{\perp,\parallel}^{(1,2)}=\mu$ as required, agreeing with our final result Eq.~\ref{eq:finalform}. 

These results point to an extension of the principle of thermalization to operator space.
Recall that for the usual notion of thermalization, if the time evolution
$U$ is completely ergodic (save the presence of $U(1)$ symmetry),
we expect (and have indeed argued in previous sections for random $U$) that local observables equilibrate according to
\begin{align}
\left\langle \hat{O}\left(t\rightarrow\infty\right)\right\rangle _{\psi} & =\frac{\text{tr}\left(e^{-\mu_{\psi}\hat Q}\hat{O}\right)}{\text{tr}\left(e^{-\mu_{\psi}\hat Q}\right)}\label{eq:regular_ergodicity}
\end{align}
in the thermodynamic limit. Here $\mu_{\psi}$ is determined for a
given state $\psi$ by balancing this equation for $\hat O=\hat Q$. As we have
found above, a similar notion of ETH occurs in operator space. One
uses the ansatz Eq.~\eqref{eq:ansatzforETH}, and determines
the chemical potentials $\eta_{\parallel,\perp}^{(1,2)}$ by ensuring
that the superoperator charge densities in the initial state agree
with that of the final state (see Eq.~\eqref{eq:genETHetaperp} and Eq.~\eqref{eq:genETHetaperp-1}). The analogy is especially apparent for the $\perp$ terms, where the ensemble is precisely a Gibbs distribution with respect to the superoperators.

\section{Solution of $\sigma^+ \sigma^+$ OTOC in the $\mu = \infty$ limit}\label{App:2pwalk}

In this appendix we show how the OTOC $\mathcal{F}^{\sigma_0^+ \sigma_r^+}_{\mu=\infty}$ (the only non-trivial OTOC in the $\mu\to\infty$ limit) can be understood in terms of a two-particle random walk of absorbing particles, and how this description gives rise to the two important qualitative features (double plateau structure and lack of ballistic light cone) shown in Fig.~\ref{fig:OTOC_pp_exact}. The considerations of this appendix apply also for higher dimensional random circuits, which should therefore also exhibit the same qualitative features.

As described in~\secref{Sec:mubig}, the OTOC $\mathcal{F}^{\sigma_0^+ \sigma_r^+}_{\mu=\infty}$, which is the zeroth order term in the perturbative expansion, is given by a process wherein the partition function is evaluated between boundary conditions that contain 2 particles. These boundary conditions are the following (using the notation of Fig.~\ref{fig:keldysh}):
\begin{itemize}
	\item At time $0$ there is a particle on site $0$ on layers ${\color{red} 2}{\color{blue} 1}$ and a second particle on site $s \neq 0$ on layers ${\color{red} 2}{\color{blue} 2}$
	\item At time $t$ there is a particle on site $0$ on layers ${\color{red} 2}{\color{blue} 2}$ and a second particle on site $s' \neq r$ on layers ${\color{red} 2}{\color{blue} 1}$. 
\end{itemize}

As long as the two particles in the initial state do not meet they each perform a random walk process of the type described in Eq.~\eqref{eq:randomwalk}. Upon meeting each other the two particles annihilate, since there is no matrix element with this specific set of incoming particles (see, e.g., Fig~\ref{fig:linetypes}). This means that the computation of the OTOC reduces to the following problem:

\emph{Given two random walkers, one that has to start at site $0$, and another which has to end up at site $r$ at time $t$, what is the probability that their paths avoid each other?}

The solution to this problem can be easily formulated in terms of single-particle transition probabilities, by noting that there is a one-to-one mapping between crossing paths of the two particles with a fixed set of starting and endpoints and \emph{arbitrary} paths where the two endpoints at time $t$ are interchanged. This mapping is simply given by reinterpreting the paths of the two particle by changing the last crossing into a reflection or vice versa (this is a simple case of the Lindstr{\"o}m-Gessel-Viennot Lemma, see Ref. \onlinecite{DeLuca16} and the references therein)~\cite{BCNote}. Using this trick, the solution is given by
\begin{equation}\label{eq:noncrossing}
\overline{\mathcal{F}^{\sigma_0^+ \sigma_r^+}_{\mu=\infty}}(t) = \left( \sum_{s'<r}\sum_{s>0} + \sum_{s'>r}\sum_{s<0} \right) \big[ K_{0,s'}(\tau) K_{s,r}(\tau) - K_{0,r}(\tau) K_{s,s'}(\tau) \big],
\end{equation}
where $K_{r_1,r_2}(t)$ is the probability of a single random walker travelling from site $r_1$ to $r_2$ in time $t$

The problem of calculating the OTOC thus reduces to solving a single-particle diffusion problem. This is easily done in an infinite system, with the result already stated in Eq.~\eqref{eq:diffusiononlattice}. Plugging this formula into Eq.~\eqref{eq:noncrossing} yields a solution shown in the right panel of Fig.~\ref{fig:OTOC_pp_exact}, which is a function of $r / \sqrt{t}$ and saturates to the value $\frac{1}{2} - \frac{1}{\pi}$ as $t^{-1/2}$. This saturation value is non-zero because in an infinite systems there is a finite probability that the two particles avoid each other for arbitrarily long times, i.e. by travelling in opposite directions. Note that the mapping of the $\mu=\infty$ OTOC to the random walk problem defined above is not restricted to 1D and we would end up with a similar counting of non-crossing paths for random circuits in higher dimensions. This means that the saturation value (which equals the probability of non-crossing paths) is even larger in those cases, as random walkers in higher dimensions have a larger probability of avoiding each other.

To get the full form of the OTOC, with eventual saturation to the second plateau at zero, one needs to solve the diffusion problem in a finite system with either periodic or reflecting boundaries. For a finite system of size $L$, and for times $t \gg L^2/D$ (where $D$ is the diffusion constant which is of $\mathcal{O}(1)$ in our case) the paths of the two particles have to cross eventually, and as a result the OTOC decays to zero. Here we focus on the case of reflecting boundaries, where the above way of counting crossing paths remains valid, although we checked numerically that the the results are similar for closed boundaries (the time signalling the end of the prethermal plateau is numerically larger in the case with open boundaries, reflecting that fact that particles can evade each other for longer). Instead of giving an exact solution on the lattice (which is nevertheless possible), we solve the same problem in the continuum, substituting the single particle transition probabilities with the solution of the continuum diffusion equation with reflecting boundaries,
\begin{align}
\partial_{t}\,K(x,t) & = D \partial_{x}^{2}\,K(x,t); \\
\partial_x\,K(x,t) &|_{x=0,L} = 0 \nonumber,
\end{align}
where we defined $K(x,t) \equiv K_{0,x}(t)$. This equation can be solved by doing an eigendecomposition of the operator $-D\partial_x^2$, using eigenstates with the appropriate boundary conditions, resulting in the single-particle propagator
\begin{equation}
K_{x,y}(t) = \frac{1}{L} \sum_{n\in\mathbb{Z}} e^{-\frac{\pi^2 Dtn^2}{2L^2}} \cos{\frac{\pi n x}{L}} \, \cos{\frac{\pi n y}{L}}.
\end{equation}
We can then approximate the OTOC by plugging this formula into Eq.~\eqref{eq:noncrossing}. At short times, when $Dt \ll L^2$ the resulting curve follows the result in an infinite system (which can be seen explicitly ba applying the Poisson summation formula the the above expression and then looking at the lowest order term in $\frac{L^2}{Dt}$) while at times $Dt \gg L^2$ it goes to zero as $\propto \exp(-\frac{\pi^2 D t}{2L^2})$.

\section{$\mathcal{F}^{Q_0 Q_r}$ and $\mathcal{F}^{Q_0\sigma_r^+}$ in the $\mu\gg 1$ limit}\label{App:Qp_lowT}

\begin{figure}[h!]
	\centering
	\includegraphics[width=0.37\columnwidth]{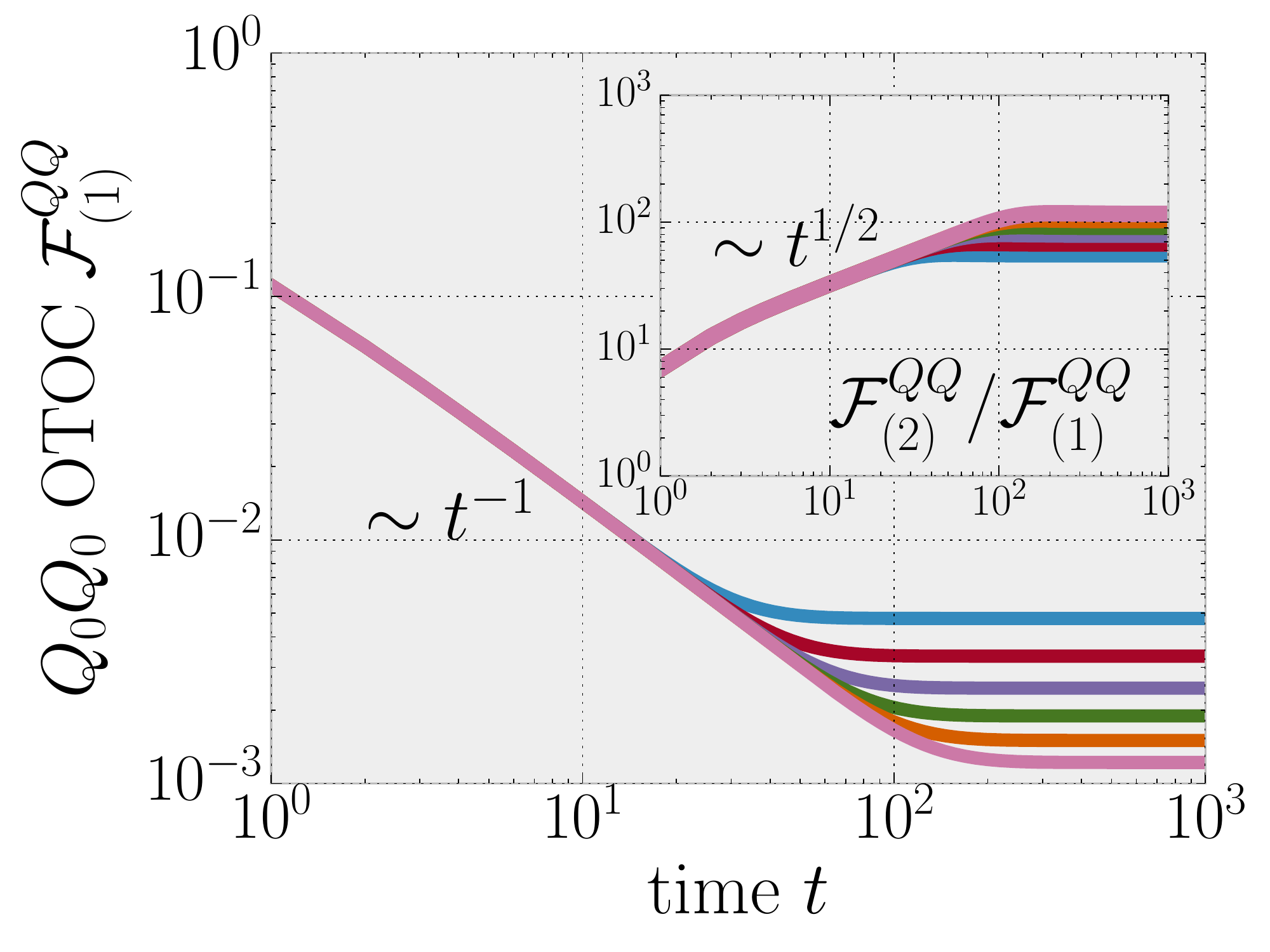}
	\includegraphics[width=0.3\columnwidth]{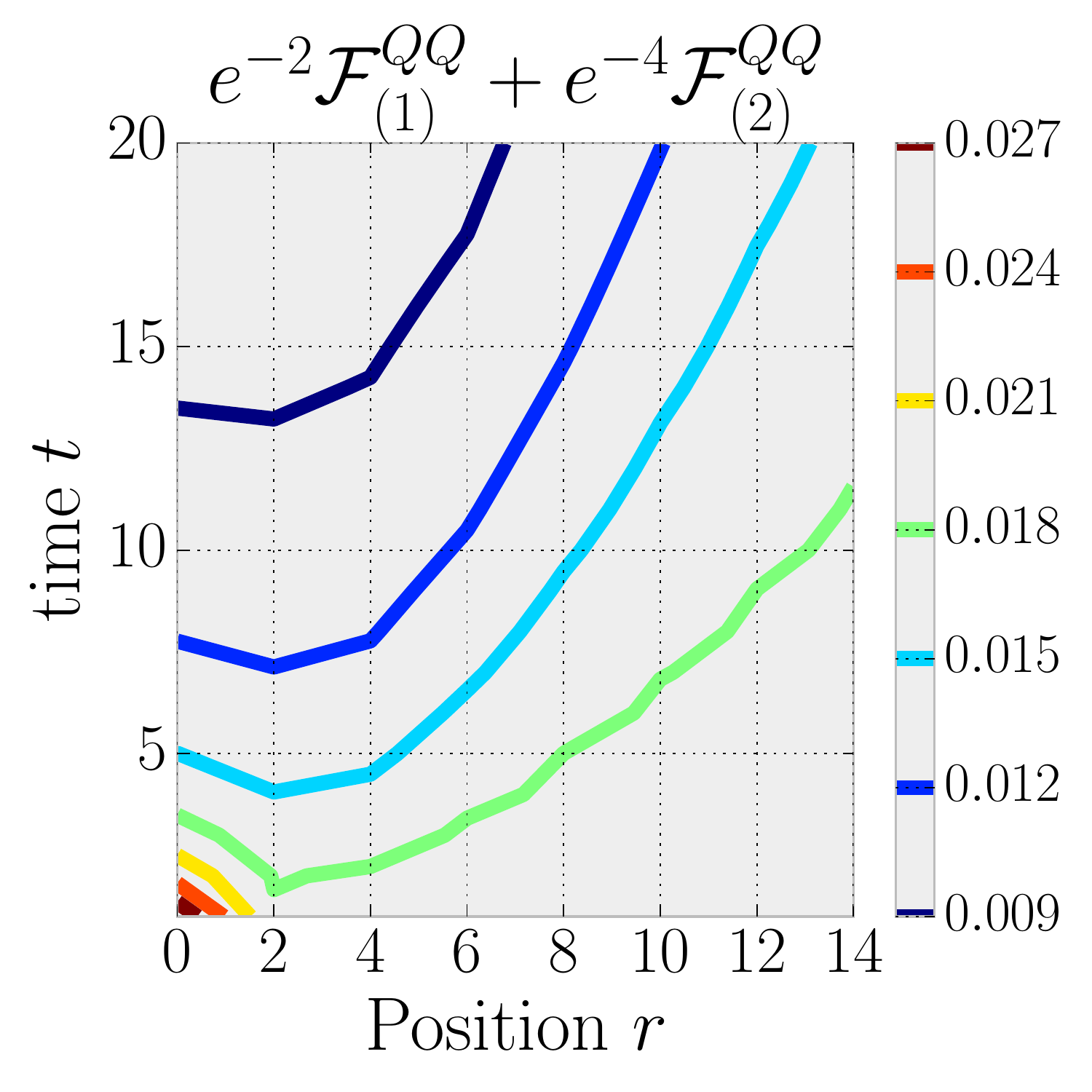}
	\caption{Left: First non-vanishing contribution to the $\hat Q_0\hat Q_0$ OTOC at order $e^{-\mu}$ for system sizes $L = 20, 24, \ldots ,40$. This term decreases as $1/t$ until saturation. Inset: ratio of the second and first order contributions increases as $\sqrt{t}$ and saturates to an $\mathcal{O}(L)$ value. Right: Contour lines of the $\hat Q\hat Q$ OTOC truncated at $\mathcal{O}(e^{-2\mu})$ for $\mu=2$, consistent with diffusively spreading OTOC front.}
	\label{fig:OTOC_qq_pert}
\end{figure}

Here we complement the results, presented in~\secref{Sec:mubig} for $\mu\gg 1$ behavior of the $\hat \sigma^+\hat \sigma^+$ OTOC, with analogous results for the $\hat Q\hat Q$ and $\hat Q\hat \sigma^+$ OTOCs. Looking at the first two terms in the expansion~\eqnref{eq:otoc_pert} for $\mathcal{F}^{QQ}_{\mu}$, evaluated as the partition function~\eqnref{eq:otoc_as_partfunc} with the appropriate boundary conditions, we find that both terms decay algebraically: the $\mathcal{O}(e^{-\mu})$ term as $t^{-1/2}$ and the $\mathcal{O}(e^{-2\mu})$ term as $t^{-1}$, such that the relative size of the second to the first term increases as $\sqrt{t}$, similarly to the $\hat \sigma^+\hat \sigma^+$ case presented in the main text. Eventually the ratio saturates to a value that is linear in system size. These results are shown in Fig.~\ref{fig:OTOC_qq_pert}. 

\begin{figure}[h!]
	\centering
	\includegraphics[width=0.37\columnwidth]{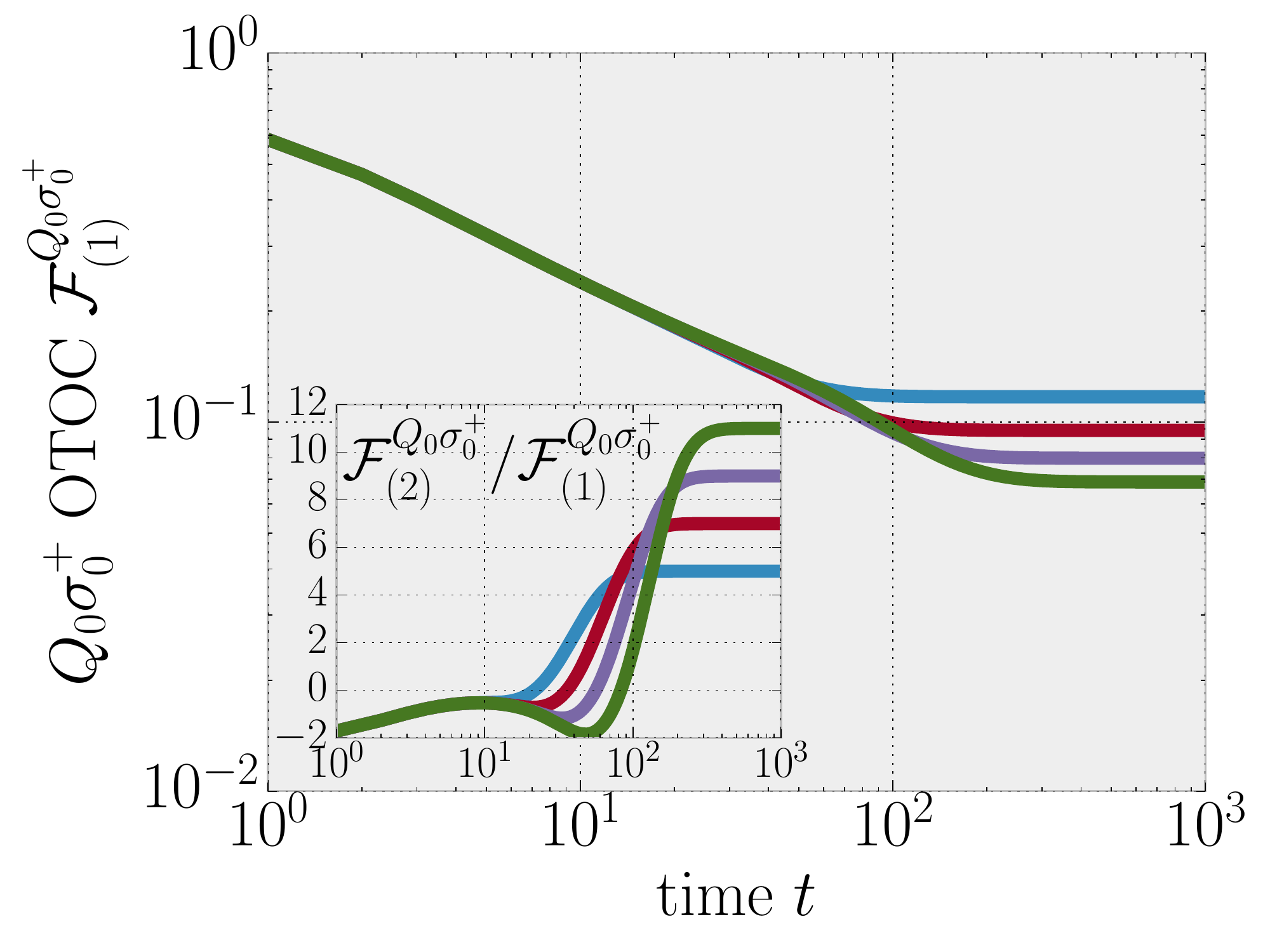}
	\includegraphics[width=0.3\columnwidth]{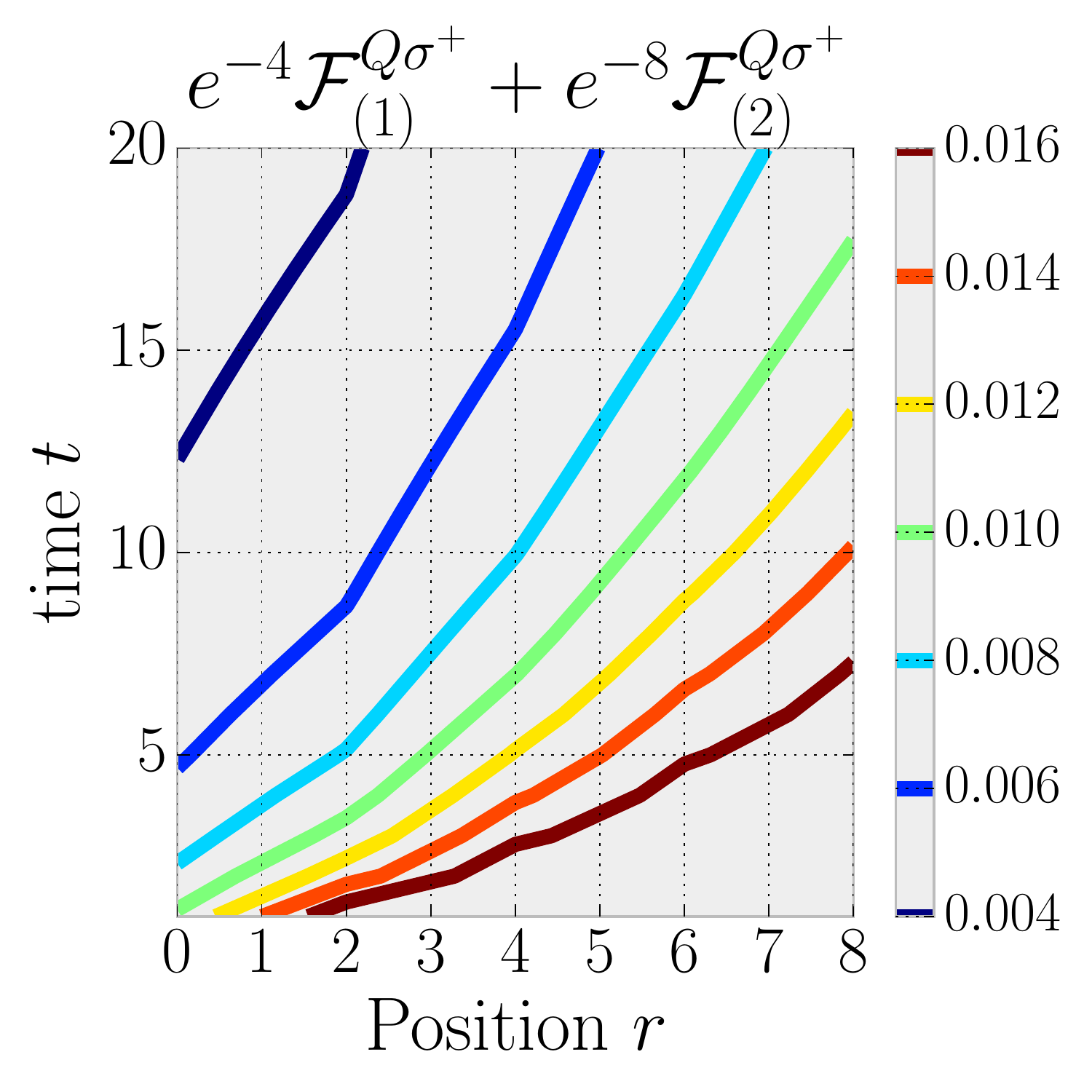}
	\caption{Left: First non-vanishing contribution to the $\hat Q_0\hat \sigma^+_0$ OTOC at order $e^{-\mu}$ for system sizes $L = 16,20,24,28$. This term decreases approximately as $t^{-0.4}$ until saturation. Inset: ratio of the second and first order contributions shows non-monotonic behavior. Right: Contour lines of the $\hat Q\hat \sigma^+$ OTOC truncated at $\mathcal{O}(e^{-2\mu})$ for $\mu=4$.}
	\label{fig:OTOC_qp_pert}
\end{figure}

Considering the $\mathcal{F}^{Q\sigma^+}_{\mu}$ OTOC, shown in the left panel of Fig.~\ref{fig:OTOC_qp_pert}, we observe that while the first order term has a simple algebraic decay, the second term has a somewhat more complicated structure than the ones presented in the main text. Rather than the ratio of the two terms simply increasing monotonically in time as a power law, it has an initial increase, a maximum and then a decreasing part. At an even later time scale, $t\sim L^2$, finite size effects become prominent which leads to an eventual increase to an $\mathcal{O}(L)$ value. Looking at the spatial structure (right panel of Figs.~\ref{fig:OTOC_qq_pert} and~\ref{fig:OTOC_qp_pert}) we observe diffusive spreading of both OTOCs, similarly to $\hat \sigma^+\hat \sigma^+$ and $QQ$ discussed in the main text.

\end{document}

%% file: OSWS_resubmit.bbl
%

%% file: OSWS_resubmit.bbl
\begin{thebibliography}{63}
\bibitem [{Dif()}]{DiffusionNote}%
  \BibitemOpen
  \href@noop {} {}\bibinfo {note} {This is in contrast with the biased
  diffusion process describing the dynamics of out-of-time-ordered correlators
  in random circuits without conserved quantities as described in previous
  work\cite{RvK17,Nahum17}. In that case the diffusion constant was found to be
  a decreasing function of the local Hilbert space dimension $q$. We emphasize
  that the diffusive broadening of OTOC front is different from the diffusion
  of conserved charge discussed in this section.}\BibitemShut {Stop}%
\bibitem [{Wea()}]{WeaklyCoupledNote}%
  \BibitemOpen
  \href@noop {} {}\bibinfo {note} {Here by weakly coupled we mean that a large
  $N$ limit is used to control the perturbation theory, sometimes implicitly by
  simply ignoring certain diagrams as in
  Ref.~\onlinecite{Aleiner16}.}\BibitemShut {Stop}%
\bibitem [{Dav()}]{DavidNote}%
  \BibitemOpen
  \href@noop {} {}\bibinfo {note} {Note that this is different from the
  operator entanglement of the time evolution operator itself, wich is another
  proposed measure of chaotic behavior~\cite{Zhou17}.}\BibitemShut {Stop}%
\bibitem [{Tim()}]{TimeOrderedNote}%
  \BibitemOpen
  \href@noop {} {}\bibinfo {note} {In fact, for $q=2$, even when the
  time-ordered part is non-trivial (i.e., time dependent), it can be computed
  exactly on average using Eq.~\eqref{eq:diffusiononlattice}}\BibitemShut
  {NoStop}%
\bibitem [{Non()}]{NonLocalOpNote}%
  \BibitemOpen
  \href@noop {} {}\bibinfo {note} {We expect that more generally these
  conclusions hold for any traceless operator $\hat V$ with a finite support
  that satisfies $\text{tr}(\hat Q^m \hat V) \neq 0$ for some positive integer
  $m$.}\BibitemShut {Stop}%
\bibitem [{Fac()}]{FactorizeNote}%
  \BibitemOpen
  \href@noop {} {}\bibinfo {note} {This is in contrast with the case without
  symmetries, where the sums in Eq.~\eqref{eq:def_gate_projectors} would go
  over all states in the two-site Hilbert space and thus
  factorize~\cite{Nahum17}.}\BibitemShut {Stop}%
\bibitem [{VBN()}]{VBNote}%
  \BibitemOpen
  \href@noop {} {}\bibinfo {note} {Note that the velocity $v_{B} \approx 0.45$
  observed here numerically is smaller than the butterfly velocity of a random
  circuit with the no symmetries and the same local Hilbert space dimension,
  derived to be 0.6 in Refs. \onlinecite{RvK17,Nahum17}.}\BibitemShut {Stop}%
\bibitem [{Cal()}]{CalligraphicNote}%
  \BibitemOpen
  \href@noop {} {}\bibinfo {note} {Note that the OTOCs, $\mathcal{C}$ and
  $\mathcal{F}$ are expectation values of superoperators.}\BibitemShut {Stop}%
\bibitem [{Sup()}]{SupopDiffusionNote}%
  \BibitemOpen
  \href@noop {} {}\bibinfo {note} {The presence of hydrodynamic tails can be
  made even more explicit in the case of the $\hat Z \hat \sigma^+$ OTOC by
  noting that the state $\ket{\hat \sigma_0^+}$ has an initial imbalance of the
  conserved quantities of the form $\braket{\hat \sigma_0^+ |\mathcal{L}_{Q_r}
  - \mathcal{R}_{Q_r}|\hat \sigma_0^+} = \delta_{r0}$. Since both
  $\mathcal{L}_Q$ and $\mathcal{R}_Q$ obey diffusion, this imbalance also
  relaxes diffusively, a fact picked up by the OTOC.}\BibitemShut {Stop}%
\bibitem [{BCN()}]{BCNote}%
  \BibitemOpen
  \href@noop {} {}\bibinfo {note} {Note that this mapping works for an infinite
  system or a finite system with open boundaries. The calculation is more
  complicated if we take closed boundary conditions, although this does not
  affect the results qualitatively.}\BibitemShut {Stop}%
\end{thebibliography}
